\documentclass[fleqn]{2026SCGE}
\usepackage{bm}
\usepackage{algorithm}
\usepackage{amsmath}
\usepackage{multirow}    
\usepackage{amssymb}
\usepackage{graphicx}
\usepackage{enumitem}
\usepackage{booktabs}
\usepackage{array}
\usepackage{subfigure}
\usepackage{soul}

\usepackage{ulem}

\begin{document}
\ensubject{subject}

\ArticleType{Article}
\SpecialTopic{SPECIAL TOPIC: }
\Year{2026}
\Month{}
\Vol{}
\No{}
\DOI{}
\ArtNo{000000}
\ReceiveDate{}
\AcceptDate{}

\title{Prospects for probing supermassive binary black holes in strongly lensed active galactic nuclei via microlensing effects}{Prospects for probing supermassive binary black holes in strongly lensed active galactic nuclei via microlensing effects}

\author[1,2]{Zekun Li}{}%
\author[1,2]{Changshuo Yan}{}%
\author[1,2]{Youjun Lu}{{luyj@nao.cas.cn}}%
\author[1]{Zhiwei Chen}{}%

\AuthorMark{Li Z K et al.}

\AuthorCitation{Li Z K, Yan C S, Lu Y J, Chen Z W}

\address[1]{National Astronomical Observatories, Chinese Academy of Sciences, Beijing 100101, China;}
\address[2]{School of Astronomy and Space Sciences, University of Chinese Academy of Sciences, Beijing 100049, China}


\abstract{Gravitational microlensing in strongly lensed active galactic nuclei (AGNs) provides a powerful probe of accretion-disk structure and may offer a unique opportunity to identify sub-parsec supermassive binary black holes (SMBBHs) through distinctive signatures. Upcoming wide-field surveys are expected to discover a large sample of strongly lensed AGNs, a fraction of which may host SMBBHs detectable via microlensing. In this paper, we employ semi-analytic models to assess the detection prospects of sub-parsec SMBBHs in lensed AGNs, focusing on systems with resolvable multiple images suitable for microlensing analysis. 
Accounting for galaxy-scale strong-lensing statistics, and including the cluster-lensing contribution as an optimistic statistical extension, we estimate that up to approximately 87 such systems could be spatially resolved by individual upcoming surveys, including the Vera C. Rubin Observatory (Rubin), Euclid, China's Space Survey Telescope (CSST), and the Nancy Grace Roman Space Telescope (Roman). For the Rubin Wide–Fast–Deep (WFD) survey, we estimate that periodic oscillations in microlensing light curves could identify $\sim0.12-0.68$ short-period ($<5$\,yr) binaries, whereas diagnostics based on the reconstructed size–wavelength relation apply to a broader population, potentially identifying $\sim0.4-5.3$ systems, depending on the model assumptions. CSST, Euclid, and Roman may also reveal up to $10.6$, $8.9$, and $1.2$ systems exhibiting size--wavelength features, and $2.3$, $1.9$, $0.3$ systems showing periodic oscillations, which should be regarded as imaging-selected candidates whose microlensing-based identification requires adequate time-domain follow-up, including cadence, monitoring baseline, photometric precision, and reliable time-delay measurements.
Our results show that strongly lensed AGNs provide a promising, though observationally demanding, pathway for finding candidate SMBBHs via microlensing.}



\maketitle


\twocolumn

\section{Introduction} 
\label{sec:intro}

Mergers of galaxies, typically hosting a central supermassive black hole (SMBH) \cite{2013ARA&A..51..511K}, naturally lead to the formation of supermassive binary black holes (SMBBHs) in the center of remnant galaxies \cite{1980Natur.287..307B, 2002MNRAS.331..935Y}. Gas-rich galaxy mergers can funnel substantial amounts of gas into regions around each of the two SMBHs in the center of progenitors or SMBBHs in the remnants and trigger nuclear activity. Dual-active galactic nuclei (AGNs) formed through galaxy mergers have been found with various observational techniques on kpc scales \cite{2003ApJ...582L..15K,2010ApJ...715L..30L,2011ApJ...735...48S, 2022NatAs...6.1185M, 2023Natur.616...45C, 2025CQGra..42k3001L}, which may evolve to SMBBHs on sub-pc scale and finally coalesce. However, conclusive observational evidence that AGNs host sub-pc SMBBHs remains elusive and difficult to obtain \cite{2022LRR....25....3B, 2023arXiv231016896D}. 

Over the years, a wide range of techniques have been proposed and utilized to hunt for SMBBHs \cite{yi+25}, including double-peaked or asymmetric broad emission lines (BELs) \cite{2012ApJS..201...23E, 2014ApJ...789..140L, 2021ApJ...910..101J}, periodicity in light curves \cite{1988ApJ...325..628S,2008Natur.452..851V,2015MNRAS.453.1562G,2015Natur.525..351D,2025arXiv250506656F,2025arXiv250516884H}, periodic self-lensing signals \cite{2018MNRAS.474.2975D}, and characteristic spectral energy distribution (SED) features induced by the unique triple-disk scenario \cite{1996ApJ...467L..77A,2002ApJ...567L...9A,2012ApJ...761...90G,2015ApJ...809..117Y,2023ARA&A..61..517L,lyl+23}. Among these, variability in AGN electromagnetic emission has attracted particular attention in recent years, studied by theoretical estimations \cite{2019MNRAS.485.1579K,2025arXiv250821510C} and fueled by observations \cite{2015Natur.525..351D, 2015MNRAS.453.1562G,2016MNRAS.463.2145C,2023MNRAS.526.1588Z,2025arXiv250506656F,2025arXiv250516884H}. However, confirming candidates as genuine SMBBHs against intrinsic AGN variability or complex accretion physics remains difficult \cite{2018ApJ...856...42S,2023arXiv231016896D,2025arXiv250910601E}. This motivates the exploration of alternative approaches.

Gravitational lensing offers a fundamentally different avenue for revealing close SMBBHs. In lensed AGNs with multiple images, strong lensing magnifies and spatially separates the emission from the background source, enabling precise comparisons of flux and variability between different images. Owing to the high microlensing optical depth, gravitational microlensing by stars in the lensing galaxy is expected to affect at least one image of a lensed AGN \cite{1986ApJ...301..503P,2012A&A...544A..62S,2024SSRv..220...14V}, and can be used to identify SMBBHs \cite{2014ApJ...784..100Y,2022A&A...668A..77M}. By isolating microlensing-induced variations through comparisons among macro images, microlensing light curves provide a sensitive probe of the inner accretion-disk structure of the source \cite{1990ApJ...358L..33W, 2004ApJ...605...58K, 2014ApJ...784..100Y, 2021A&A...653A.121S, 2024SSRv..220...14V}. Ref.~\cite{2014ApJ...784..100Y} first proposed that microlensing light curves could probe sub-pc SMBBHs, either through periodic variations induced by the binary's orbital motion or through departures from the standard size-wavelength relation. Recent identification of the close SMBBH candidate Q J0158-4325 \cite{2022A&A...668A..77M} based on microlensing periodicity highlights the potential of this approach. The number of SMBBHs identifiable with this method depends on both the total number of lensed AGNs that can be detected by galaxy surveys and the fraction of SMBBH systems among those detectable lensed AGNs, which can be monitored over long terms for identifying microlensing events. 

Next-generation wide-field surveys will uncover unprecedented numbers of strongly gravitationally lensed AGNs, dramatically expanding the discovery space for these lensed SMBBHs. The Vera C. Rubin Observatory (Rubin; i.e., the Large Synoptic Survey Telescope; LSST) \cite{2019ApJ...873..111I}, the China's Space Survey Telescope (CSST; \cite{Zhan2021}), Euclid \cite{2022A&A...662A.112E} and the Nancy Grace Roman Space Telescope (Roman) \cite{2019BAAS...51c.341D} are expected to identify thousands of strongly lensed AGNs \cite{2010MNRAS.405.2579O, 2015ApJ...811...20C, 2024MNRAS.533.1960C}. A fraction of these lensed AGNs are expected to host sub-pc SMBBHs and frequently experience microlensing, providing a unique opportunity to identify the signatures discussed above. However, the expected abundance of strongly lensed SMBBHs and the detectability of their multiple images given realistic survey sensitivities and resolutions remain unconstrained. Quantifying the detection prospects of such systems is therefore crucial for estimating the number and  abundance of close SMBBHs that can be identified through microlensing signatures in upcoming surveys. 

In this paper, we investigate the prospects of detecting SMBBHs in strongly lensed AGNs, with a  focus on systems resolvable by future facilities like Rubin, CSST, Euclid, and Roman. The paper is organized as follows. In Section~\ref{sec:method}, we describe our modeling of the SMBBH population, their radiative properties, and the adopted strong-lensing statistics. In Section~\ref{sec:results}, we present our predictions for SMBBHs in lensed AGNs, focusing especially on systems with two or more resolvable strong-lensing images that are detectable in upcoming surveys. Specifically, we discuss the probability that these SMBBHs in lensed AGNs can be probed via microlensing. We conclude with a summary and discussion in Section~\ref{sec:discussion}. 

Throughout the paper, a flat $\Lambda$CDM cosmology with parameters $\Omega_{\rm{m}}= 0.3$, $\Omega_{\Lambda}= 0.7$, and $H_0=70$\,km\,s$^{-1}$\,Mpc$^{-1}$ \cite{2016A&A...594A..13P} is adopted. 

\section{Methods} 
\label{sec:method}

In this section we describe our procedure for modeling the population and radiative properties of sub-parsec SMBBHs. We then outline the framework for estimating their strong-lensing rates, focusing particularly on systems whose multiple images are resolvable by future surveys.

\subsection{Close SMBBH population} 
\label{subsec:mocksamples}

We construct a mock population of SMBBHs by adapting the methodology of Ref.~\cite{lyl+23}, in which the formation and evolution of active SMBBHs are modelled analytically, starting from gas-rich galaxy merger, followed by the orbital evolution of SMBBHs in merged galaxies. This approach yields a sample spanning a broad range of orbital periods and evolutionary stages. 

Our model begins with the comoving number density of merging galaxies. We combine the merger rate per galaxy (MRPG), $N_{\rm{mrg}}(M_{\rm{tot}}(M_{*}),q_{\rm{gal}},z)$, from the Illustris simulations \cite{2015MNRAS.449...49R}, with the observed and simulated galaxy stellar mass function (GSMF), $\Phi(M_{*},z)$ \cite{2017MNRAS.471.3098L,2023A&A...677A.184W}. The merger rate density per unit cosmic time per redshift is then obtained by convolving the MRPG with the GSMF and multiplying by the comoving volume element:
\begin{equation}
\begin{aligned}
\frac{d^4 N_{\mathrm{gal}}(z,M_{*},q_{\mathrm{gal}})}{dz dM_{*} dq_{\mathrm{gal}} dt}=& \frac{d^2N_{\mathrm{mrg}}(M_{\mathrm{tot}}(M_{*}),q_{\mathrm{gal}},z)}{dq_{\mathrm{gal}} dt}\\
&\times\Phi(M_{*},z)\times\frac{dV}{dz},
\label{eq:dNdt}
\end{aligned}
\end{equation}
where $V$ denotes the comoving volume, $M_{\rm{tot}}(M_{*})\simeq(1+q_{\rm{gal}})M_{*}$ is the mass of the merger remnant galaxy with $q_{\rm gal}$ representing the mass ratio of the two progenitor galaxies. 

To model SMBBHs in a gas-rich environment, we incorporate galaxy morphology according to the observed distribution in the general galaxy population \cite{2010MNRAS.404.2087B,2011ApJ...738...92Y}. The morphological classification is crucial as it dictates the dependence of the SMBH–host galaxy relation on galaxy type. We relate the galaxy stellar mass $M_{*}$ to bulge mass $M_{\rm{bulge}}$ assuming a log-normal distribution with a constant scatter of 0.1 dex in log $M_{\rm{bulge}}$ for different galaxy types \cite{2011ApJ...738...92Y}; see also \cite{2009ApJ...696..411W} and \cite{2015MNRAS.447.2772R}, as
\begin{equation}
\bar M_{\rm{bulge}}\left(M_*\right)= \begin{cases} M_*, & \text { for elliptical, } \\
0.7 M_*, & \text { for S0, } \\
0.2 M_*, & \text { for Sab, } \\
0.05M_*, & \text { for Scd/Irr. }\end{cases} 
\label{eq:bulgemass}
\end{equation}

For the black hole-bulge mass relation ($M_{\rm{BH}}-M_{\rm{bulge}}$), we consider both a redshift-independent form and a redshift-dependent form to explore plausible black hole growth histories and assess modeling uncertainties. The relation can be expressed as
\begin{equation}
\log M_{\rm{BH}}=\tilde{\gamma}+\tilde{\omega} \log (1+z)+\tilde{\alpha} \log M_{\rm{bulge}, 11}, 
\label{eq:massrelation}
\end{equation}
where $M_{\rm{bulge}, 11}$ is the bulge mass in units of $10^{11} M_{\odot}$, and an intrinsic scatter of $\tilde{\epsilon}$\,dex of the relation is assumed. For the redshift-dependent case, we adopt the local empirical relation of \cite{2013ARA&A..51..511K}, with $(\tilde{\gamma},\tilde{\omega},\tilde{\alpha},\tilde{\epsilon})=(8.69,1.99,1.00,0.31)$; for the redshift-independent case, we adopt $(8.69,0,1.17,0.29)$.  
In the redshift-dependent case, the redshift evolution slope $\omega=1.99$ is adopted according to Ref.~\cite{2024ApJ...974..261C} obtained from the constraints by the gravitational wave background signals \cite{2023ApJ...951L...8A, 2023A&A...678A..50E, 2023ApJ...951L...6R, 2023RAA....23g5024X,2025MNRAS.536.1489M}. The resulting populations derived from these two scenarios are labeled BBH-nonevol and BBH-zevol, respectively. The redshift evolution of the $M_{\rm BH}$–$M_{\rm bulge}$ relation has been widely studied in recent works \cite{2025MNRAS.536.1489M}, and our results based on the two models provide a broad range of predictions consistent with these studies. We note that this paper primarily considers the impact of uncertainties in the $M_{\rm BH}$–$M_{\rm bulge}$ relation on the predicted SMBBH numbers, while other sources of uncertainty, including the galaxy merger rate, binary evolution, and black hole mass growth, are discussed in Section~\ref{sec:discussion}.

From the above relations and the galaxy merger rate density, we derive the formation rate density of SMBBHs by marginalizing over $M_{*}$ and $q_{\rm{gal}}$. The Eddington ratio distribution, $P(f_{\mathrm{Edd}})$, follows the Illustris simulation results \cite{2015MNRAS.452..575S}, truncated at a maximum $f_{\rm{Edd,max}}=1$ for consistency. Because this Illustris-based distribution is derived from a cosmological simulation including galaxy mergers, black-hole growth through gas accretion and coalescence, and AGN feedback, the mock SMBBHs are assigned different activity levels through their sampled Eddington ratios, rather than assuming that all systems are equally active.

We convert the SMBBH producing rate into the comoving number density per unit logarithmic semi-major axis, $\Delta \rm{log}a$, by modeling the binary evolution and calculating the shrinking speed of binary separation $da/dt$ \cite{2009ApJ...700.1952H}. Generally speaking, SMBBHs are driven toward coalescence by a combination of processes such as dynamical friction, stellar scattering, gaseous disk interactions, and gravitational wave (GW) radiation \cite{1980Natur.287..307B,2002MNRAS.331..935Y,2017MNRAS.464.3131K,2020MNRAS.495.4681I,2020ApJ...897...86C,2022LRR....25....3B}. 

At sub-parsec separations, activated SMBBHs are expected to be surrounded by a circumbinary disk with two inner mini-disks around each black hole—a triple-disk configuration particularly relevant for producing distinctive radiative and lensing signatures. Therefore, we focus on sub-parsec systems with $a < 10^4 R_{\rm{g}}$ (where the circumbinary disk remains stable; see \cite{2009ApJ...700.1952H}), which are typically in the viscous-driven or GW-driven regime. 

For the binary evolution driven by interaction with viscous disk and GW emission, we adopt the framework of Ref.~\cite{2009ApJ...700.1952H}. In this model, orbital decay due to viscous interactions, $t_{\mathrm{visc}}(a)$, is estimated by analogous with Type 2 planetary migration (see Eqs. 26a, b, c in \cite{2009ApJ...700.1952H}). The evolution of binary separation $da/dt$ is determined by the competing effects of viscous interactions and GW emission in dominating the energy dissipation, i.e.,
\begin{equation}
\frac{1}{t_{\mathrm{evol}}(a)} \simeq \frac{1}{t_{\mathrm{GW}}(a)} +\frac{1}{t_{\mathrm{visc}}(a)} .
\end{equation}

Given significant uncertainties in gas-driven orbital decay \cite{2002ApJ...567L...9A,2009ApJ...700.1952H,2012MNRAS.420..705T,2012MNRAS.427.2660K,2017MNRAS.469.4258T,2020ApJ...900...43T,2021MNRAS.507.1458F,2022A&A...660A.101P,2023MNRAS.522.2707S,2024MNRAS.534.2609S}, we also consider a comparison model where evolution is driven solely by GW emission. This GW-only case is used as a reference model for the limiting case in which gas-driven hardening is inefficient. In this model,
\begin{equation}
t_{\rm{GW}}(a) = \frac{5}{64} \frac{c^5 a^4}{G^3 \mu_{12} M^2_{\rm{BBH}}},
\end{equation}
where $\mu_{12}$ represents the reduced mass of the binary system.

We further introduce a conservative CBD-filtered scenario to account for the possibility that some binaries may slow down, stall, or even migrate outward during the disk-dominated stage, and therefore fail to reach the small separations relevant to our diagnostics (e.g., \cite{2020ApJ...899..126S,2023LRR....26....2A}). This treatment is motivated by hydrodynamical simulations showing that circular binaries can experience positive disk torques and outward migration for a range of mass ratios (e.g., \cite{2019ApJ...875...66M,2020ApJ...889..114M}). In particular, \cite{2020ApJ...901...25D} found positive torques for circular binaries with $q\gtrsim 0.05$ in relatively warm/thick circumbinary discs, while \cite{2023MNRAS.522.2707S} showed that circular binaries with $q\gtrsim0.2$ can expand rather than inspiral in part of their parameter space. Therefore, in the CBD-filtered scenario, we adopt a moderate threshold and assume that systems with $q\gtrsim0.2$ do not contribute to the short-period close-SMBBH sample. The remaining systems are evolved using the same prescription as in the fiducial viscous+GW model. This scenario is a conservative test of how inefficient inward migration may reduce the number of close SMBBHs accessible to our diagnostics.

Using the evolutionary residence time, we convert the SMBBH merger rate, derived from galaxy merger rates and black hole-host scaling relations into the comoving number density of SMBBHs per unit logarithmic semi-major axis $\log a$, i.e.,
\begin{equation}
\begin{aligned}
&\frac{d^5 \mathcal{N}(z, M_{\mathrm{BBH}}, q, a, f_{\rm Edd})} 
{dz d\log M_{\mathrm{BBH}} d\log q d\log a df_{\rm Edd}} \\
& = \frac{dt}{d\log a} \frac{d^4 N(z, M_{\mathrm{BBH}}, q)}
{dz d\log M_{\mathrm{BBH}} d\log q dt} P(f_{\mathrm{Edd}}).
\label{eq:dNdlogr}
\end{aligned}
\end{equation}
The SMBBH merger rate $\frac{d^4 N(z, M_{\mathrm{BBH}}, q)}{dz\, d\log M_{\mathrm{BBH}}\, d\log q\, dt}$ here is derived by mapping the galaxy merger rate in Equation~\eqref{eq:dNdt} into the BBH parameter space using the $M_{\rm BH}$–$M_{\rm bulge}$ relation together with the probabilistic connection between bulge and total galaxy stellar mass for different galaxy types as described earlier (Eqs.~\ref{eq:bulgemass} and \ref{eq:massrelation}). 

Using a Monte Carlo approach, we sample the resulting distribution over the following ranges: redshift $z \in (0,7)$, total binary mass $M_{\rm{BBH}} \in (10^6, 10^{11}) M_{\odot}$, mass ratio $q \in (0.01,1)$, Eddington ratio $f_{\rm{Edd}} \in (0.01,1)$, semi-major axis $a \in (10, 10^4) R_{\rm{g}}$, and inclination $\cos i$ uniformly distributed. This mock population of close SMBBHs serves as the input for calculating their radiative signatures, strong-lensing probabilities, and detectability. 

\subsection{SMBBH radiation} 
\label{subsec:BBHradiation}

With the mock population of close SMBBHs established, we now introduce the accretion and radiation model adopted for SMBBHs. We employ the triple-disk model, which is widely used to model SMBBH accretion flows \cite{2012ApJ...761...90G,2012MNRAS.420..860S,2015MNRAS.446L..36F,2015ApJ...809..117Y,2023arXiv231016896D}. The model consists of two mini-disks, each surrounding one black hole, and a circumbinary disk. The secondary black hole opens a gap/cavity between the mini-disks and the circumbinary disk, with accretion streams bridging the cavity but contributing little to the total luminosity \cite{1996ApJ...467L..77A,2008ApJ...672...83M,2014ApJ...783..134F,2016ApJ...832...22S,2020ApJ...901...25D,2023ARA&A..61..517L}. The total emission of the SMBBH system is taken as the sum of the contributions from the circumbinary disk and the two mini-disks. 

We set the outer radius of the mini-disks as $R_{\rm{out}}(x) = 0.5 R_{\rm{RL}}(x)$, where $R_{\rm{RL}}(x)$ is the Roche lobe radius \cite{1983ApJ...268..368E} and $x = q$ or $1/q$ for the secondary and primary black holes, respectively. The inner edge of the circumbinary disk is approximated as $R_{\rm{c}} \approx a/(1+q) + 2 R_{\rm{H}}$ \cite{2014ApJ...784..100Y}, where $R_{\rm{H}} \approx 0.69 q^{1/3} a$ is the Hill radius of the secondary black hole. 
We adopt a fixed inner boundary for each mini-disk, $R_{\rm in,k}=3.5GM_k/c^2$ for $k=1,2$, corresponding to a radiative efficiency $\epsilon\simeq0.1$ (suggested by observations, e.g., \cite{2002MNRAS.335..965Y,2008ApJ...689..732Y}), in the standard thin disk model. 

The accretion allocation follows the commonly adopted prescription from hydrodynamic simulations \cite{2020ApJ...901...25D}: 
$
\dot{M}_{\rm{BH},1}=\dot{M}_{\rm{BH},2}(0.1+0.9q),
$
valid for $q > 0.01$. If the secondary BH (which receives a larger mass supply under our assumption) exceeds the Eddington limit, we modify the disk emission based on Ref.~\cite{2006ApJ...648..523W} that covers the supercritical accretion adopting slim disk scenario, rather than the standard thin disk \cite{1973A&A....24..337S} in the remaining cases. We also account for the radiatively inefficient regime of the primary disk (due to the preferential accretion allocation) when its accretion rate falls below $0.01$ \cite{2014ARA&A..52..529Y,2017MNRAS.467..898Q}. Under these assumptions, we calculate the temperature distribution of the three disk components based on either the standard thin disk or slim disk model. 

The effective temperature profile $T_{\rm eff}(R)$ is computed for each disk component according to the chosen accretion model. Disk emission is treated as blackbody radiation as
\begin{equation}
F_{\rm{\lambda}}=\frac{1}{ D_{\rm{L}}^2}\int_{R_{\rm in}}^{R_{\rm out}} \frac{2 h c^2 \cos i / \lambda^5}{\exp \left[h c / \lambda k_{\rm{B}} T_{\rm eff}(R)\right]-1} 2\pi R dR, 
\end{equation}
where $h$ is the Planck constant, $k_{\rm{B}}$ is the Boltzmann constant and $i$ is the inclination angle sampled uniformly in $\cos i$. 
The integration limits $R_{\rm in}$ and $R_{\rm out}$ are taken from the corresponding disk component described above.
Cosmological redshift is included when integrating the flux to obtain the observed $r$-band magnitudes in the observer’s frame. Applying this model, we compute the $r$-band magnitudes for our mock population. 
For illustration, the distribution of close SMBBHs brighter than $r=25$\,mag resulting from the BBH-nonevol model is shown in Figure~\ref{fig:fig1}. 

\subsection{Lensing statistics} 
\label{subsec:stronglensing}

\begin{figure}
\centering
\includegraphics[width=1\linewidth]{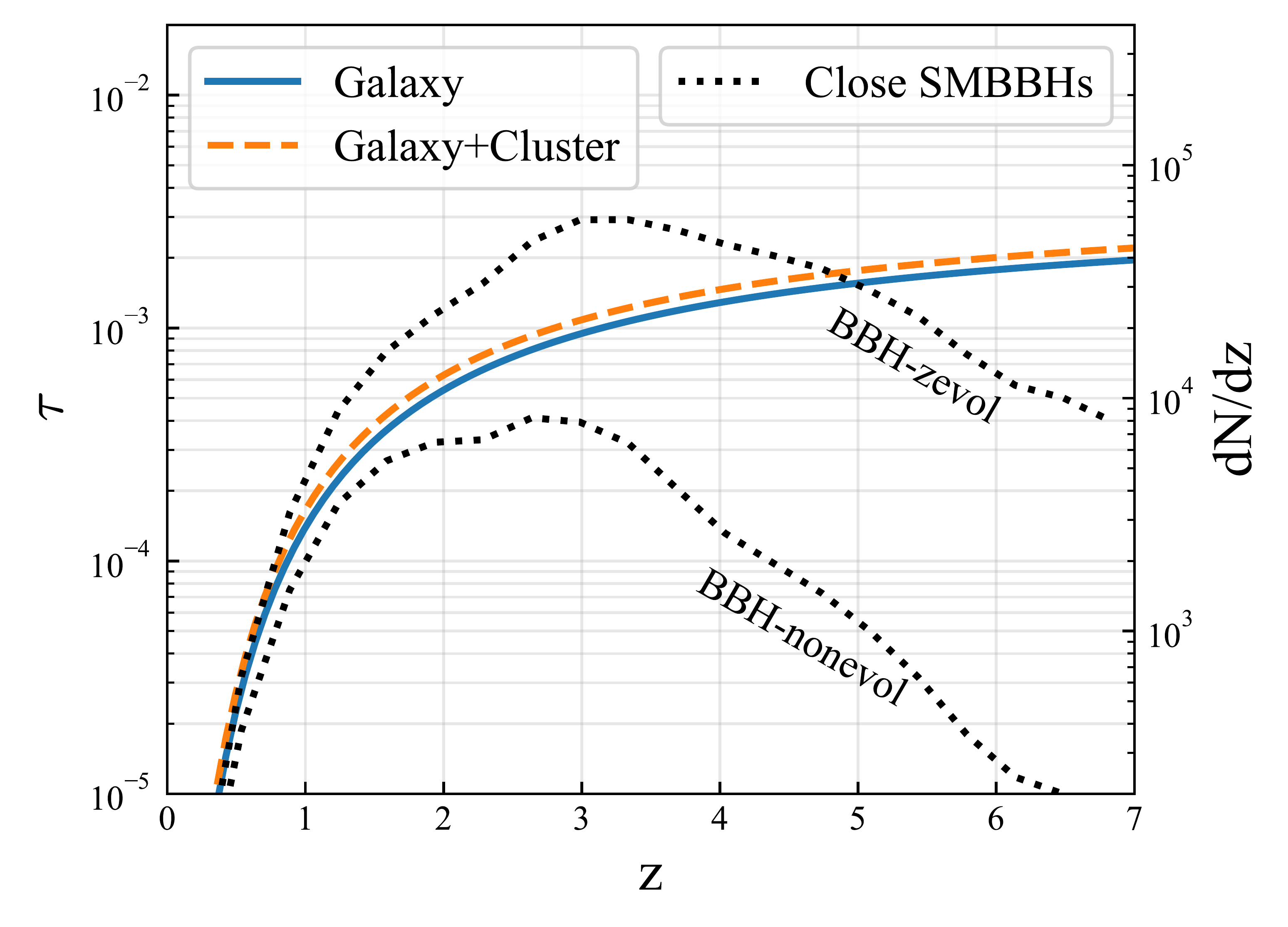}
\caption{
Optical depth $\tau$ and the number distribution of close SMBBHs against redshift. The blue solid curve shows the optical depth for galaxy-galaxy strong lensing modeled with SIE \cite{2024ApJ...962....3C}. The orange dashed curve represents the optical depth after including the cluster lensing contribution estimated with a halo-mass-function-based spherical NFW+SIS proxy following Ref. \cite{2024ApJ...977...64C}. The black dotted curves illustrate the number distributions of close SMBBHs (with $r$-band magnitude $<25$) against redshift resulting from the BBH-nonevol and BBH-zevol models, respectively. 
}
\label{fig:fig1}
\end{figure}

To evaluate the probability that close SMBBHs generate resolvable multiple images, we perform forward strong-lensing simulations following the procedures introduced in Refs.~\cite{2022ApJ...940...17C, 2024ApJ...962....3C}. In our framework, close SMBBHs can be strongly lensed by foreground galaxies, with the probability characterized by the optical depth $\tau(z)$. Galaxy–galaxy strong lensing is mainly caused by elliptical galaxies, e.g., \cite{2010MNRAS.405.2579O,2015ApJ...811...20C}, making the singular isothermal ellipsoid (SIE) an appropriate choice for the lens model. 
The differential lensing probability is jointly determined by the redshift of the foreground galaxy, $z_{\rm l}$, its velocity dispersion, $\sigma_{\rm v}$, the lens axis ratio, and the external shear. Following Refs.~\cite{2022ApJ...940...17C,2024ApJ...962....3C}, we write

\begin{equation}
\label{eq:dtau}
\begin{aligned}
\frac{d^{2}\tau(z_{\rm s})}{d\sigma_{\rm v}\,dz_{\rm l}}
=&\,
\frac{1}{4\pi}
\frac{dn(\sigma_{\rm v}, z_{\rm l})}{d\sigma_{\rm v}}\,
\frac{dV(z_{\rm l})}{dz_{\rm l}}
\int de\,P_e(e) \\
&\times
\int d\gamma_1\,d\gamma_2\,P_\gamma(\gamma_1,\gamma_2)\,
S_{\rm cr}(\sigma_{\rm v},z_{\rm l},z_{\rm s},e,\gamma_1,\gamma_2).
\end{aligned}
\end{equation}

Here, $S_{\rm cr}$ denotes the multiple-imaging cross section for the adopted SIE plus external shear lens model. It depends on the lens velocity dispersion $\sigma_{\rm v}$, the lens and source redshifts $(z_{\rm l},z_{\rm s})$, the lens axis ratio $e$, and the two components of the external shear $(\gamma_1,\gamma_2)$. The distributions $P_e(e)$ and $P_\gamma(\gamma_1,\gamma_2)$ describe the lens-axis-ratio and external-shear distributions, respectively.
The term $dn(\sigma_{\rm v}, z_{\rm l})/d\sigma_{\rm v}$ represents the velocity dispersion function (VDF) of the foreground galaxy lenses, which quantifies the comoving number density of lensing galaxies with velocity dispersions between $\sigma_{\rm v}$ and $\sigma_{\rm v}+d\sigma_{\rm v}$. $d V\left(z_{\rm l}\right)$ denotes the comoving volume within the redshift interval $z_{\rm l}$ to $z_{\rm l}+dz_{\rm l}$.

We adopt the VDF described by a simple Schechter function, e.g., \cite{2007ApJ...658..884C, 2013JCAP...10..022P},
\begin{equation}
\frac{d n\left(\sigma_{\rm{v}}, z_{\rm l}\right)}{d \ln \sigma_v}
=n_{\rm{z}}(z_{\rm l}) \frac{\beta}{\Gamma(\alpha / \beta)}
\left(\frac{\sigma_{\rm{v}}}{\sigma_{\rm{z}}(z_{\rm l})}\right)^\alpha
\exp \left[-\left(\frac{\sigma_{\rm{v}}}{\sigma_{\rm{z}}(z_{\rm l})}\right)^\beta\right],
\end{equation}
and
\begin{equation}
n_{\rm{z}}(z_{\rm l}) = n_0 (1 + z_{\rm l})^{\kappa_n}, \quad
\sigma_{\rm{z}}(z_{\rm l}) = \sigma_{\rm{v}0} (1 + z_{\rm l})^{\kappa_v},
\end{equation}
where $\sigma_{\rm v0}$ is the characteristic velocity dispersion, $\alpha$ is the low-velocity power-law index, $\beta$ is the high-velocity exponential cutoff index, and $\Gamma(\alpha/\beta)$ is the Gamma function. The adopted parameter values are $(n_0, \sigma_{\rm v0}, \alpha, \beta) = (0.008 \, h^3 \, \text{Mpc}^{-3}, 161 \, \text{km} \, \text{s}^{-1}, 2.32, 2.67)$. The evolution parameters $\kappa_n = -1.18$ and $\kappa_v = 0.18$ follow the fitting results given in Ref.~\cite{2021MNRAS.503.1319G}. 
The lens-axis-ratio distribution $P_e(e)$ and the external-shear distribution $P_\gamma(\gamma_1,\gamma_2)$ are adopted following Ref.~\cite{2024ApJ...962....3C}. In the Monte Carlo lensing realizations below, these quantities are sampled for each lens and passed to the lensing realization.
The resulting galaxy–galaxy strong-lensing optical depth as a function of source redshift, $\tau(z_{\rm s})$, is shown in Figure~\ref{fig:fig1}.

We map the mock SMBBH population constructed in Section~\ref{subsec:mocksamples} onto potentially observable lensed systems through the following procedure. First, we compute the strong-lensing probability for each SMBBH (adopting an $r$-band limit of 30 mag) at its redshift using the optical depth $\tau(z)$, and perform a Bernoulli trial to determine whether the source is strongly lensed. Second, for each lensed SMBBH at redshift $z$, we assign the lens properties including $\sigma_{\rm v}$ and $z_{\rm l}$, by sampling from the differential optical depth given in Equation~\eqref{eq:dtau} under the adopted lensing model. This step yields a population of strongly lensed close SMBBHs prior to any detectability constraints. The corresponding source magnitudes are also computed for this sample and used in the subsequent analysis. 

We next generate specific image configurations. We randomly assign the source position ($x_{\rm{s}},y_{\rm{s}}$) uniformly over the area within the multiple-imaging region to produce a lensing realization for each lensed SMBBH. We employ the \textsc{Lenstronomy} package \cite{2018PDU....22..189B} to solve the lens equation for each realization and compute the corresponding image positions ($x_{\rm{im}},y_{\rm{im}}$) and image magnitudes. Finally, we compare the image separations and magnitudes with given instrumental capabilities to identify lensed close SMBBHs with resolvable two-image/effective-double configurations. For quad-image systems, contributing a small fraction to the total events, we retain only the two brightest images and treat them as an effective double, since these two images dominate the observational detectability relevant to our microlensing analysis. In the selected samples under different survey configurations, quad systems treated as effective doubles account for only $\sim 3$--$7\%$, while the remaining systems are genuine doubles.
The same resolvability and magnitude criteria are then applied to the two images of genuine doubles and to the retained two brightest images of quads.
Only systems with two resolvable (image separation larger than two times the FWHM) and detectable images (brighter than the limiting magnitude) are selected as observable two-image/effective-double lensed SMBBHs. 

For Rubin, we adopt $r$-band parameters, including a PSF FWHM of $0.77$\,arcsec over a sky coverage of $18,000$\,deg$^2$ \cite{2019ApJ...873..111I}. The Rubin Wide–Fast–Deep (WFD) survey, with a median revisit time of $\sim3$\,days and a limiting magnitude of $24.9$\,mag, is well suited for monitoring variable sources, such as the microlensing systems considered here. We also consider a deeper Rubin co-added depth of $27.5$\,mag for identifying lensed systems. Although high-cadence epoch-level photometry at this depth is extremely challenging, it provides an approximate limit for discovering and resolving macro-images, serving as a proxy for the discovery potential of future deep-field surveys. Similarly, we evaluate the detectability of lensed SMBBHs for CSST, Euclid, and Roman. For CSST, we adopt $r$-band specifications from Ref.~\cite{Zhan2021}, including a limiting magnitude of $m_{\rm{lim}}=26.3$, a PSF FWHM of $0.064$\,arcsec, and a total sky coverage of $17,500$ deg$^2$. For Euclid, we adopt the VIS Wide Survey parameters \cite{2022A&A...662A.112E}: a depth of $26.2$\,mag, a sky coverage of $15,000$\,deg$^2$, and a PSF FWHM of $0.18$\,arcsec. Finally, for Roman, we assume a point-source depth of $\sim26.5$ AB mag in the F062 band over a survey area of $\sim2000$\,deg$^2$, with a PSF FWHM of $\sim0.058$\,arcsec \cite{2019BAAS...51c.341D}.\footnote{See more
descriptions from the Roman Wide Field Instrument technical page: \url{https://science.nasa.gov/mission/roman-space-telescope/wfi-technical/}.}

We also include galaxy clusters as strong lenses estimated using a halo-mass-function-based spherical NFW+SIS proxy, following the cluster lensing framework of Ref.~\cite{2024ApJ...977...64C}, as also shown in Figure~\ref{fig:fig1}. 
The cluster mass distribution is modeled as a spherical NFW halo plus a central SIS component associated with the BCG.
For simplicity, we assume that the fraction of detectable two-image/effective-double SMBBHs resulting from lensing effects is similar for both galaxy lenses and cluster lenses. Consequently, we apply the detection fraction derived from galaxy lensing realizations to the estimates for cluster lenses. 
This treatment is intended only as a statistical approximation, motivated by the larger characteristic macro-lensing angular scales of cluster lenses, e.g., \cite{2006MNRAS.367.1241O,2012ApJ...749...38M,2020MNRAS.495.3727R,2024SSRv..220...23L}. The cluster-lensing contribution should be regarded as an optimistic extension to the galaxy-lens baseline.

Finally, given the small number of strongly lensed SMBBHs produced in a single realization, we perform $1000$ Monte Carlo realizations for each model. This substantially reduces stochastic fluctuations and yields converged estimates for both the total number of detectable systems and the associated parameter distributions. By adopting two BBH models with different black hole mass relations, we are able to bracket the range of our predictions by considering the main uncertainty that is induced by the uncertainties in the redshift evolution of SMBH-host galaxy relationship.

\section{Results} 
\label{sec:results}

\subsection{Strongly lensed close SMBBHs} 
\label{sec:lenBBHs}

\begin{figure}
\includegraphics[width=0.48\textwidth]{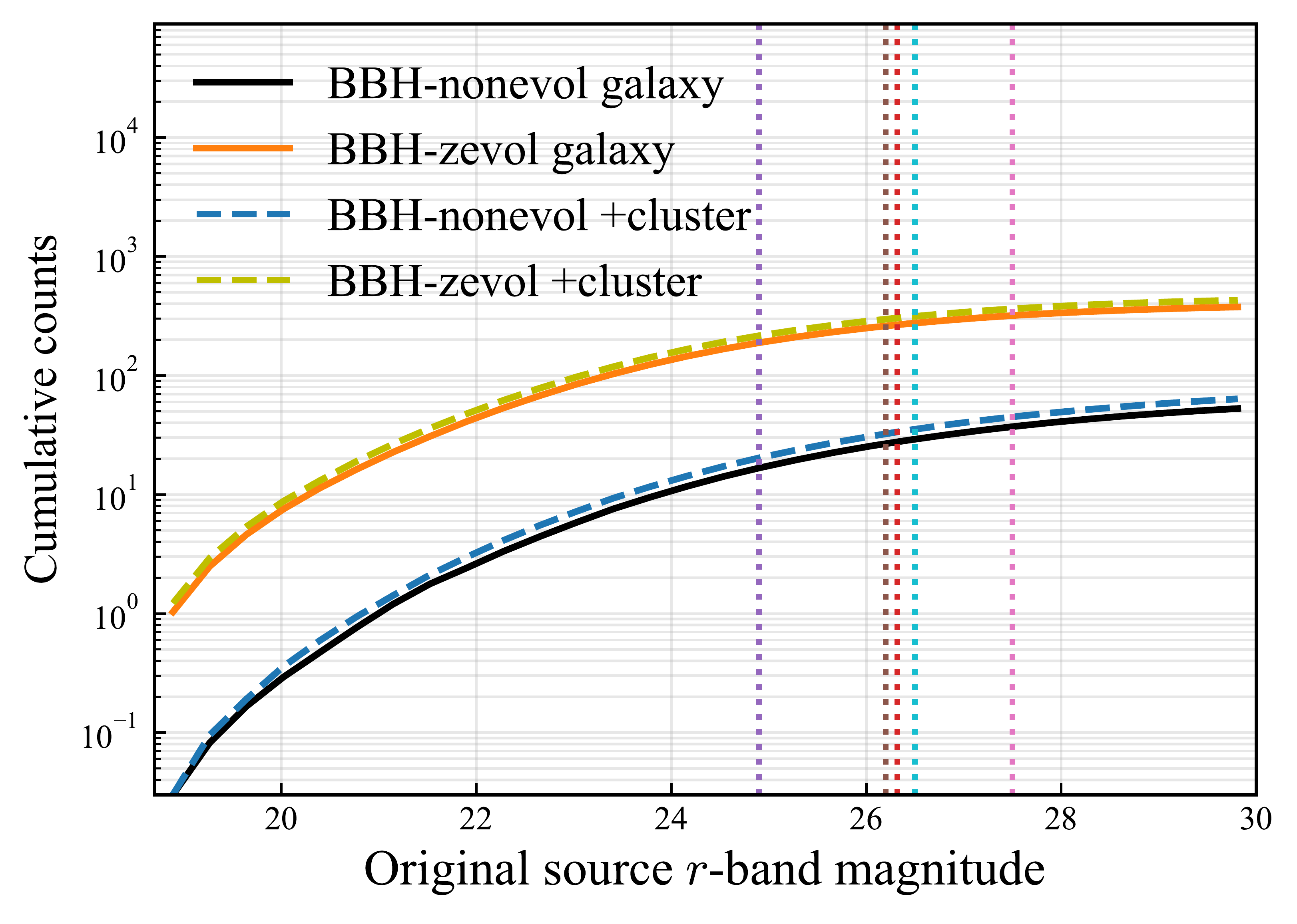}  
\caption{
Cumulative number distribution of close SMBBHs expected to undergo strong lensing based on the lensing optical depth (across all sky) after 1000 Monte Carlo realizations, shown as a function of the original-source $r$-band magnitude before lensing realizations.
The black solid curve denotes the cumulative counts under the BBH-nonevol model, while the blue dashed curve shows the corresponding results when cluster lenses are included. The orange solid and yellow dashed curves indicate the predictions of the BBH-zevol model considering galaxy lensing only and galaxy+cluster lensing, respectively. The vertical dotted lines show, from left to right, the reference $r$-band limiting magnitude of Rubin WFD ($24.9$\,mag; purple), Euclid (brown), CSST (red), Roman (cyan), and the deeper Rubin imaging case ($27.5$\,mag; pink).
} 
\label{fig:fig2}
\end{figure}

We first present the predicted population of close SMBBHs statistically expected to undergo strong lensing. Figure~\ref{fig:fig2} shows the cumulative number of such systems over the full sky as a function of 
the original source magnitude, before performing lensing realizations and applying image-level cuts. The vertical survey reference lines are shown only as references for the apparent $r$-band magnitude of the original source. The cumulative curves are plotted up to a source magnitude of $m_{\rm{lim}}\simeq 30$.
Considering only galaxy-galaxy strong lensing, we find approximately $53$ and $379$ close ($a < 10^4 R_{\rm{g}}$) SMBBHs expected to undergo strong lensing resulting from the BBH-nonevol and BBH-zevol models, respectively. 
When the cluster-lensing contribution is statistically included, the corresponding optimistic estimates over the whole sky increase to 60 and 431. Although BBH-nonevol and BBH-zevol differ in the local slope, this difference partly counteracts the redshift-evolution effect at the high-mass end, so the larger BBH-zevol population is mainly driven by the positive $\tilde{\omega}\log(1+z)$ term. 

We note here that extending the source-magnitude limit from $\sim26$ to $\sim27.5$\,mag does not significantly boost the number of lensed systems, suggesting that the lensed population is relatively well-sampled at magnitudes brighter than $26$\,mag. In addition, the BBH-zevol model preferentially produces more massive SMBBHs, especially at high redshifts where the lensing probability is also higher. This naturally leads to a significantly larger number of bright lensed close SMBBHs predicted by the BBH-zevol model in Figure~\ref{fig:fig2}.

To put our results into the context of upcoming large-scale surveys, we estimate the total observable AGN population, as well as the subset of strongly lensed cases. Following Ref.~\cite{2024MNRAS.531.3991M}, we adopt the quasar luminosity function (QLF) in the $r$ band from Ref.~\cite{2016A&A...587A..41P} and convolve it with the lensing probability given by the optical depth (see Figure~\ref{fig:fig1}) to obtain the detection prospects for various facilities, based on their specific sky coverages and limiting magnitudes. As an intermediate validation of this calculation, we compare the magnitude distributions of the lensed AGN population with those of the lensed close SMBBHs in Appendix~\ref{app:lensing_properties}.

For the Rubin Observatory, we obtain an expected number of $\sim 10^4$ lensed AGNs by galaxies, consistent with literature forecasts \cite{2010MNRAS.405.2579O}. Comparing this to our lensed SMBBH predictions (scaled to Rubin's survey area), we find that roughly a fraction of $\sim10^{-3}$--$10^{-2}$ of these lensed AGNs detected by Rubin may host close SMBBH systems. We further extend this analysis to other space-based surveys. Considering nominal sky coverages of Euclid ($\sim 15,000\,\mathrm{deg}^2$), CSST ($\sim 17,500\,\mathrm{deg}^2$), and Roman ($\sim 2,000\,\mathrm{deg}^2$), we estimate total detections of approximately $3.6\times10^7$, $4.3\times10^7$, and $5.8\times10^6$ photometric AGNs, respectively. Our estimate for Euclid is broadly consistent with a rough extrapolation of $1.4\times10^7$, based on the recent Euclid Q1 data release for a small sky area \cite{2025arXiv250315318E}.\footnote{Our estimate of the AGN population is based on a simplified approach, whereas spectroscopic forecasts would typically yield smaller AGN samples, e.g., \cite{2024MNRAS.531.3991M}.} By applying the galaxy+cluster lensing probability, we predict $2.0\times10^4$, $2.6\times10^4$, and $3.2\times10^3$ lensed AGNs for Euclid, CSST, and Roman, respectively. For direct comparison with the predicted lensed close-SMBBH counts, the expected numbers of detectable lensed AGNs for both the galaxy-only and galaxy+cluster lensing cases are also listed in Table~\ref{tab:table1}. Our estimate for Roman is also in agreement with the $\gtrsim10^3$ value estimated by Ref.~\cite{2023Galax..11...71H}. 

\subsection{Lensed SMBBHs with resolvable images} 
\label{sec:lenvariBBHs}

\begin{table*}[t]
\centering
\caption{
Predicted numbers of mock lensed SMBBHs selected under various model settings and survey configurations. 
}
\label{tab:table1}
\resizebox{\textwidth}{!}{
\begin{tabular}{c c c c c c c c c} \hline
 \multirow{2}{*}{Lens}  & \multirow{2}{*}{Detector [$m_{\rm{lim}}$,FWHM]} &
 \multirow{2}{*}{Lensed AGNs} &
 \multirow{2}{*}{BBH-zevol} &  \multicolumn{2}{c}{microlensing SMBBHs} & \multirow{2}{*}{BBH-nonevol} &  \multicolumn{2}{c}{microlensing SMBBHs} \\
 & & & & $P_{\rm{orb}} \leq 5$ yr  & $r$-$\lambda$ feature  & &$P_{\rm{orb}} \leq 5$ yr  & $r$-$\lambda$ feature  \\
 \hline
 \multirow{5}{*}{Galaxy}
   & Rubin [24.9,$0.77^{\prime\prime}$]    & $1.0\times10^4$ &  38.7 & $0.61$  & 4.8    & 4.1 & $0.12$ & $0.39 $   \\
   & Rubin  [27.5,$0.77^{\prime\prime}$]   & $4.4\times10^4$ &  68.7 & 2.7  & 8.8 & 9.6 & $0.58$ & 1.1     \\
   & CSST [26.3,$0.064^{\prime\prime}$]    & $2.2\times10^4$ & 76.8 & 2.1  & 9.5 & 9.8 & $0.41$ & 1.0     \\
   & Euclid [26.2,$0.18^{\prime\prime}$]   & $1.8\times10^4$ & 64.2 & 1.7  & 8.0     & 8.1 & $0.34$ & $0.83$     \\
   & Roman [26.5,$0.058^{\prime\prime}$] & $2.8\times10^3$ &  9.2 & $0.26$  & 1.1     & 1.2 & $0.05$ & $0.12$    \\
 \hline
 \multirow{5}{*}{Gal+Cl}
   & Rubin [24.9,$0.77^{\prime\prime}$]    & $1.2\times10^4$ &  43.7 & 0.68  & 5.3    & 4.7 & 0.14 & 0.44     \\
   & Rubin  [27.5,$0.77^{\prime\prime}$]   & $5.1\times10^4$ &  77.6 & 3.0  & 9.8     & 10.9    & 0.66 & 1.2   \\
   & CSST [26.3,$0.064^{\prime\prime}$]    & $2.6\times10^4$ &  86.8 & 2.3  & 10.6     & 11.2    & 0.46 & 1.1   \\
   & Euclid [26.2,$0.18^{\prime\prime}$]   & $2.0\times10^4$ &  72.5 & 1.9  & 8.9    & 9.2 & 0.38 & 0.94   \\
   & Roman [26.5,$0.058^{\prime\prime}$] & $3.2\times10^3$ &  10.4 & 0.29 & 1.2     & 1.4 & 0.06 & 0.14   \\
 \bottomrule
\end{tabular}
}
\smallskip
\begin{minipage}{0.95\linewidth}
\smallskip
\textbf{Notes:} The reported SMBBH numbers represent averages over 1000 Monte Carlo realizations. We include both strongly lensed SMBBH systems with at least two resolvable images and two microlensing-relevant subsets. The first column specifies the type of lenses, considering either galaxies only or galaxies $+$ clusters as lenses. For the Gal+Cl rows, the cluster contribution is included statistically through the NFW+SIS cluster-lensing optical-depth proxy and should be regarded as an optimistic extension to the galaxy-lens baseline. The second column lists the adopted survey characteristics, including both the limiting magnitude and PSF FWHM of each facility  (Section~\ref{subsec:stronglensing}). 
The third column lists the expected number of parent AGNs statistically expected to undergo strong lensing for each survey configuration, without applying an image-resolution requirement.
The fourth column lists the predicted number of resolvable two-image/effective-double SMBBHs by adopting the BBH-zevol model, with consideration of the sky coverage of each survey. Values in the fifth and sixth columns summarize two microlensing-relevant subsets: microlensing SMBBH systems with $P_{\rm{orb}} \leq 5$ yr in the observer's frame to be detected with periodicity, and systems with separations that satisfy the criterion for producing detectable size–wavelength signatures. For CSST, Euclid, Roman, and the deeper Rubin imaging case, these subsets should be interpreted as imaging-selected candidates; microlensing-based identification requires adequate time-domain follow-up. The last three columns provide, for comparison, the corresponding numbers predicted by adopting the BBH-nonevol model. The numbers for different surveys indicate per-survey imaging/discovery capabilities and should not be summed directly, because the survey footprints overlap. 
\end{minipage}
\end{table*}

\begin{figure}
\centering
\includegraphics[width=1\linewidth]{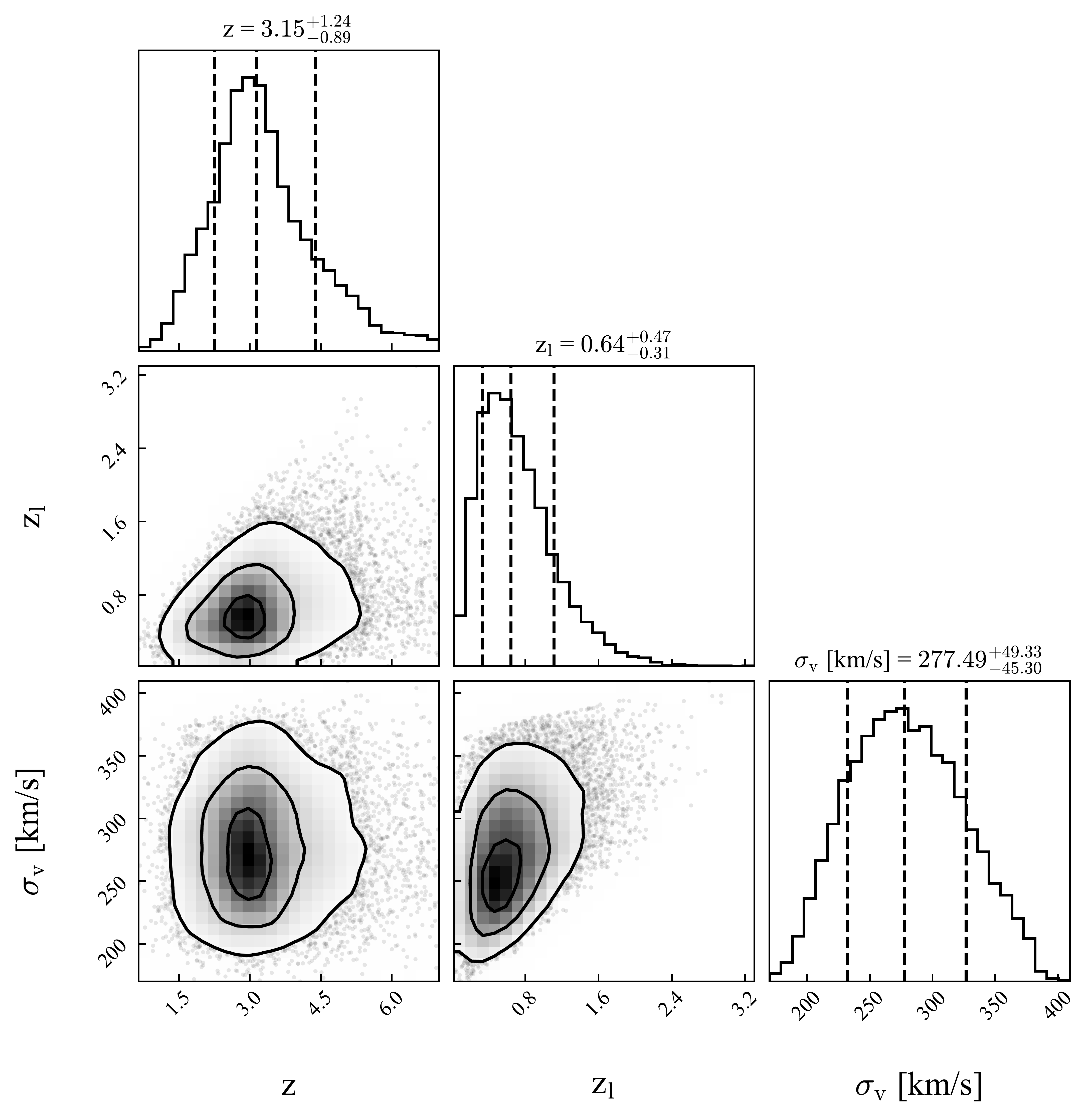}
\caption{
Source redshift, lens redshift and lens velocity dispersion distributions for lensed mock SMBBHs by galaxies that can be potentially detected with resolvable two-image/effective-double images by Rubin WFD. All distributions are obtained from $1000$ realizations to mitigate statistical fluctuations. In the top, middle-right and bottom-right panels, the black curves show the one dimensional distribution of the corresponding parameters. In the remaining panels, individual lens systems are marked by black dots, with the contour curves showing the $16\%, 50\%, 84\%$ probability regions. Vertical dashed lines indicate the mean values and the 1-$\sigma$ intervals of the corresponding distributions. 
}
\label{fig:fig3}
\end{figure}

\begin{figure}
\includegraphics[width=0.48\textwidth]{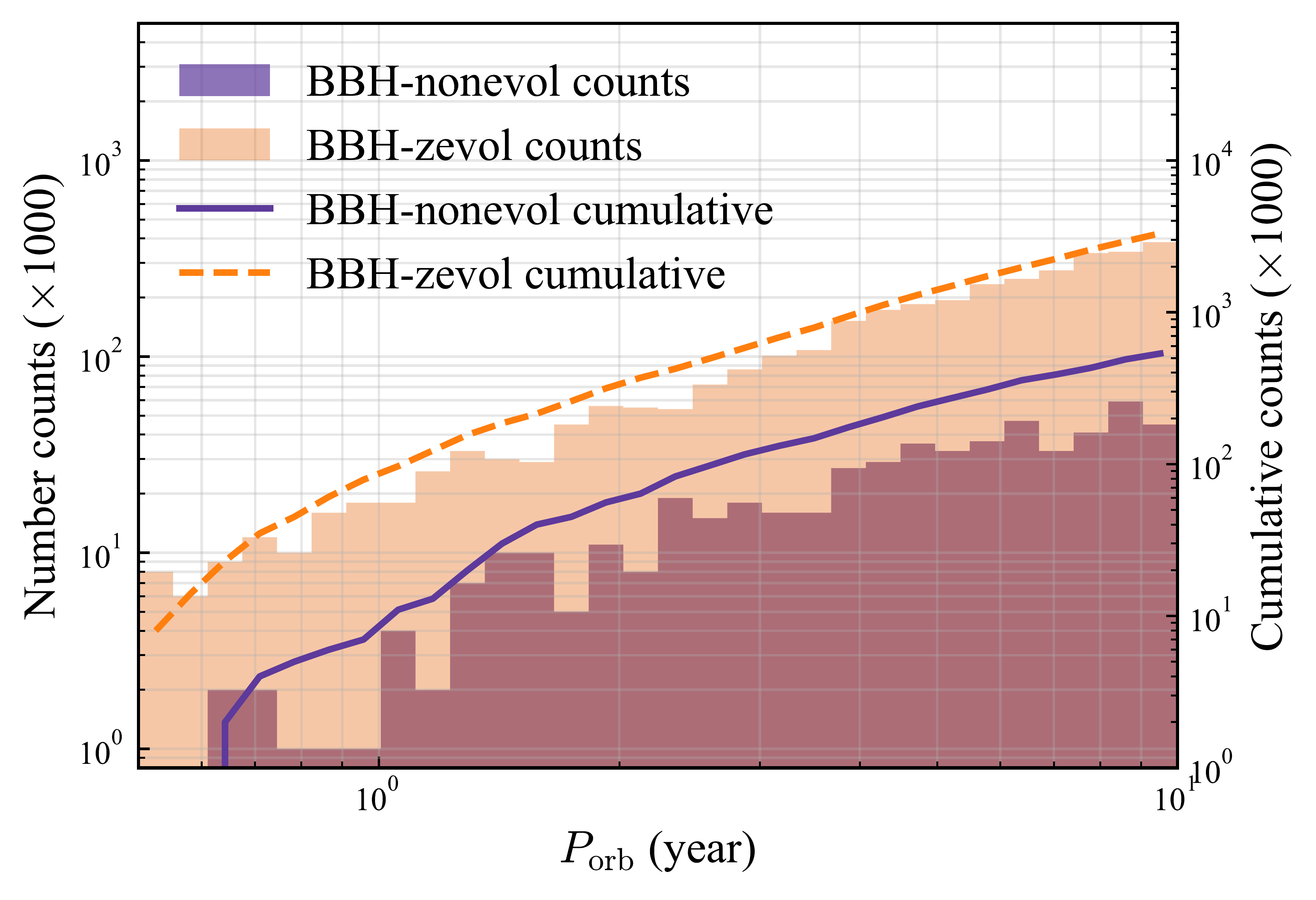}  
\caption{
Number distribution of resolvable lensed SMBBHs as a function of orbital period, combining 1000 realizations and adopting Rubin WFD detection capabilities. The orange dashed curve shows the cumulative counts predicted by the BBH-zevol model, with the orange bars indicating the corresponding binned distribution. The purple curve and bars represent the cumulative number and binned distributions for the BBH-nonevol model. The left and right vertical axes correspond to the number counts in each bin and the cumulative counts, respectively.
} 
\label{fig:fig4}
\end{figure}

We obtain mock SMBBH systems with two-image/effective-double images in strongly lensed AGNs that satisfy the detection criteria outlined in Section~\ref{subsec:stronglensing}, i.e., systems with two images that are both detectable and spatially resolvable by future facilities. Figure~\ref{fig:fig3} shows the distributions of source redshift, lens redshift, and lens velocity dispersion for the Rubin-detectable systems. The redshift and velocity dispersion distributions of the lens galaxy closely follow those of the general galaxy strong-lensing population, e.g., \cite{2010MNRAS.405.2579O,2015ApJ...811...20C,2024MNRAS.533.1960C}, while the source redshift distribution declines rapidly beyond $z\gtrsim4$ due to the decreasing abundance of close SMBBHs at high redshift. 

Rubin, with its $10$-year observation baseline and $3$-day revisit cadence on average, is well suited for time-domain monitoring down to its WFD depth. Therefore, Rubin can naturally provide substantial light curves for microlensing events and help identify SMBBHs through periodic microlensing modulations or the reconstruction of distinctive size–wavelength relations (as discussed later). Table~\ref{tab:table1} summarizes the numbers of mock resolvable lensed SMBBHs under different lensing configurations (galaxy only vs. galaxy $+$ cluster) and different detection settings. Here $P_{\rm orb}$ denotes the orbital period in the observer frame.  For the BBH-nonevol population, the predicted detection rates remain modest, yielding only $\sim4$ SMBBH systems over the $18,000$\,deg$^2$ sky coverage of Rubin. In contrast, the BBH-zevol model suggests a more optimistic prediction, revealing $\sim 44$ detectable systems when both galaxy and cluster strong lensing are included. We simulate the $\sim10^4$ detectable lensed AGNs expected from Rubin using the lensing realization procedure described in Section~\ref{subsec:stronglensing}, and find that about 40\% of these systems satisfy the criteria for having resolvable images. Therefore, among lensed SMBBHs, systems with resolvable images correspond to roughly $\sim 1\times 10^{-3}-9\times10^{-3}$ of all resolvable lensed AGNs by Rubin. The uncertainties of these estimates depend on the underlying modeling of galaxy mergers and SMBBH evolution, e.g., \cite{2019MNRAS.485.1579K}, which can alter the abundance of close SMBBHs and thus the corresponding lensing predictions (see discussion in Section~\ref{sec:discussion}). 

Different facilities yield markedly different numbers of SMBBHs, primarily due to differences in sky coverage, limiting magnitude, and spatial resolution. BBH-zevol model may reveal up to 87, 73, 10, and 78 detectable systems for CSST, Euclid, Roman, and Rubin 27.5 mag. CSST provides the largest number of spatially resolvable lensed AGNs that host SMBBHs due to its superior angular resolution. Euclid yields comparable numbers under favorable conditions, while the result for Roman is limited by its sky coverage ($\sim 2,000$\,deg$^2$). Rubin is less effective at resolving multiple images due to its larger PSF FWHM. As a result, Rubin is expected to resolve only about half as many resolvable lensed AGNs that host SMBBHs as CSST and Euclid do (adopting the WFD depth of $24.9$\,mag; see Table~\ref{tab:table1}). Although we test Rubin with a deeper imaging depth ($27.5$\,mag), which would yield a larger number of lensed SMBBH systems, achieving a similar sensitivity in a single epoch would require hour-scale exposure times that are hardly compatible with current Rubin observing strategy. Thus it is difficult to obtain light curves for those faint lensed AGNs with $r$-band magnitude larger than $24.9$, which are required to identify SMBBHs via microlensing effects. Deeper observations (by Rubin or other facilities discussed above) beyond the Rubin WFD detection increase the number of resolvable lensed macro-images, but again extracting periodic microlensing modulations from the faintest sources in this subset remains observationally challenging. 

We emphasize that the numbers in Table~\ref{tab:table1} represent per-survey imaging/discovery capabilities under the corresponding limiting magnitude, angular resolution, and sky coverage, and should not be interpreted as mutually exclusive populations or summed directly. Using the published footprint-overlap estimates, e.g., \cite{2023A&A...669A.128L,2024ApJ...970...49C}, we obtain a rough footprint-scaled estimate of about $\sim(1.2$--$1.5)\times10^{2}$ unique imaging-detectable SMBBHs in the optimistic BBH-zevol galaxy+cluster case.
We further distinguish between imaging discovery and microlensing-based identification. The first-level selection in Table~\ref{tab:table1} depends mainly on imaging depth, angular resolution, and sky coverage, for which space-based facilities such as CSST, Euclid, and Roman are particularly valuable. Microlensing-based identification, however, requires long-term time-domain monitoring with adequate cadence and baseline. Rubin WFD naturally provides such monitoring, whereas the microlensing-relevant subsets listed for CSST, Euclid, Roman, and the deeper Rubin imaging case should be interpreted as candidate numbers conditional on dedicated follow-up or overlap with Rubin-like time-domain data.

The orbital period distribution of Rubin-detectable SMBBH systems after lensing realization is shown in Figure~\ref{fig:fig4}. For both SMBBH models, the number of detectable systems increases with orbital period, primarily due to the longer residence time associated with earlier SMBBH evolution stages at these periods, specifically $\delta t \propto P_{\rm orb}^{5/3}\delta P_{\rm orb}$ when GW radiation dominates the orbital decay. However, this increment tends to be slightly slower for longer periods near $P_{\rm{orb}} \simeq 10$ yr compared to short periods $P_{\rm{orb}} \lesssim 1$ yr, as a larger fraction of SMBBHs resides in the viscous-driven evolution stage, rather than in the GW-dominated stage which is more sensitive to $P_{\rm{orb}}$. 

\begin{figure}[t]
\includegraphics[width=1\linewidth]{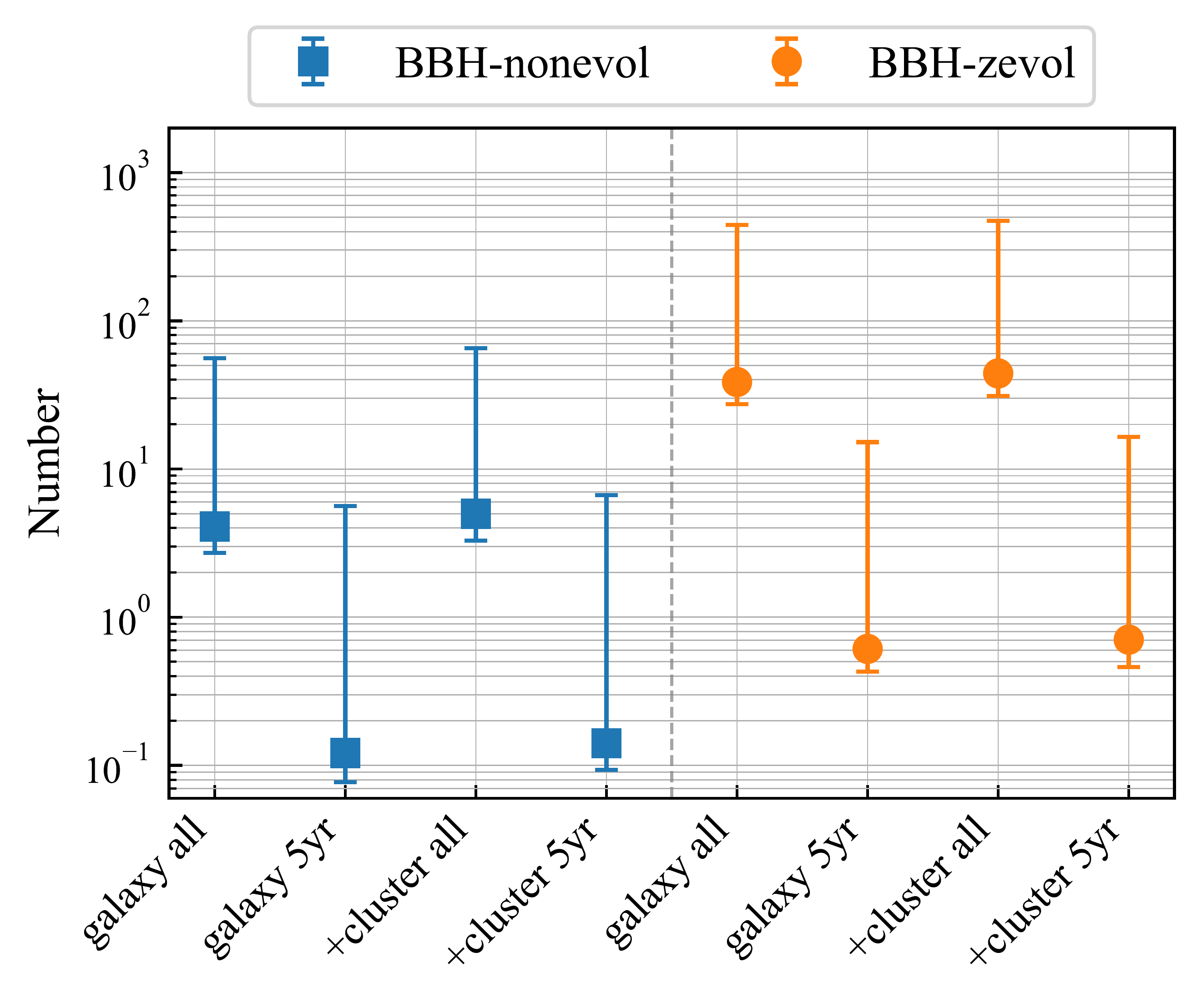}
\caption{
Comparison of the predicted detection numbers of resolvable lensed SMBBH systems under different model and detection configurations, assuming Rubin sky coverage and WFD capabilities.
The symbols show the fiducial GW+viscous model, while the vertical ranges indicate the model-dependent uncertainty associated with different assumptions about late-stage binary evolution. The lower end of each range corresponds to the CBD-filtered case, 
while the upper end corresponds to the GW-only case.
Blue squares show the predictions from the BBH-nonevol model in the first four configurations, while orange circles correspond to the BBH-zevol model in the last four configurations.
Results are shown separately for galaxy-only lensing (labels 0, 1, 4, 5) and for the combined galaxy + cluster scenario (labels 2, 3, 6, 7).
Among the x-axis labels, odd-numbered labels denote the subset of short-period systems with $P_{\rm orb} \leq 5\mathrm{yr}$, whereas even-numbered labels represent the full close SMBBH population.
} 
\label{fig:fig5}
\end{figure}

We further examine the detection prospects for Rubin-like detectors under different model assumptions as shown in Figure~\ref{fig:fig5}. As expected, incorporating cluster strong lensing consistently leads to an increase in the predicted number of lensed active SMBBHs. 
The cluster contribution is included statistically through the cluster-lensing optical depth rather than through full cluster-lens realizations. The angular-scale validation presented in \ref{app:lensing_properties}(Figure~\ref{fig:app2}) shows that the corresponding cluster-lens proxy has larger characteristic critical radii than the galaxy-scale SIE lenses. However, these values do not include additional observational difficulties associated with identifying and modeling cluster-lensed AGNs, and should therefore be regarded as optimistic statistical estimates.

Among all such systems, only a small fraction are expected to exhibit observable periodic oscillations, specifically those with observed orbital periods $P_{\rm orb} \leq 5\mathrm{yr}$. In a limiting scenario in which the SMBBH orbital decay is driven purely by GW emission (as described in Section~\ref{subsec:mocksamples}), the predicted detection numbers reach their maximum, up to $5\times10^2$ systems across the sky coverage of Rubin. This GW-only case gives the upper end of the model-dependent range shown in Figure~\ref{fig:fig5}.
In contrast, the conservative CBD-filtered case, in which systems with $q>0.2$ are assumed not to efficiently enter the short-period close-SMBBH sample, gives the lower end of the range. Compared with the fiducial model, this filtering reduces the predicted numbers by roughly 30\%-35\% for both BBH-nonevol and BBH-zevol models. Therefore, the vertical ranges in Figure~\ref{fig:fig5} illustrate the sensitivity of the predicted detection numbers to uncertain late-stage SMBBH orbital evolution.

\subsection{Microlensing implications}
\label{sec:microlensing discussion}

Microlensing in strongly lensed AGNs provides a powerful probe for identifying sub-pc SMBBHs. The signatures can appear either as periodic modulations in the microlensing light curves \cite{2014ApJ...784..100Y,2022A&A...668A..77M} or as systematic deviations in the disk size–wavelength relation compared to single black hole systems \cite{2014ApJ...784..100Y}. Our analysis of strongly lensed close SMBBHs shows that the size of their characteristic emission region (including triple disk) is comparable to the projected Einstein radius of typical foreground microlensing stars (Section~\ref{sec:lenBBHs}). Consequently, the unique source structure can be properly scanned during a single Einstein-radius crossing event, which is a necessary condition for resolving disk-scale features through microlensing.

In strongly lensed systems the microlensing optical depth is often of order unity, making it likely that at least one image of a lensed AGN is significantly affected by microlensing, e.g., \cite{1986ApJ...301..503P,2012A&A...544A..62S,2024SSRv..220...14V}. Although this implies a high incidence of microlensing, the characteristic timescale of microlensing events in SMBBHs must still be quantified. The caustic-crossing timescale is determined by the Einstein radius $R_{\rm{E}}$ and the effective relative transverse velocity $v_{\rm{e}}$, e.g., \cite{2004ApJ...605...58K,2011ApJ...738...96M,2014ApJ...784..100Y}. From the lens-galaxy properties in our realizations (Figure~\ref{fig:fig4}), we infer a one-dimensional stellar velocity dispersion of $\sigma_{\rm v} = 277^{+49}_{-45}\rm{km s^{-1}}$. Given the source and lens redshifts, we compute $v_{\rm{e}}$ by combining $\sigma_{\rm v}$ with the bulk motions of the observer, lens and source following Ref.~\cite{2004ApJ...605...58K} and Ref.~\cite{2011ApJ...738...96M}. Assuming a mean stellar mass of $\langle m \rangle \simeq 0.3 M_{\odot}$ and adopting the distribution of source and lens redshifts from our resolvable systems, we estimate a characteristic Einstein-radius crossing time of $t_{\rm{E}} = R_{\rm{E}}/v_{\rm{e}}\approx 10-33$ yr, where $v_{\rm{e}}$ spans the range $\sim 200$–$600~\mathrm{km s^{-1}}$. This is comparable to the typical microlensing timescale inferred for normal lensed AGNs, which is statistically estimated to be $\sim20$ yr from a sample of 87 lensed quasars \cite{2011ApJ...738...96M}. Moreover, Ref.~\cite{2011ApJ...738...96M} showed that over a 10-year monitoring baseline, roughly half of lensed AGNs reside in demagnified regions, with the other half is located in or near caustic regions and undergo active microlensing. Thus, at any given time a substantial fraction of images are significantly microlensed, and the associated caustic-crossing events last long enough to enable multi-year monitoring and the identification of SMBBH signatures.

Next, we evaluate the feasibility of these two microlensing-based identification methods by relating them to the orbital properties of our lensed SMBBH sample. These approaches probe complementary regions of parameter space: periodic microlensing favors short-period systems, whereas disk-structure diagnostics remain effective for wider binaries. We further assess the detection prospects for each method.

\subsubsection{Periodic oscillation}
\label{sec:microlensing variation}

For lensed SMBBHs with relatively short orbital periods (e.g., $P_{\rm{orb}} \lesssim 5$ yr), the binary orbital motion can imprint periodic modulations onto the microlensing light curves. Such systems may complete multiple orbits during a single microlensing event, making the periodic signal detectable, as reported for the lensed quasar Q~J0158-4325 by Ref.~\cite{2022A&A...668A..77M}. 
In this scenario, the periodicity is most naturally produced when the primary or secondary accretion components repeatedly approach or cross caustic structures as the binary orbits. More generally, repeated direct caustic crossings are not strictly required: orbital motion across a non-uniform microlensing magnification field, especially near caustics where the magnification gradient is large, can also generate periodic flux-ratio variations.

As in standard quasar microlensing, the effective source size $R_{\rm S}$ controls the microlensing amplitude, e.g., \cite{2011ApJ...738...96M,2024SSRv..220...14V}. Hereafter, $R_S$ denotes the effective source size in the microlensing diagnostic, defined as the monochromatic half-light radius of the emitting region at the wavelength of interest, as commonly adopted in quasar microlensing studies, e.g., \cite{2014ApJ...784..100Y,2024SSRv..220...14V}. Observationally, variability amplitudes are typically of order $\sim0.1$-$0.5$ mag, with larger excursions $\sim1$\,mag occurring during caustic-crossing events, e.g., \cite{2018ApJ...869..132F}. In theoretical point of view, smaller $R_{\rm S}$ leads to larger amplitudes, and SMBBHs naturally have smaller $R_{\rm S}$ because each mini-disk is truncated at an outer radius set by the binary separation. They are therefore expected to exhibit relatively large-amplitude microlensing oscillations. As estimated in Ref.~\cite{2014ApJ...784..100Y}, this magnitude variation can be $\gtrsim 1$\,mag. To quantify this effect, we compute the half-light radius of each mini-disk by integrating the disk emission out to its truncation radius. We find characteristic effective source sizes of $R_{\rm S}\approx30$–$150 R_{\rm g}$ for Rubin-detectable systems under the BBH-zevol model, which supports the discussion above. For Rubin observations, the photometric accuracy is roughly $10$\,mmag. \footnote{\url{https://rubinobservatory.org/for-scientists/rubin-101/key-numbers}}
This value represents a general photometric performance specification. Since the microlensing signal is extracted from flux-ratio light curves, at least two lensed images need to be measured with sufficient signal-to-noise ratios.
Large-amplitude microlensing variations, such as those discussed above, are therefore expected to be detectable for sufficiently bright image pairs.
As a result, the small number of SMBBHs with $P_{\rm orb}\leq5 \mathrm{yr}$ shown in Table~\ref{tab:table1} provides an initial indication of the feasibility of detecting SMBBHs using this method. 

Detecting such periodic microlensing signatures requires sustained time-domain monitoring with adequate cadence. In practice, sampling intervals of order weeks (e.g., $\sim1–2$ weeks) over multi-year baselines are typically needed, making the Rubin WFD observations naturally suited for searching for such signatures. 
Candidates discovered or resolved by other facilities may also be identified through periodic microlensing signatures, provided that dedicated follow-up or overlapping time-domain data with adequate cadence and baseline are available.

We calculate the detection prospect of microlensed SMBBHs with potential periodic oscillations that can be detected by Rubin, CSST, Euclid, and Roman. As summarized in Table~\ref{tab:table1}, only $\sim 0.1-0.7$ of such systems are expected to be identifiable by Rubin, underscoring the intrinsically low probability of identifying SMBBHs through periodic microlensing signatures. 
For completeness, we further apply conservative photometric-quality cuts requiring the fainter image of a two-image system to be brighter than $m=22$, 23, and 24 mag, respectively. For the Rubin WFD BBH-zevol galaxy-lens case, these cuts retain 5.9\%, 18.3\%, and 47.8\% of the original sample, corresponding to 0.04, 0.11, and 0.29 expected sources, respectively.
The BBH mass is concentrated around $10^{8.2} M_\odot$ (see Figure~\ref{fig:fig6}) and correlates with the binary separation, with more massive systems preferentially found at smaller separations. This trend is determined by both the residence time of binary evolution governed by the competing timescales of gas-driven migration and GW radiation induced orbital decay, and that the typical optical emission regions correspond to smaller radii for more massive BBHs. We further examine systems with even shorter orbital periods that would allow more cycles to be monitored within a single microlensing event. For instance, imposing an observed-frame period cutoff of $2.5$\,yr still yields about $0.19$ such systems, monitoring all lensed AGNs for Rubin WFD survey coverage. In practice, monitoring and identifying such lensed systems remains challenging.

Ref.~\cite{2022A&A...668A..77M} reported periodic oscillations during a high magnification microlensing event in the light curve of lensed quasar Q J0158-4325, with an observed period of $172.6$\,days. Among several possible explanations, the SMBBH scenario provides the most consistent interpretation, making this source the first SMBBH candidate identified via microlensing features. For high mass-ratio binaries, periodic modulation tends to occur at half the orbital period, thus the SMBBH orbital period in the observer's frame would be about 345 days. Based on our modeling for Rubin WFD survey, systems with $P_{\rm{orb}} < 345$\,days constitute about 5\% of all resolvable lensed SMBBHs with $P_{\rm orb} \leq 5\mathrm{yr}$ under the BBH-zevol model. The corresponding detection probability of systems with $P_{\rm{orb}} < 345$\,days among all resolvable strongly lensed AGNs that can be detected by Rubin is about $3.7 \times 10^{-6}$, assuming an active microlensing fraction of $50$\%, e.g., \cite{2011ApJ...738...96M}. Or if the orbital period was $172.6$\,days in the observer's frame \cite{2026MNRAS.tmp..388Y}, the fraction of systems with $P_{\rm orb} \leq172.6$\,days in all microlensed SMBBHs with $P_{\rm orb} \leq 5\mathrm{yr}$ decreases to $1.6$\%. These results confirm that the probability of observing such systems remains extremely small, consistent with the discussion in Ref.~\cite{2022A&A...668A..77M}, which may challenge the SMBBH interpretation of this system. 

\subsubsection{Disk size versus wavelength feature}
\label{sec:size-wavelength}

\begin{figure}[t]
\centering
\includegraphics[width=1\linewidth]{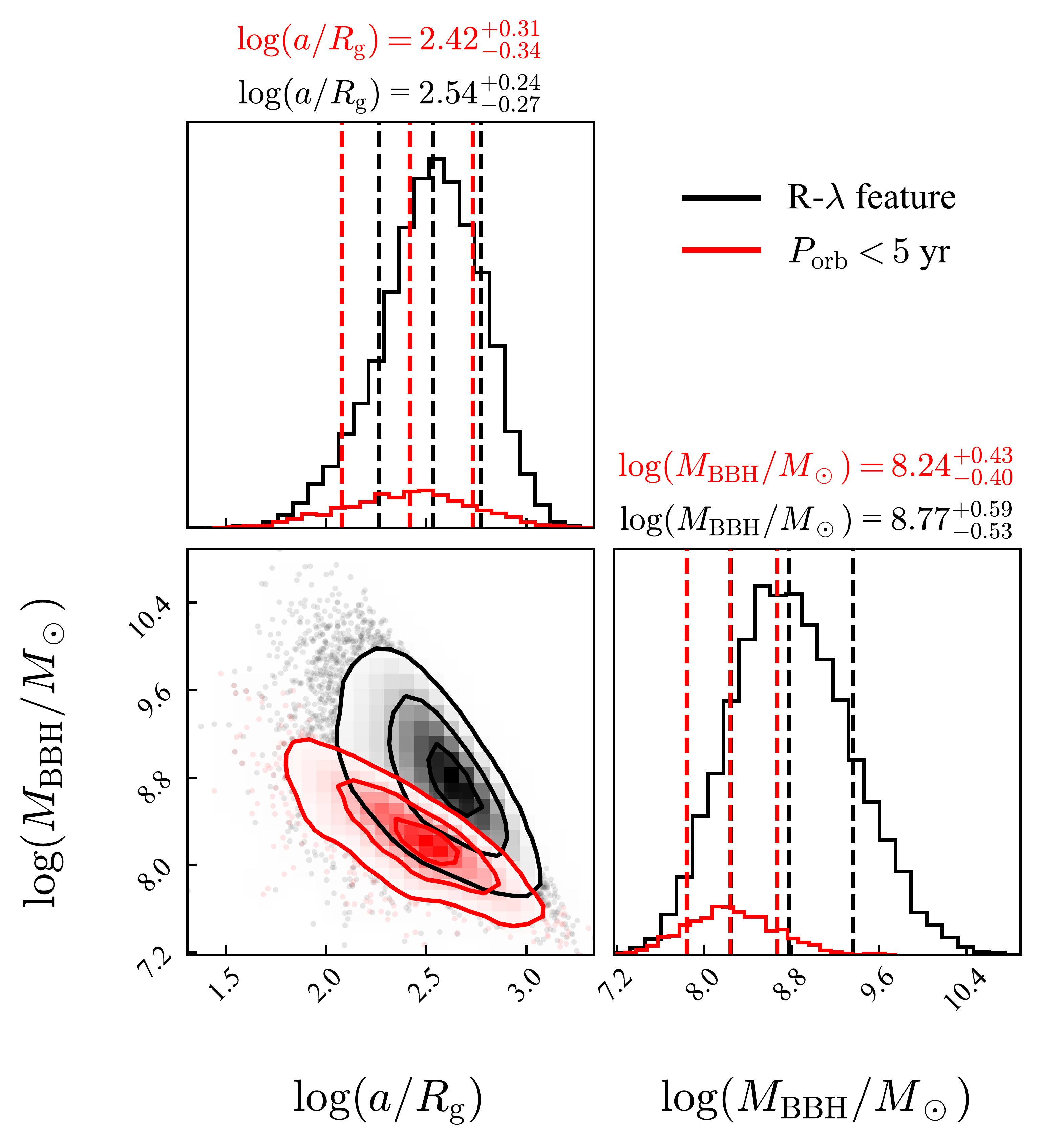}
\caption{
Number distributions of the semi-major axis and total BBH mass for microlensed SMBBH systems identified through two different approaches with Rubin WFD configurations. Black points show the result for identifiable SMBBHs via the size-wavelength approach, while red points correspond to systems with periodic oscillations. The black and red annotations indicate the median values and the 16–84 percentile ranges for the two samples, respectively. All distributions are obtained from $1000$ lensing realizations. 
}
\label{fig:fig6}
\end{figure}

For lensed close SMBBHs with longer orbital periods, the disk size versus effective wavelength relation inferred from microlensing light curves can help identify SMBBHs through the unique disk structure \cite{2014ApJ...784..100Y}. In general, larger binary separations facilitate the spatial separation of the individual accretion disks, enhancing the detectability of such structural signatures through microlensing. However, the applicability of the disk size–wavelength relation is sensitive to the location of the circumbinary gap. As the binary separation increases, the associated spectral feature shifts toward longer wavelengths and may move beyond the optical bands accessible to instruments such as Rubin, thereby reducing the effectiveness of this method. For a typical SMBBH system with total mass $\sim 10^{8}M_{\odot}$, this approach is expected to remain observable only for separations $a_{\rm{BBH}}\lesssim 1000r_{\rm{g}}$, see Figure~13 in \cite{2014ApJ...784..100Y}, where the characteristic features lie within the optical regime. 

The locations of the gap can be empirically approximated using the half-light radius $r_{1/2}$, see Figure~13 and Section 4.2.3 in \cite{2014ApJ...784..100Y}. For SMBBHs with total mass $10^7 M_{\odot} \lesssim M_{\rm{BBH}}\lesssim 10^9 M_{\odot}$, we find that the presence of the size-wavelength feature approximately requires: $a_{\rm{BBH}}\lesssim 1000 \left( \frac{M_{\rm{BBH}}}{10^8 M_\odot} \right)^{-1/3} r_{\rm{g}}$, which is consistent with the fact that the half-light radius scales with black hole mass as $r_{1/2} \propto M_{\rm{BBH}}^{-1/3}$. Adopting this empirical relation, we estimate the number of lensed SMBBHs in our sample that satisfy the criterion for exhibiting this diagnostic size–wavelength signature. We note that this signature also requires sustained time-domain monitoring with proper cadence for extracting microlensing light curves. 
 
As shown in Table~\ref{tab:table1}, several to about ten microlensed SMBBHs exhibiting characteristic size-wavelength features are expected to be detectable across different facilities. Under the BBH-zevol model, Rubin may detect about $5.3$ SMBBH systems via this microlensing feature in its survey sky area. This corresponds to a fraction of identifiable microlensed SMBBH of order $10^{-4}$ among all lensed AGNs to be detected by Rubin, and accounts for 10\% of lensed close SMBBHs of which the fraction in all resolvable lensed AGNs is of order $1\times 10^{-3}-9\times 10^{-3}$, as discussed in Section~\ref{sec:lenBBHs}.

The expected numbers of SMBBH systems are approximately $10$, $11$, $9$, and $1.2$ for Rubin (with a much deeper limiting magnitude of $27.5$\,mag), CSST, Euclid, and Roman, if a ten-year long monitoring of these AGNs with cadence of 1-2 weeks can be done. As shown in Figure~\ref{fig:fig6}, the binary separation of these SMBBHs peaks at about $400 r_{\rm{g}}$, with total masses concentrated in the range $10^8-10^9 M_\odot$. Compared to the short-period microlensing systems discussed in Section~\ref{sec:microlensing variation}, the size-wavelength approach yields several times more SMBBHs, particularly toward the high-mass end. 

\section{Conclusions and Discussions} 
\label{sec:discussion}

In this work, we investigate the prospects for detecting sub-pc SMBBHs in strongly lensed AGNs. We focus on “close” systems which are expected to accrete via a triple-disk structure and exhibit distinctive microlensing signatures. We construct the SMBBH population and compute the total emission from their accretion disks. We then perform strong-lensing statistics for both galaxy-scale and galaxy+cluster-scale lenses to estimate the population of lensed close SMBBHs. In particular, we select lensed systems with two resolvable images by computing the properties of each macro-image, enabling predictions for upcoming surveys such as Rubin. We further predict the detection prospects for SMBBHs through microlensing based on our resolvable lensed binaries. Our main conclusions are summarized as follows.

\begin{enumerate}
\item Future facilities like Rubin are expected to detect a modest number of multi-image lensed close SMBBHs among all lensed AGNs. 
For the Rubin WFD configuration, we predict approximately 4-39 detectable systems for galaxy lenses alone, while the optimistic galaxy+cluster statistical extension gives approximately 5-44 systems.
Under the same lensing assumptions, the BBH-zevol model yields up to 87, 73, 10, and 78 detectable systems for CSST, Euclid, Roman, and Rubin 27.5 mag, considering the sky coverage designed for each survey. These moderate numbers highlight the intrinsic difficulty of identifying lensed SMBBHs, even with next-generation wide-field surveys.
\item In our statistical treatment, including cluster-scale lenses can increase the predicted numbers by about 10\%-16\% compared to galaxy lenses alone. These galaxy+cluster results should be interpreted as optimistic estimates, because full cluster-lens reconstructions and observational complications are not modeled.
Moreover, the BBH-zevol model produces nearly an order of magnitude more detectable systems than the BBH-nonevol model, primarily because it contains a larger population of massive binaries. 
\item We further demonstrate that microlensing provides a physically motivated and observationally viable pathway for probing close SMBBHs in strongly lensed AGNs. The triple-disk structure produces source sizes comparable to stellar Einstein radii and microlensing timescales similar to those of normal lensed AGNs, enabling disk-scale structures to be revealed. Periodic microlensing oscillations arise only in a small subset of short-period binaries and are intrinsically rare, yielding fewer than one detectable system for Rubin (WFD). In contrast, disk size versus wavelength diagnostics apply to a broader subset of wider systems, providing up to 5.3 SMBBHs under Rubin WFD configurations, mainly subject to wavelength-dependent limitations. For candidates discovered or resolved by CSST, Euclid, Roman, or deeper imaging observations, such microlensing-based identification would require dedicated long-term follow-up or overlap with Rubin-like time-domain data with adequate cadence, baseline, photometric precision, and reliable time-delay measurements.
\end{enumerate}

\subsection{Observational requirements and lensing limitations}

Close SMBBHs with identifiable microlensing signatures constitute only a small fraction of all lensed AGNs, at the order of $\sim1\times 10^{-4}$. In real observations, identifying such systems through microlensing light curves is further challenged by substantial contamination from the much larger population of lensed AGNs. The Rubin survey is designed to obtain a large number of light curves through its WFD observations, making it naturally well suited for searching for such SMBBH signatures. In contrast, facilities like CSST and Euclid do not inherently provide time-domain light curves, although they are capable of finding a larger number of resolvable lensed AGNs. If future follow-up programs can monitor these lensed systems with sufficient time baselines and appropriate cadence, the SMBBHs predicted under the configurations of Rubin (27.5 mag), CSST, Euclid, and Roman may become identifiable. 

Our analysis is intentionally simplified and omits several complex effects. For instance, to estimate the number of resolvable SMBBHs in cluster-lens systems, we simply adopt the detectable fraction derived from galaxy-lens realizations as a proxy, instead of performing full cluster-lens reconstructions. This approximation may be imperfect, especially given that cluster Einstein radii are typically larger, e.g., \cite{2021A&A...646A..83R}. 
This assumption is more uncertain for cluster lenses, where the more complex and inhomogeneous mass and stellar distributions make the microlensing optical depth strongly dependent on the local stellar surface density and image positions. Therefore, the galaxy+cluster predictions should be regarded as approximate and potentially optimistic in terms of microlensing detectability. 
We also neglect observational complications such as dust extinction and host-galaxy obscuration, which can further reduce detectability. If we approximate the net flux attenuation from lens-galaxy extinction and reddening as $\sim 0.3$ mag, e.g., \cite{1999ApJ...523..617F,2006ApJS..166..443E,2011ApJ...737..103S}, the number of detectable systems under the BBH-zevol model with galaxy lenses alone decreases by a factor of 14\%. In addition, our adopted resolvability criterion has only a minor influence: tightening the requirement from image separations exceeding two to three times the PSF FWHM decreases the selected SMBBH numbers by only $\sim 1\%$.

We also note that microlensing-based identification requires reliable inter-image time-delay measurements. Since quasars are intrinsically variable, the macro-image light curves must be shifted by their time delays before flux-ratio or difference light curves can be used to suppress intrinsic AGN variability and isolate microlensing-induced variations \cite{2024SSRv..220...14V}. In practice, measuring such time delays is itself non-trivial: long-term monitoring studies have shown that microlensing-induced extrinsic variability, photometric noise, temporal sampling, seasonal gaps, insufficient intrinsic variability, and limited overlap between the shifted light curves can all affect the robustness of time-delay measurements \cite{2020A&A...640A.105M}. The requirement of accurate time-delay measurements may further reduce the number of practically identifiable systems, as also indicated in lensing forecasts and time-delay studies \cite{2010MNRAS.405.2579O,2023MNRAS.524.5446T}. We therefore do not impose an explicit time-delay-measurement efficiency in our calculation. The predicted numbers should be interpreted as resolvable and sufficiently bright microlensing-relevant candidates, rather than completeness-corrected practical identifications.

This caveat is especially important for cluster-lensed systems. 
Applying microlensing diagnostics in cluster fields is more challenging than for galaxy-scale lenses. Cluster-scale lenses are dominated by dark matter, and the stellar surface density that produces microlensing can be low and highly inhomogeneous, except near the BCG or cluster member galaxies. In addition, crowded fields, contamination from cluster members, complex mass distributions, uncertain time delays, and the intrinsic variability of lensed quasars \cite{2023MNRAS.524.5446T} can complicate the construction of flux-ratio light curves and the extraction of microlensing-induced signals. 
Thus, the galaxy-only results provide a conservative baseline, whereas the galaxy+cluster case should be regarded as an optimistic statistical extension for well-identified and well-monitored cluster-lensed systems.

\subsection{Interpretation of microlensing diagnostics}

We further clarify the interpretation of the size--wavelength diagnostic. The diagnostic should not be understood as treating any departure from the standard thin-disk scaling, $R\propto\lambda^{4/3}$, as a unique SMBBH signature. The global slope of the size--wavelength relation depends on the temperature structure and detailed accretion physics. Instead, the key feature proposed by \cite{2014ApJ...784..100Y} is a structural anomaly caused by the binary-induced cavity/gap. In this scenario, the residuals of the inferred half-light-radius--wavelength relation relative to a smooth single-MBH disk show a localized dip or flattening over the wavelength range where the gap limits the emitting area, while outside this range the relation can recover that of a corresponding single-disk system. Therefore, the relevant SMBBH indicator is not simply a non-$4/3$ spectral size index, but a wavelength-dependent residual feature associated with the cavity scale. Nevertheless, such features are not by themselves definitive proof of an SMBBH. Normal AGN physics may still mimic or obscure parts of the signal, and robust identification would require additional evidence.

We only consider the $r$-band variability for illustration. In principle, one could consider the variability at other bands. 
Since the orbital-decay timescale scales as $t_{\rm GW}\sim a/|\dot{a}|\propto a^4$ when GW radiation dominates, the differential residence time per unit separation is $dt/da=t_{\rm GW}/a\propto a^3$. If the gap location is approximated by the half-light radius $r_{1/2}$, with $a\propto r_{1/2}\propto \lambda^{4/3}$, e.g., \cite{1973A&A....24..337S,2004ApJ...605...58K,2014ApJ...784..100Y}, then the differential residence time per unit wavelength is $dt/d\lambda=(dt/da)(da/d\lambda)\propto \lambda^{13/3}$. If monitoring SMBBH systems at a wavelength twice the $r$-band wavelength and the typical gap radius is $\sim 2.5$ times of that for the $r$-band case, then the number of the resolvable SMBBHs via the diagnostic size-wavelength signature may increase substantially by a factor of $\sim 20$. However at such a radius, the viscosity driven orbital decay may dominate and the decay timescale may change to about $\propto a^{9/4}$, and thus the residence time is increased by a factor of 3. Therefore, monitoring at a longer wavelength does increase the number of resolvable SMBBHs via the size-wavelength diagnostic but not as significant as in the GW-dominated case. However, it does lead to an enhancement of resolvable SMBBHs via the periodic oscillation signature as larger gap means longer orbital periods, which is limited by the typical search baseline of $10$-years as mentioned earlier in Section~\ref{sec:microlensing variation}. If monitoring SMBBH systems at bluer bands, it leads to smaller number of resolvable SMBBHs mainly caused by the sharp decrease of the SMBBH residence time.

\subsection{Population-model uncertainties}

The modeling of SMBBH orbital evolution is crucial to our predictions, yet it remains subject to substantial uncertainties. If viscous dissipation is inefficient and the orbital evolution of close SMBBHs is driven predominantly by GW emission, the SMBBH residence timescales can be enhanced by several times to more than an order of magnitude, depending on binary parameters. This ``GW-only'' scenario represents a reference case in which gas-driven hardening is minimal, potentially yielding up to thousands of detectable SMBBH systems across the sky under our calculations. In contrast, the CBD-filtered scenario adopted in this work is a simple representative case for inefficient inward migration rather than a detailed model. Current hydrodynamical simulations have not yet reached a unique conclusion on the sign and magnitude of the resulting orbital evolution \cite{2022ApJ...929L..13F}. Our CBD-filtered case is therefore intended only to illustrate how stalling or outward migration may reduce the number of systems that enter the close, short-period regime relevant to our diagnostics. 
Our predictions also depend on the adopted galaxy merger rate, which is subject to substantial uncertainty and may increase by a factor of several, e.g., \cite{2010ApJ...724..915H}. Considering the uncertainties in both galaxy merger modeling and SMBBH evolution, the predicted number of resolvable SMBBHs could increase by a factor of ten or more in the optimistic cases. Our results should be interpreted as a model-dependent range rather than a single robust number.

We further assess the impact of binary mass growth on our results, which might be significant due to sustained accretion, e.g., \cite{2023ARA&A..61..517L,2023MNRAS.518.5059S,2023MNRAS.520.4463L}. Using the widely adopted accretion framework of Ref.~\cite{2020ApJ...901...25D}, we estimate the final masses of SMBBH systems based on their post-accretion mass ratios. Since the binaries in our sample are already near coalescence, we assume a near-equal mass configuration with $q_{\rm{final}}\simeq 0.95$. After including the mass growth of each SMBBH in our lensed sample, using Equations (17--22) in \cite{2020ApJ...901...25D}, we yield 187\% and 254\% more microlensing SMBBH systems for the periodic oscillation approach and the size-wavelength approach, respectively. This simple experiment further highlights the strong dependence of our detection forecasts on the specific assumptions underlying SMBBH evolution and accretion physics. 

Despite the inherent challenges, our research illustrates that strong gravitational lensing, when combined with microlensing monitoring, presents a promising yet observationally demanding approach to identifying close SMBBHs. The forthcoming high-resolution, wide-field surveys will play a pivotal role in resolving the multiple images of these systems. Additionally, sustained time-domain monitoring campaigns will be essential for detecting microlensing signatures that can confirm their binary nature. As the sample of strongly lensed AGNs grows to the level of thousands, targeted follow-up observations of the most promising candidates will provide a valuable opportunity to better understand the abundance and characteristics of sub-parsec SMBBHs in the universe.

\Acknowledgements{This work is partly supported by the National Astronomical Observatory of China (grant no. E4TG660101), the National SKA Program of China and the National Key Program for Science and Technology Research and Development (grant nos. 2022YFC2205201, 2020YFC2201400), the Strategic Priority Program of the Chinese Academy of Sciences (grant no. XDB0550300), and the China Manned Space Program with grant no. CMS-CSST-2025-A07. This work was performed in part at the Aspen Center for Physics, which is supported by National Science Foundation grant PHY-2210452.}

\InterestConflict{The authors declare that they have no conflict of interest.}


\bibliographystyle{scpma}   %
\bibliography{refs}

@ARTICLE{lyl+23,
       author = {{Li}, Zekun and {Yan}, Changshuo and {Lu}, Youjun},
        title = "{A Population Study for Searching Supermassive Binary Black Holes in Active Galactic Nuclei: Continuum Spectral Features and Periodic Variabilities}",
      journal = {MNRAS},
         year = {2026},
       volume = {551},
       number = {1},
          eid = {stag1409},
          doi = {10.1093/mnras/stag1409},
       adsurl = {},
}

@ARTICLE{2024ApJ...962....3C,
       author = {{Chen}, Zhiwei and {Lu}, Youjun and {Chen}, Yunfeng},
        title = "{Detectability of Strongly Gravitationally Lensed Tidal Disruption Events}",
      journal = {\apj},
         year = 2024,
        month = feb,
       volume = {962},
       number = {1},
          eid = {3},
        pages = {3},
          doi = {10.3847/1538-4357/ad19d3},
archivePrefix = {arXiv},
       eprint = {2401.00992},
 primaryClass = {astro-ph.HE},
       adsurl = {https://ui.adsabs.harvard.edu/abs/2024ApJ...962....3C}
}

@ARTICLE{2022A&A...668A..77M,
       author = {{Millon}, M. and {Dalang}, C. and {Lemon}, C. and {Sluse}, D. and {Paic}, E. and {Chan}, J.~H.~H. and {Courbin}, F.},
        title = "{Evidence for a milliparsec-separation supermassive binary black hole with quasar microlensing}",
      journal = {\aap},
         year = 2022,
        month = dec,
       volume = {668},
          eid = {A77},
        pages = {A77},
          doi = {10.1051/0004-6361/202244440},
archivePrefix = {arXiv},
       eprint = {2207.00598},
 primaryClass = {astro-ph.GA},
       adsurl = {https://ui.adsabs.harvard.edu/abs/2022A&A...668A..77M}
}

@ARTICLE{2019ApJ...873..111I,
       author = {{Ivezi{\'c}}, {\v{Z}}eljko and {Kahn}, Steven M. and {Tyson}, J. Anthony and {Abel}, Bob and {Acosta}, Emily and {Allsman}, Robyn and {Alonso}, David and {AlSayyad}, Yusra and {Anderson}, Scott F. and {Andrew}, John and {Angel}, James Roger P. and {Angeli}, George Z. and {Ansari}, Reza and {Antilogus}, Pierre and {Araujo}, Constanza and {Armstrong}, Robert and {Arndt}, Kirk T. and {Astier}, Pierre and {Aubourg}, {\'E}ric and {Auza}, Nicole and {Axelrod}, Tim S. and {Bard}, Deborah J. and {Barr}, Jeff D. and {Barrau}, Aurelian and {Bartlett}, James G. and {Bauer}, Amanda E. and {Bauman}, Brian J. and {Baumont}, Sylvain and {Bechtol}, Ellen and {Bechtol}, Keith and {Becker}, Andrew C. and {Becla}, Jacek and {Beldica}, Cristina and {Bellavia}, Steve and {Bianco}, Federica B. and {Biswas}, Rahul and {Blanc}, Guillaume and {Blazek}, Jonathan and {Blandford}, Roger D. and {Bloom}, Josh S. and {Bogart}, Joanne and {Bond}, Tim W. and {Booth}, Michael T. and {Borgland}, Anders W. and {Borne}, Kirk and {Bosch}, James F. and {Boutigny}, Dominique and {Brackett}, Craig A. and {Bradshaw}, Andrew and {Brandt}, William Nielsen and {Brown}, Michael E. and {Bullock}, James S. and {Burchat}, Patricia and {Burke}, David L. and {Cagnoli}, Gianpietro and {Calabrese}, Daniel and {Callahan}, Shawn and {Callen}, Alice L. and {Carlin}, Jeffrey L. and {Carlson}, Erin L. and {Chandrasekharan}, Srinivasan and {Charles-Emerson}, Glenaver and {Chesley}, Steve and {Cheu}, Elliott C. and {Chiang}, Hsin-Fang and {Chiang}, James and {Chirino}, Carol and {Chow}, Derek and {Ciardi}, David R. and {Claver}, Charles F. and {Cohen-Tanugi}, Johann and {Cockrum}, Joseph J. and {Coles}, Rebecca and {Connolly}, Andrew J. and {Cook}, Kem H. and {Cooray}, Asantha and {Covey}, Kevin R. and {Cribbs}, Chris and {Cui}, Wei and {Cutri}, Roc and {Daly}, Philip N. and {Daniel}, Scott F. and {Daruich}, Felipe and {Daubard}, Guillaume and {Daues}, Greg and {Dawson}, William and {Delgado}, Francisco and {Dellapenna}, Alfred and {de Peyster}, Robert and {de Val-Borro}, Miguel and {Digel}, Seth W. and {Doherty}, Peter and {Dubois}, Richard and {Dubois-Felsmann}, Gregory P. and {Durech}, Josef and {Economou}, Frossie and {Eifler}, Tim and {Eracleous}, Michael and {Emmons}, Benjamin L. and {Fausti Neto}, Angelo and {Ferguson}, Henry and {Figueroa}, Enrique and {Fisher-Levine}, Merlin and {Focke}, Warren and {Foss}, Michael D. and {Frank}, James and {Freemon}, Michael D. and {Gangler}, Emmanuel and {Gawiser}, Eric and {Geary}, John C. and {Gee}, Perry and {Geha}, Marla and {Gessner}, Charles J.~B. and {Gibson}, Robert R. and {Gilmore}, D. Kirk and {Glanzman}, Thomas and {Glick}, William and {Goldina}, Tatiana and {Goldstein}, Daniel A. and {Goodenow}, Iain and {Graham}, Melissa L. and {Gressler}, William J. and {Gris}, Philippe and {Guy}, Leanne P. and {Guyonnet}, Augustin and {Haller}, Gunther and {Harris}, Ron and {Hascall}, Patrick A. and {Haupt}, Justine and {Hernandez}, Fabio and {Herrmann}, Sven and {Hileman}, Edward and {Hoblitt}, Joshua and {Hodgson}, John A. and {Hogan}, Craig and {Howard}, James D. and {Huang}, Dajun and {Huffer}, Michael E. and {Ingraham}, Patrick and {Innes}, Walter R. and {Jacoby}, Suzanne H. and {Jain}, Bhuvnesh and {Jammes}, Fabrice and {Jee}, M. James and {Jenness}, Tim and {Jernigan}, Garrett and {Jevremovi{\'c}}, Darko and {Johns}, Kenneth and {Johnson}, Anthony S. and {Johnson}, Margaret W.~G. and {Jones}, R. Lynne and {Juramy-Gilles}, Claire and {Juri{\'c}}, Mario and {Kalirai}, Jason S. and {Kallivayalil}, Nitya J. and {Kalmbach}, Bryce and {Kantor}, Jeffrey P. and {Karst}, Pierre and {Kasliwal}, Mansi M. and {Kelly}, Heather and {Kessler}, Richard and {Kinnison}, Veronica and {Kirkby}, David and {Knox}, Lloyd and {Kotov}, Ivan V. and {Krabbendam}, Victor L. and {Krughoff}, K. Simon and {Kub{\'a}nek}, Petr and {Kuczewski}, John and {Kulkarni}, Shri and {Ku}, John and {Kurita}, Nadine R. and {Lage}, Craig S. and {Lambert}, Ron and {Lange}, Travis and {Langton}, J. Brian and {Le Guillou}, Laurent and {Levine}, Deborah and {Liang}, Ming and {Lim}, Kian-Tat and {Lintott}, Chris J. and {Long}, Kevin E. and {Lopez}, Margaux and {Lotz}, Paul J. and {Lupton}, Robert H. and {Lust}, Nate B. and {MacArthur}, Lauren A. and {Mahabal}, Ashish and {Mandelbaum}, Rachel and {Markiewicz}, Thomas W. and {Marsh}, Darren S. and {Marshall}, Philip J. and {Marshall}, Stuart and {May}, Morgan and {McKercher}, Robert and {McQueen}, Michelle and {Meyers}, Joshua and {Migliore}, Myriam and {Miller}, Michelle and {Mills}, David J.},
        title = "{LSST: From Science Drivers to Reference Design and Anticipated Data Products}",
      journal = {\apj},
         year = 2019,
        month = mar,
       volume = {873},
       number = {2},
          eid = {111},
        pages = {111},
          doi = {10.3847/1538-4357/ab042c},
archivePrefix = {arXiv},
       eprint = {0805.2366},
 primaryClass = {astro-ph},
       adsurl = {https://ui.adsabs.harvard.edu/abs/2019ApJ...873..111I}
}

@ARTICLE{2014ApJ...784..100Y,
       author = {{Yan}, Chang-Shuo and {Lu}, Youjun and {Yu}, Qingjuan and {Mao}, Shude and {Wambsganss}, Joachim},
        title = "{Microlensing of Sub-parsec Massive Binary Black Holes in Lensed QSOs: Light Curves and Size-Wavelength Relation}",
      journal = {\apj},
         year = 2014,
        month = apr,
       volume = {784},
       number = {2},
          eid = {100},
        pages = {100},
          doi = {10.1088/0004-637X/784/2/100},
archivePrefix = {arXiv},
       eprint = {1402.2504},
 primaryClass = {astro-ph.CO},
       adsurl = {https://ui.adsabs.harvard.edu/abs/2014ApJ...784..100Y}
}

@ARTICLE{2021A&A...653A.121S,
       author = {{Shalyapin}, V.~N. and {Gil-Merino}, R. and {Goicoechea}, L.~J.},
        title = "{Fast simulations of extragalactic microlensing}",
      journal = {\aap},
         year = 2021,
        month = sep,
       volume = {653},
          eid = {A121},
        pages = {A121},
          doi = {10.1051/0004-6361/202140527},
archivePrefix = {arXiv},
       eprint = {2107.07222},
 primaryClass = {astro-ph.IM},
       adsurl = {https://ui.adsabs.harvard.edu/abs/2021A&A...653A.121S}
}

@ARTICLE{2019MNRAS.485.1579K,
       author = {{Kelley}, Luke Zoltan and {Haiman}, Zolt{\'a}n and {Sesana}, Alberto and {Hernquist}, Lars},
        title = "{Massive BH binaries as periodically variable AGN}",
      journal = {\mnras},
         year = 2019,
        month = may,
       volume = {485},
       number = {2},
        pages = {1579-1594},
          doi = {10.1093/mnras/stz150},
archivePrefix = {arXiv},
       eprint = {1809.02138},
 primaryClass = {astro-ph.HE},
       adsurl = {https://ui.adsabs.harvard.edu/abs/2019MNRAS.485.1579K}
}

@ARTICLE{2023arXiv231016896D,
       author = {{D'Orazio}, Daniel J. and {Charisi}, Maria},
        title = "{Observational Signatures of Supermassive Black Hole Binaries}",
      journal = {arXiv e-prints},
         year = 2023,
        month = oct,
          eid = {arXiv:2310.16896},
        pages = {arXiv:2310.16896},
          doi = {10.48550/arXiv.2310.16896},
archivePrefix = {arXiv},
       eprint = {2310.16896},
 primaryClass = {astro-ph.HE},
       adsurl = {https://ui.adsabs.harvard.edu/abs/2023arXiv231016896D}
}

@ARTICLE{2023ApJ...951L...8A,
       author = {{Agazie}, Gabriella and {Anumarlapudi}, Akash and {Archibald}, Anne M. and {Arzoumanian}, Zaven and {Baker}, Paul T. and {B{\'e}csy}, Bence and {Blecha}, Laura and {Brazier}, Adam and {Brook}, Paul R. and {Burke-Spolaor}, Sarah and {Burnette}, Rand and {Case}, Robin and {Charisi}, Maria and {Chatterjee}, Shami and {Chatziioannou}, Katerina and {Cheeseboro}, Belinda D. and {Chen}, Siyuan and {Cohen}, Tyler and {Cordes}, James M. and {Cornish}, Neil J. and {Crawford}, Fronefield and {Cromartie}, H. Thankful and {Crowter}, Kathryn and {Cutler}, Curt J. and {Decesar}, Megan E. and {Degan}, Dallas and {Demorest}, Paul B. and {Deng}, Heling and {Dolch}, Timothy and {Drachler}, Brendan and {Ellis}, Justin A. and {Ferrara}, Elizabeth C. and {Fiore}, William and {Fonseca}, Emmanuel and {Freedman}, Gabriel E. and {Garver-Daniels}, Nate and {Gentile}, Peter A. and {Gersbach}, Kyle A. and {Glaser}, Joseph and {Good}, Deborah C. and {G{\"u}ltekin}, Kayhan and {Hazboun}, Jeffrey S. and {Hourihane}, Sophie and {Islo}, Kristina and {Jennings}, Ross J. and {Johnson}, Aaron D. and {Jones}, Megan L. and {Kaiser}, Andrew R. and {Kaplan}, David L. and {Kelley}, Luke Zoltan and {Kerr}, Matthew and {Key}, Joey S. and {Klein}, Tonia C. and {Laal}, Nima and {Lam}, Michael T. and {Lamb}, William G. and {Lazio}, T. Joseph W. and {Lewandowska}, Natalia and {Littenberg}, Tyson B. and {Liu}, Tingting and {Lommen}, Andrea and {Lorimer}, Duncan R. and {Luo}, Jing and {Lynch}, Ryan S. and {Ma}, Chung-Pei and {Madison}, Dustin R. and {Mattson}, Margaret A. and {McEwen}, Alexander and {McKee}, James W. and {McLaughlin}, Maura A. and {McMann}, Natasha and {Meyers}, Bradley W. and {Meyers}, Patrick M. and {Mingarelli}, Chiara M.~F. and {Mitridate}, Andrea and {Natarajan}, Priyamvada and {Ng}, Cherry and {Nice}, David J. and {Ocker}, Stella Koch and {Olum}, Ken D. and {Pennucci}, Timothy T. and {Perera}, Benetge B.~P. and {Petrov}, Polina and {Pol}, Nihan S. and {Radovan}, Henri A. and {Ransom}, Scott M. and {Ray}, Paul S. and {Romano}, Joseph D. and {Sardesai}, Shashwat C. and {Schmiedekamp}, Ann and {Schmiedekamp}, Carl and {Schmitz}, Kai and {Schult}, Levi and {Shapiro-Albert}, Brent J. and {Siemens}, Xavier and {Simon}, Joseph and {Siwek}, Magdalena S. and {Stairs}, Ingrid H. and {Stinebring}, Daniel R. and {Stovall}, Kevin and {Sun}, Jerry P. and {Susobhanan}, Abhimanyu and {Swiggum}, Joseph K. and {Taylor}, Jacob and {Taylor}, Stephen R. and {Turner}, Jacob E. and {Unal}, Caner and {Vallisneri}, Michele and {van Haasteren}, Rutger and {Vigeland}, Sarah J. and {Wahl}, Haley M. and {Wang}, Qiaohong and {Witt}, Caitlin A. and {Young}, Olivia and {Nanograv Collaboration}},
        title = "{The NANOGrav 15 yr Data Set: Evidence for a Gravitational-wave Background}",
      journal = {\apjl},
         year = 2023,
        month = jul,
       volume = {951},
       number = {1},
          eid = {L8},
        pages = {L8},
          doi = {10.3847/2041-8213/acdac6},
archivePrefix = {arXiv},
       eprint = {2306.16213},
 primaryClass = {astro-ph.HE},
       adsurl = {https://ui.adsabs.harvard.edu/abs/2023ApJ...951L...8A}
}

@ARTICLE{2017MNRAS.464.3131K,
       author = {{Kelley}, Luke Zoltan and {Blecha}, Laura and {Hernquist}, Lars},
        title = "{Massive black hole binary mergers in dynamical galactic environments}",
      journal = {\mnras},
         year = 2017,
        month = jan,
       volume = {464},
       number = {3},
        pages = {3131-3157},
          doi = {10.1093/mnras/stw2452},
archivePrefix = {arXiv},
       eprint = {1606.01900},
 primaryClass = {astro-ph.HE},
       adsurl = {https://ui.adsabs.harvard.edu/abs/2017MNRAS.464.3131K}
}

@ARTICLE{1980Natur.287..307B,
       author = {{Begelman}, M.~C. and {Blandford}, R.~D. and {Rees}, M.~J.},
        title = "{Massive black hole binaries in active galactic nuclei}",
      journal = {\nat},
         year = 1980,
        month = sep,
       volume = {287},
       number = {5780},
        pages = {307-309},
          doi = {10.1038/287307a0},
       adsurl = {https://ui.adsabs.harvard.edu/abs/1980Natur.287..307B}
}

@ARTICLE{2020ApJ...897...86C,
       author = {{Chen}, Yunfeng and {Yu}, Qingjuan and {Lu}, Youjun},
        title = "{Dynamical Evolution of Cosmic Supermassive Binary Black Holes and Their Gravitational-wave Radiation}",
      journal = {\apj},
         year = 2020,
        month = jul,
       volume = {897},
       number = {1},
          eid = {86},
        pages = {86},
          doi = {10.3847/1538-4357/ab9594},
archivePrefix = {arXiv},
       eprint = {2005.10818},
 primaryClass = {astro-ph.HE},
       adsurl = {https://ui.adsabs.harvard.edu/abs/2020ApJ...897...86C}
}

@ARTICLE{2013ARA&A..51..511K,
       author = {{Kormendy}, John and {Ho}, Luis C.},
        title = "{Coevolution (Or Not) of Supermassive Black Holes and Host Galaxies}",
      journal = {\araa},
         year = 2013,
        month = aug,
       volume = {51},
       number = {1},
        pages = {511-653},
          doi = {10.1146/annurev-astro-082708-101811},
archivePrefix = {arXiv},
       eprint = {1304.7762},
 primaryClass = {astro-ph.CO},
       adsurl = {https://ui.adsabs.harvard.edu/abs/2013ARA&A..51..511K}
}

@ARTICLE{2017MNRAS.471.3098L,
       author = {{Lopes}, Amanda R. and {Gruppioni}, C. and {Ribeiro}, M.~B. and {Pozzetti}, L. and {February}, S. and {Ilbert}, O. and {Pozzi}, F.},
        title = "{Effect of different cosmologies on the galaxy stellar mass function}",
      journal = {\mnras},
         year = 2017,
        month = nov,
       volume = {471},
       number = {3},
        pages = {3098-3111},
          doi = {10.1093/mnras/stx1799},
       adsurl = {https://ui.adsabs.harvard.edu/abs/2017MNRAS.471.3098L}
}

@ARTICLE{2015ApJ...809..117Y,
       author = {{Yan}, Chang-Shuo and {Lu}, Youjun and {Dai}, Xinyu and {Yu}, Qingjuan},
        title = "{A Probable Milli-parsec Supermassive Binary Black Hole in the Nearest Quasar Mrk 231}",
      journal = {\apj},
         year = 2015,
        month = aug,
       volume = {809},
       number = {2},
          eid = {117},
        pages = {117},
          doi = {10.1088/0004-637X/809/2/117},
archivePrefix = {arXiv},
       eprint = {1508.06292},
 primaryClass = {astro-ph.HE},
       adsurl = {https://ui.adsabs.harvard.edu/abs/2015ApJ...809..117Y}
}

@ARTICLE{2002MNRAS.331..935Y,
       author = {{Yu}, Qingjuan},
        title = "{Evolution of massive binary black holes}",
      journal = {\mnras},
         year = 2002,
        month = apr,
       volume = {331},
       number = {4},
        pages = {935-958},
          doi = {10.1046/j.1365-8711.2002.05242.x},
archivePrefix = {arXiv},
       eprint = {astro-ph/0109530},
 primaryClass = {astro-ph},
       adsurl = {https://ui.adsabs.harvard.edu/abs/2002MNRAS.331..935Y}
}

@ARTICLE{2008ApJ...689..732Y,
       author = {{Yu}, Qingjuan and {Lu}, Youjun},
        title = "{Toward Precise Constraints on the Growth of Massive Black Holes}",
      journal = {\apj},
         year = 2008,
        month = dec,
       volume = {689},
       number = {2},
        pages = {732-754},
          doi = {10.1086/592770},
archivePrefix = {arXiv},
       eprint = {0808.3777},
 primaryClass = {astro-ph},
       adsurl = {https://ui.adsabs.harvard.edu/abs/2008ApJ...689..732Y}
}

@ARTICLE{1983ApJ...268..368E,
       author = {{Eggleton}, P.~P.},
        title = "{Aproximations to the radii of Roche lobes.}",
      journal = {\apj},
         year = 1983,
        month = may,
       volume = {268},
        pages = {368-369},
          doi = {10.1086/160960},
       adsurl = {https://ui.adsabs.harvard.edu/abs/1983ApJ...268..368E}
}

@ARTICLE{2014ApJ...783..134F,
       author = {{Farris}, Brian D. and {Duffell}, Paul and {MacFadyen}, Andrew I. and {Haiman}, Zoltan},
        title = "{Binary Black Hole Accretion from a Circumbinary Disk: Gas Dynamics inside the Central Cavity}",
      journal = {\apj},
         year = 2014,
        month = mar,
       volume = {783},
       number = {2},
          eid = {134},
        pages = {134},
          doi = {10.1088/0004-637X/783/2/134},
archivePrefix = {arXiv},
       eprint = {1310.0492},
 primaryClass = {astro-ph.HE},
       adsurl = {https://ui.adsabs.harvard.edu/abs/2014ApJ...783..134F}
}

@ARTICLE{2015MNRAS.446L..36F,
       author = {{Farris}, B.~D. and {Duffell}, P. and {MacFadyen}, A.~I. and {Haiman}, Z.},
        title = "{Characteristic signatures in the thermal emission from accreting binary black holes.}",
      journal = {\mnras},
         year = 2015,
        month = jan,
       volume = {446},
        pages = {L36-L40},
          doi = {10.1093/mnrasl/slu160},
archivePrefix = {arXiv},
       eprint = {1406.0007},
 primaryClass = {astro-ph.HE},
       adsurl = {https://ui.adsabs.harvard.edu/abs/2015MNRAS.446L..36F}
}

@ARTICLE{2012ApJ...761...90G,
       author = {{G{\"u}ltekin}, Kayhan and {Miller}, Jon M.},
        title = "{Observable Consequences of Merger-driven Gaps and Holes in Black Hole Accretion Disks}",
      journal = {\apj},
         year = 2012,
        month = dec,
       volume = {761},
       number = {2},
          eid = {90},
        pages = {90},
          doi = {10.1088/0004-637X/761/2/90},
archivePrefix = {arXiv},
       eprint = {1207.0296},
 primaryClass = {astro-ph.HE},
       adsurl = {https://ui.adsabs.harvard.edu/abs/2012ApJ...761...90G}
}

@ARTICLE{1996ApJ...467L..77A,
       author = {{Artymowicz}, Pawel and {Lubow}, Stephen H.},
        title = "{Mass Flow through Gaps in Circumbinary Disks}",
      journal = {\apjl},
         year = 1996,
        month = aug,
       volume = {467},
        pages = {L77},
          doi = {10.1086/310200},
       adsurl = {https://ui.adsabs.harvard.edu/abs/1996ApJ...467L..77A}
}

@ARTICLE{2002ApJ...567L...9A,
       author = {{Armitage}, Philip J. and {Natarajan}, Priyamvada},
        title = "{Accretion during the Merger of Supermassive Black Holes}",
      journal = {\apjl},
         year = 2002,
        month = mar,
       volume = {567},
       number = {1},
        pages = {L9-L12},
          doi = {10.1086/339770},
archivePrefix = {arXiv},
       eprint = {astro-ph/0201318},
 primaryClass = {astro-ph},
       adsurl = {https://ui.adsabs.harvard.edu/abs/2002ApJ...567L...9A}
}

@ARTICLE{2008ApJ...672...83M,
       author = {{MacFadyen}, Andrew I. and {Milosavljevi{\'c}}, Milo{\v{s}}},
        title = "{An Eccentric Circumbinary Accretion Disk and the Detection of Binary Massive Black Holes}",
      journal = {\apj},
         year = 2008,
        month = jan,
       volume = {672},
       number = {1},
        pages = {83-93},
          doi = {10.1086/523869},
archivePrefix = {arXiv},
       eprint = {astro-ph/0607467},
 primaryClass = {astro-ph},
       adsurl = {https://ui.adsabs.harvard.edu/abs/2008ApJ...672...83M}
}

@ARTICLE{2023ARA&A..61..517L,
       author = {{Lai}, Dong and {Mu{\~n}oz}, Diego J.},
        title = "{Circumbinary Accretion: From Binary Stars to Massive Binary Black Holes}",
      journal = {\araa},
         year = 2023,
        month = aug,
       volume = {61},
        pages = {517-560},
          doi = {10.1146/annurev-astro-052622-022933},
archivePrefix = {arXiv},
       eprint = {2211.00028},
 primaryClass = {astro-ph.HE},
       adsurl = {https://ui.adsabs.harvard.edu/abs/2023ARA&A..61..517L}
}

@ARTICLE{2009ApJ...700.1952H,
       author = {{Haiman}, Zolt{\'a}n and {Kocsis}, Bence and {Menou}, Kristen},
        title = "{The Population of Viscosity- and Gravitational Wave-driven Supermassive Black Hole Binaries Among Luminous Active Galactic Nuclei}",
      journal = {\apj},
         year = 2009,
        month = aug,
       volume = {700},
       number = {2},
        pages = {1952-1969},
          doi = {10.1088/0004-637X/700/2/1952},
archivePrefix = {arXiv},
       eprint = {0904.1383},
 primaryClass = {astro-ph.CO},
       adsurl = {https://ui.adsabs.harvard.edu/abs/2009ApJ...700.1952H}
}

@ARTICLE{1973A&A....24..337S,
       author = {{Shakura}, N.~I. and {Sunyaev}, R.~A.},
        title = "{Black holes in binary systems. Observational appearance.}",
      journal = {\aap},
         year = 1973,
        month = jan,
       volume = {24},
        pages = {337-355},
       adsurl = {https://ui.adsabs.harvard.edu/abs/1973A&A....24..337S}
}

@ARTICLE{2020ApJ...901...25D,
       author = {{Duffell}, Paul C. and {D'Orazio}, Daniel and {Derdzinski}, Andrea and {Haiman}, Zoltan and {MacFadyen}, Andrew and {Rosen}, Anna L. and {Zrake}, Jonathan},
        title = "{Circumbinary Disks: Accretion and Torque as a Function of Mass Ratio and Disk Viscosity}",
      journal = {\apj},
         year = 2020,
        month = sep,
       volume = {901},
       number = {1},
          eid = {25},
        pages = {25},
          doi = {10.3847/1538-4357/abab95},
archivePrefix = {arXiv},
       eprint = {1911.05506},
 primaryClass = {astro-ph.SR},
       adsurl = {https://ui.adsabs.harvard.edu/abs/2020ApJ...901...25D}
}

@ARTICLE{2015MNRAS.447.2772R,
       author = {{Ravi}, V. and {Wyithe}, J.~S.~B. and {Shannon}, R.~M. and {Hobbs}, G.},
        title = "{Prospects for gravitational-wave detection and supermassive black hole astrophysics with pulsar timing arrays}",
      journal = {\mnras},
         year = 2015,
        month = mar,
       volume = {447},
       number = {3},
        pages = {2772-2783},
          doi = {10.1093/mnras/stu2659},
archivePrefix = {arXiv},
       eprint = {1406.5297},
 primaryClass = {astro-ph.CO},
       adsurl = {https://ui.adsabs.harvard.edu/abs/2015MNRAS.447.2772R}
}

@ARTICLE{2015MNRAS.449...49R,
       author = {{Rodriguez-Gomez}, Vicente and {Genel}, Shy and {Vogelsberger}, Mark and {Sijacki}, Debora and {Pillepich}, Annalisa and {Sales}, Laura V. and {Torrey}, Paul and {Snyder}, Greg and {Nelson}, Dylan and {Springel}, Volker and {Ma}, Chung-Pei and {Hernquist}, Lars},
        title = "{The merger rate of galaxies in the Illustris simulation: a comparison with observations and semi-empirical models}",
      journal = {\mnras},
         year = 2015,
        month = may,
       volume = {449},
       number = {1},
        pages = {49-64},
          doi = {10.1093/mnras/stv264},
archivePrefix = {arXiv},
       eprint = {1502.01339},
 primaryClass = {astro-ph.GA},
       adsurl = {https://ui.adsabs.harvard.edu/abs/2015MNRAS.449...49R}
}

@ARTICLE{2022LRR....25....3B,
       author = {{Bogdanovi{\'c}}, Tamara and {Miller}, M. Coleman and {Blecha}, Laura},
        title = "{Electromagnetic counterparts to massive black-hole mergers}",
      journal = {Living Reviews in Relativity},
         year = 2022,
        month = dec,
       volume = {25},
       number = {1},
          eid = {3},
        pages = {3},
          doi = {10.1007/s41114-022-00037-8},
archivePrefix = {arXiv},
       eprint = {2109.03262},
 primaryClass = {astro-ph.HE},
       adsurl = {https://ui.adsabs.harvard.edu/abs/2022LRR....25....3B}
}

@ARTICLE{2011ApJ...738...92Y,
       author = {{Yu}, Qingjuan and {Lu}, Youjun and {Mohayaee}, Roya and {Colin}, Jacques},
        title = "{The Low Frequency of Dual Active Galactic Nuclei versus the High Merger Rate of Galaxies: A Phenomenological Model}",
      journal = {\apj},
         year = 2011,
        month = sep,
       volume = {738},
       number = {1},
          eid = {92},
        pages = {92},
          doi = {10.1088/0004-637X/738/1/92},
archivePrefix = {arXiv},
       eprint = {1105.1963},
 primaryClass = {astro-ph.CO},
       adsurl = {https://ui.adsabs.harvard.edu/abs/2011ApJ...738...92Y}
}

@ARTICLE{2015MNRAS.452..575S,
       author = {{Sijacki}, Debora and {Vogelsberger}, Mark and {Genel}, Shy and {Springel}, Volker and {Torrey}, Paul and {Snyder}, Gregory F. and {Nelson}, Dylan and {Hernquist}, Lars},
        title = "{The Illustris simulation: the evolving population of black holes across cosmic time}",
      journal = {\mnras},
         year = 2015,
        month = sep,
       volume = {452},
       number = {1},
        pages = {575-596},
          doi = {10.1093/mnras/stv1340},
archivePrefix = {arXiv},
       eprint = {1408.6842},
 primaryClass = {astro-ph.GA},
       adsurl = {https://ui.adsabs.harvard.edu/abs/2015MNRAS.452..575S}
}

@ARTICLE{2024MNRAS.534.2609S,
       author = {{Siwek}, Magdalena and {Kelley}, Luke Zoltan and {Hernquist}, Lars},
        title = "{Signatures of circumbinary disc dynamics in multimessenger population studies of massive black hole binaries}",
      journal = {\mnras},
         year = 2024,
        month = nov,
       volume = {534},
       number = {3},
        pages = {2609-2620},
          doi = {10.1093/mnras/stae2251},
archivePrefix = {arXiv},
       eprint = {2403.08871},
 primaryClass = {astro-ph.HE},
       adsurl = {https://ui.adsabs.harvard.edu/abs/2024MNRAS.534.2609S}
}

@ARTICLE{2020MNRAS.495.4681I,
       author = {{Izquierdo-Villalba}, David and {Bonoli}, Silvia and {Dotti}, Massimo and {Sesana}, Alberto and {Rosas-Guevara}, Yetli and {Spinoso}, Daniele},
        title = "{From galactic nuclei to the halo outskirts: tracing supermassive black holes across cosmic history and environments}",
      journal = {\mnras},
         year = 2020,
        month = jul,
       volume = {495},
       number = {4},
        pages = {4681-4706},
          doi = {10.1093/mnras/staa1399},
archivePrefix = {arXiv},
       eprint = {2001.10548},
 primaryClass = {astro-ph.GA},
       adsurl = {https://ui.adsabs.harvard.edu/abs/2020MNRAS.495.4681I}
}

@ARTICLE{2014ARA&A..52..529Y,
       author = {{Yuan}, Feng and {Narayan}, Ramesh},
        title = "{Hot Accretion Flows Around Black Holes}",
      journal = {\araa},
         year = 2014,
        month = aug,
       volume = {52},
        pages = {529-588},
          doi = {10.1146/annurev-astro-082812-141003},
archivePrefix = {arXiv},
       eprint = {1401.0586},
 primaryClass = {astro-ph.HE},
       adsurl = {https://ui.adsabs.harvard.edu/abs/2014ARA&A..52..529Y}
}

@ARTICLE{2023MNRAS.522.2707S,
       author = {{Siwek}, Magdalena and {Weinberger}, Rainer and {Hernquist}, Lars},
        title = "{Orbital evolution of binaries in circumbinary discs}",
      journal = {\mnras},
         year = 2023,
        month = jun,
       volume = {522},
       number = {2},
        pages = {2707-2717},
          doi = {10.1093/mnras/stad1131},
archivePrefix = {arXiv},
       eprint = {2302.01785},
 primaryClass = {astro-ph.HE},
       adsurl = {https://ui.adsabs.harvard.edu/abs/2023MNRAS.522.2707S}
}

@ARTICLE{2023Natur.616...45C,
       author = {{Chen}, Yu-Ching and {Liu}, Xin and {Foord}, Adi and {Shen}, Yue and {Oguri}, Masamune and {Chen}, Nianyi and {Di Matteo}, Tiziana and {Holgado}, Miguel and {Hwang}, Hsiang-Chih and {Zakamska}, Nadia},
        title = "{A close quasar pair in a disk-disk galaxy merger at z = 2.17}",
      journal = {\nat},
         year = 2023,
        month = apr,
       volume = {616},
       number = {7955},
        pages = {45-49},
          doi = {10.1038/s41586-023-05766-6},
archivePrefix = {arXiv},
       eprint = {2209.11249},
 primaryClass = {astro-ph.CO},
       adsurl = {https://ui.adsabs.harvard.edu/abs/2023Natur.616...45C}
}

@ARTICLE{2016A&A...587A..41P,
       author = {{Palanque-Delabrouille}, N. and {Magneville}, Ch. and {Y{\`e}che}, Ch. and {P{\^a}ris}, I. and {Petitjean}, P. and {Burtin}, E. and {Dawson}, K. and {McGreer}, I. and {Myers}, A.~D. and {Rossi}, G. and {Schlegel}, D. and {Schneider}, D. and {Streblyanska}, A. and {Tinker}, J.},
        title = "{The extended Baryon Oscillation Spectroscopic Survey: Variability selection and quasar luminosity function}",
      journal = {\aap},
         year = 2016,
        month = mar,
       volume = {587},
          eid = {A41},
        pages = {A41},
          doi = {10.1051/0004-6361/201527392},
archivePrefix = {arXiv},
       eprint = {1509.05607},
 primaryClass = {astro-ph.CO},
       adsurl = {https://ui.adsabs.harvard.edu/abs/2016A&A...587A..41P}
}

@ARTICLE{2012MNRAS.420..860S,
       author = {{Sesana}, A. and {Roedig}, C. and {Reynolds}, M.~T. and {Dotti}, M.},
        title = "{Multimessenger astronomy with pulsar timing and X-ray observations of massive black hole binaries}",
      journal = {\mnras},
         year = 2012,
        month = feb,
       volume = {420},
       number = {1},
        pages = {860-877},
          doi = {10.1111/j.1365-2966.2011.20097.x},
archivePrefix = {arXiv},
       eprint = {1107.2927},
 primaryClass = {astro-ph.CO},
       adsurl = {https://ui.adsabs.harvard.edu/abs/2012MNRAS.420..860S}
}

@ARTICLE{2015MNRAS.453.1562G,
       author = {{Graham}, Matthew J. and {Djorgovski}, S.~G. and {Stern}, Daniel and {Drake}, Andrew J. and {Mahabal}, Ashish A. and {Donalek}, Ciro and {Glikman}, Eilat and {Larson}, Steve and {Christensen}, Eric},
        title = "{A systematic search for close supermassive black hole binaries in the Catalina Real-time Transient Survey}",
      journal = {\mnras},
         year = 2015,
        month = oct,
       volume = {453},
       number = {2},
        pages = {1562-1576},
          doi = {10.1093/mnras/stv1726},
archivePrefix = {arXiv},
       eprint = {1507.07603},
 primaryClass = {astro-ph.GA},
       adsurl = {https://ui.adsabs.harvard.edu/abs/2015MNRAS.453.1562G}
}

@ARTICLE{2012MNRAS.420..705T,
       author = {{Tanaka}, Takamitsu and {Menou}, Kristen and {Haiman}, Zolt{\'a}n.},
        title = "{Electromagnetic counterparts of supermassive black hole binaries resolved by pulsar timing arrays}",
      journal = {\mnras},
         year = 2012,
        month = feb,
       volume = {420},
       number = {1},
        pages = {705-719},
          doi = {10.1111/j.1365-2966.2011.20083.x},
archivePrefix = {arXiv},
       eprint = {1107.2937},
 primaryClass = {astro-ph.CO},
       adsurl = {https://ui.adsabs.harvard.edu/abs/2012MNRAS.420..705T}
}

@ARTICLE{2012MNRAS.427.2660K,
       author = {{Kocsis}, Bence and {Haiman}, Zolt{\'a}n. and {Loeb}, Abraham},
        title = "{Gas pile-up, gap overflow and Type 1.5 migration in circumbinary discs: general theory}",
      journal = {\mnras},
         year = 2012,
        month = dec,
       volume = {427},
       number = {3},
        pages = {2660-2679},
          doi = {10.1111/j.1365-2966.2012.22129.x},
archivePrefix = {arXiv},
       eprint = {1205.4714},
 primaryClass = {astro-ph.EP},
       adsurl = {https://ui.adsabs.harvard.edu/abs/2012MNRAS.427.2660K}
}

@ARTICLE{2002MNRAS.335..965Y,
       author = {{Yu}, Qingjuan and {Tremaine}, Scott},
        title = "{Observational constraints on growth of massive black holes}",
      journal = {\mnras},
         year = 2002,
        month = oct,
       volume = {335},
       number = {4},
        pages = {965-976},
          doi = {10.1046/j.1365-8711.2002.05532.x},
archivePrefix = {arXiv},
       eprint = {astro-ph/0203082},
 primaryClass = {astro-ph},
       adsurl = {https://ui.adsabs.harvard.edu/abs/2002MNRAS.335..965Y}
}

@ARTICLE{2021MNRAS.507.1458F,
       author = {{Franchini}, Alessia and {Sesana}, Alberto and {Dotti}, Massimo},
        title = "{Circumbinary disc self-gravity governing supermassive black hole binary mergers}",
      journal = {\mnras},
         year = 2021,
        month = oct,
       volume = {507},
       number = {1},
        pages = {1458-1467},
          doi = {10.1093/mnras/stab2234},
archivePrefix = {arXiv},
       eprint = {2106.13253},
 primaryClass = {astro-ph.HE},
       adsurl = {https://ui.adsabs.harvard.edu/abs/2021MNRAS.507.1458F}
}

@ARTICLE{2020ApJ...900...43T,
       author = {{Tiede}, Christopher and {Zrake}, Jonathan and {MacFadyen}, Andrew and {Haiman}, Zoltan},
        title = "{Gas-driven Inspiral of Binaries in Thin Accretion Disks}",
      journal = {\apj},
         year = 2020,
        month = sep,
       volume = {900},
       number = {1},
          eid = {43},
        pages = {43},
          doi = {10.3847/1538-4357/aba432},
archivePrefix = {arXiv},
       eprint = {2005.09555},
 primaryClass = {astro-ph.GA},
       adsurl = {https://ui.adsabs.harvard.edu/abs/2020ApJ...900...43T}
}

@ARTICLE{2017MNRAS.469.4258T,
       author = {{Tang}, Yike and {MacFadyen}, Andrew and {Haiman}, Zolt{\'a}n},
        title = "{On the orbital evolution of supermassive black hole binaries with circumbinary accretion discs}",
      journal = {\mnras},
         year = 2017,
        month = aug,
       volume = {469},
       number = {4},
        pages = {4258-4267},
          doi = {10.1093/mnras/stx1130},
archivePrefix = {arXiv},
       eprint = {1703.03913},
 primaryClass = {astro-ph.HE},
       adsurl = {https://ui.adsabs.harvard.edu/abs/2017MNRAS.469.4258T}
}

@ARTICLE{2022A&A...660A.101P,
       author = {{Penzlin}, Anna B.~T. and {Kley}, Wilhelm and {Audiffren}, Hugo and {Sch{\"a}fer}, Christoph M.},
        title = "{Binary orbital evolution driven by a circumbinary disc}",
      journal = {\aap},
         year = 2022,
        month = apr,
       volume = {660},
          eid = {A101},
        pages = {A101},
          doi = {10.1051/0004-6361/202141399},
archivePrefix = {arXiv},
       eprint = {2202.06681},
 primaryClass = {astro-ph.SR},
       adsurl = {https://ui.adsabs.harvard.edu/abs/2022A&A...660A.101P}
}

@ARTICLE{2016ApJ...832...22S,
       author = {{Shi}, Ji-Ming and {Krolik}, Julian H.},
        title = "{How Bright are the Gaps in Circumbinary Disk Systems?}",
      journal = {\apj},
         year = 2016,
        month = nov,
       volume = {832},
       number = {1},
          eid = {22},
        pages = {22},
          doi = {10.3847/0004-637X/832/1/22},
archivePrefix = {arXiv},
       eprint = {1609.07110},
 primaryClass = {astro-ph.HE},
       adsurl = {https://ui.adsabs.harvard.edu/abs/2016ApJ...832...22S}
}

@article{Zhan2021,
  author = {Hu Zhan},
  title = {The wide-field multiband imaging and slitless spectroscopy survey to be carried out by the Survey Space Telescope of China Manned Space Program},
  journal = {Chinese Science Bulletin},
  year = {2021},
  volume = {66},
  number = {11},
  pages = {1290-1298},
  url = {http://www.sciengine.com/publisher/Science China Press/journal/Chinese Science Bulletin/66/11/10.1360/TB-2021-0016},
  doi = {10.1360/TB-2021-0016}
}

@ARTICLE{2010MNRAS.404.2087B,
       author = {{Bernardi}, M. and {Shankar}, F. and {Hyde}, J.~B. and {Mei}, S. and {Marulli}, F. and {Sheth}, R.~K.},
        title = "{Galaxy luminosities, stellar masses, sizes, velocity dispersions as a function of morphological type}",
      journal = {\mnras},
         year = 2010,
        month = jun,
       volume = {404},
       number = {4},
        pages = {2087-2122},
          doi = {10.1111/j.1365-2966.2010.16425.x},
archivePrefix = {arXiv},
       eprint = {0910.1093},
 primaryClass = {astro-ph.CO},
       adsurl = {https://ui.adsabs.harvard.edu/abs/2010MNRAS.404.2087B}
}

@ARTICLE{2009ApJ...696..411W,
       author = {{Weinzirl}, Tim and {Jogee}, Shardha and {Khochfar}, Sadegh and {Burkert}, Andreas and {Kormendy}, John},
        title = "{Bulge n and B/T in High-Mass Galaxies: Constraints on the Origin of Bulges in Hierarchical Models}",
      journal = {\apj},
         year = 2009,
        month = may,
       volume = {696},
       number = {1},
        pages = {411-447},
          doi = {10.1088/0004-637X/696/1/411},
archivePrefix = {arXiv},
       eprint = {0807.0040},
 primaryClass = {astro-ph},
       adsurl = {https://ui.adsabs.harvard.edu/abs/2009ApJ...696..411W}
}

@ARTICLE{2015Natur.525..351D,
       author = {{D'Orazio}, Daniel J. and {Haiman}, Zolt{\'a}n and {Schiminovich}, David},
        title = "{Relativistic boost as the cause of periodicity in a massive black-hole binary candidate}",
      journal = {\nat},
         year = 2015,
        month = sep,
       volume = {525},
       number = {7569},
        pages = {351-353},
          doi = {10.1038/nature15262},
archivePrefix = {arXiv},
       eprint = {1509.04301},
 primaryClass = {astro-ph.HE},
       adsurl = {https://ui.adsabs.harvard.edu/abs/2015Natur.525..351D}
}

@ARTICLE{2021ApJ...910..101J,
       author = {{Ji}, Xiang and {Lu}, Youjun and {Ge}, Junqiang and {Yan}, Changshuo and {Song}, Zihao},
        title = "{Variation of Broad Emission Lines from QSOs with Optical/UV Periodicity to Test the Interpretation of Supermassive Binary Black Holes}",
      journal = {\apj},
         year = 2021,
        month = apr,
       volume = {910},
       number = {2},
          eid = {101},
        pages = {101},
          doi = {10.3847/1538-4357/abe386},
archivePrefix = {arXiv},
       eprint = {2103.16448},
 primaryClass = {astro-ph.GA},
       adsurl = {https://ui.adsabs.harvard.edu/abs/2021ApJ...910..101J}
}

@ARTICLE{2016MNRAS.463.2145C,
       author = {{Charisi}, M. and {Bartos}, I. and {Haiman}, Z. and {Price-Whelan}, A.~M. and {Graham}, M.~J. and {Bellm}, E.~C. and {Laher}, R.~R. and {M{\'a}rka}, S.},
        title = "{A population of short-period variable quasars from PTF as supermassive black hole binary candidates}",
      journal = {\mnras},
         year = 2016,
        month = dec,
       volume = {463},
       number = {2},
        pages = {2145-2171},
          doi = {10.1093/mnras/stw1838},
archivePrefix = {arXiv},
       eprint = {1604.01020},
 primaryClass = {astro-ph.GA},
       adsurl = {https://ui.adsabs.harvard.edu/abs/2016MNRAS.463.2145C}
}

@ARTICLE{2022NatAs...6.1185M,
       author = {{Mannucci}, F. and {Pancino}, E. and {Belfiore}, F. and {Cicone}, C. and {Ciurlo}, A. and {Cresci}, G. and {Lusso}, E. and {Marasco}, A. and {Marconi}, A. and {Nardini}, E. and {Pinna}, E. and {Severgnini}, P. and {Saracco}, P. and {Tozzi}, G. and {Yeh}, S.},
        title = "{Unveiling the population of dual and lensed active galactic nuclei at sub-arcsec separations}",
      journal = {Nature Astronomy},
         year = 2022,
        month = aug,
       volume = {6},
        pages = {1185-1192},
          doi = {10.1038/s41550-022-01761-5},
archivePrefix = {arXiv},
       eprint = {2203.11234},
 primaryClass = {astro-ph.GA},
       adsurl = {https://ui.adsabs.harvard.edu/abs/2022NatAs...6.1185M}
}

@ARTICLE{2023ApJ...951L...6R,
       author = {{Reardon}, Daniel J. and {Zic}, Andrew and {Shannon}, Ryan M. and {Hobbs}, George B. and {Bailes}, Matthew and {Di Marco}, Valentina and {Kapur}, Agastya and {Rogers}, Axl F. and {Thrane}, Eric and {Askew}, Jacob and {Bhat}, N.~D. Ramesh and {Cameron}, Andrew and {Cury{\l}o}, Ma{\l}gorzata and {Coles}, William A. and {Dai}, Shi and {Goncharov}, Boris and {Kerr}, Matthew and {Kulkarni}, Atharva and {Levin}, Yuri and {Lower}, Marcus E. and {Manchester}, Richard N. and {Mandow}, Rami and {Miles}, Matthew T. and {Nathan}, Rowina S. and {Os{\l}owski}, Stefan and {Russell}, Christopher J. and {Spiewak}, Ren{\'e}e and {Zhang}, Songbo and {Zhu}, Xing-Jiang},
        title = "{Search for an Isotropic Gravitational-wave Background with the Parkes Pulsar Timing Array}",
      journal = {\apjl},
         year = 2023,
        month = jul,
       volume = {951},
       number = {1},
          eid = {L6},
        pages = {L6},
          doi = {10.3847/2041-8213/acdd02},
archivePrefix = {arXiv},
       eprint = {2306.16215},
 primaryClass = {astro-ph.HE},
       adsurl = {https://ui.adsabs.harvard.edu/abs/2023ApJ...951L...6R}
}

@ARTICLE{2023A&A...678A..50E,
       author = {{EPTA Collaboration} and {InPTA Collaboration} and {Antoniadis}, J. and {Arumugam}, P. and {Arumugam}, S. and {Babak}, S. and {Bagchi}, M. and {Bak Nielsen}, A. -S. and {Bassa}, C.~G. and {Bathula}, A. and {Berthereau}, A. and {Bonetti}, M. and {Bortolas}, E. and {Brook}, P.~R. and {Burgay}, M. and {Caballero}, R.~N. and {Chalumeau}, A. and {Champion}, D.~J. and {Chanlaridis}, S. and {Chen}, S. and {Cognard}, I. and {Dandapat}, S. and {Deb}, D. and {Desai}, S. and {Desvignes}, G. and {Dhanda-Batra}, N. and {Dwivedi}, C. and {Falxa}, M. and {Ferdman}, R.~D. and {Franchini}, A. and {Gair}, J.~R. and {Goncharov}, B. and {Gopakumar}, A. and {Graikou}, E. and {Grie{\ss}meier}, J. -M. and {Guillemot}, L. and {Guo}, Y.~J. and {Gupta}, Y. and {Hisano}, S. and {Hu}, H. and {Iraci}, F. and {Izquierdo-Villalba}, D. and {Jang}, J. and {Jawor}, J. and {Janssen}, G.~H. and {Jessner}, A. and {Joshi}, B.~C. and {Kareem}, F. and {Karuppusamy}, R. and {Keane}, E.~F. and {Keith}, M.~J. and {Kharbanda}, D. and {Kikunaga}, T. and {Kolhe}, N. and {Kramer}, M. and {Krishnakumar}, M.~A. and {Lackeos}, K. and {Lee}, K.~J. and {Liu}, K. and {Liu}, Y. and {Lyne}, A.~G. and {McKee}, J.~W. and {Maan}, Y. and {Main}, R.~A. and {Mickaliger}, M.~B. and {Ni{\c{t}}u}, I.~C. and {Nobleson}, K. and {Paladi}, A.~K. and {Parthasarathy}, A. and {Perera}, B.~B.~P. and {Perrodin}, D. and {Petiteau}, A. and {Porayko}, N.~K. and {Possenti}, A. and {Prabu}, T. and {Quelquejay Leclere}, H. and {Rana}, P. and {Samajdar}, A. and {Sanidas}, S.~A. and {Sesana}, A. and {Shaifullah}, G. and {Singha}, J. and {Speri}, L. and {Spiewak}, R. and {Srivastava}, A. and {Stappers}, B.~W. and {Surnis}, M. and {Susarla}, S.~C. and {Susobhanan}, A. and {Takahashi}, K. and {Tarafdar}, P. and {Theureau}, G. and {Tiburzi}, C. and {van der Wateren}, E. and {Vecchio}, A. and {Venkatraman Krishnan}, V. and {Verbiest}, J.~P.~W. and {Wang}, J. and {Wang}, L. and {Wu}, Z.},
        title = "{The second data release from the European Pulsar Timing Array. III. Search for gravitational wave signals}",
      journal = {\aap},
         year = 2023,
        month = oct,
       volume = {678},
          eid = {A50},
        pages = {A50},
          doi = {10.1051/0004-6361/202346844},
archivePrefix = {arXiv},
       eprint = {2306.16214},
 primaryClass = {astro-ph.HE},
       adsurl = {https://ui.adsabs.harvard.edu/abs/2023A&A...678A..50E}
}

@ARTICLE{2023RAA....23g5024X,
       author = {{Xu}, Heng and {Chen}, Siyuan and {Guo}, Yanjun and {Jiang}, Jinchen and {Wang}, Bojun and {Xu}, Jiangwei and {Xue}, Zihan and {Caballero}, R. Nicolas and {Yuan}, Jianping and {Xu}, Yonghua and {Wang}, Jingbo and {Hao}, Longfei and {Luo}, Jingtao and {Lee}, Kejia and {Han}, Jinlin and {Jiang}, Peng and {Shen}, Zhiqiang and {Wang}, Min and {Wang}, Na and {Xu}, Renxin and {Wu}, Xiangping and {Manchester}, Richard and {Qian}, Lei and {Guan}, Xin and {Huang}, Menglin and {Sun}, Chun and {Zhu}, Yan},
        title = "{Searching for the Nano-Hertz Stochastic Gravitational Wave Background with the Chinese Pulsar Timing Array Data Release I}",
      journal = {Research in Astronomy and Astrophysics},
         year = 2023,
        month = jul,
       volume = {23},
       number = {7},
          eid = {075024},
        pages = {075024},
          doi = {10.1088/1674-4527/acdfa5},
archivePrefix = {arXiv},
       eprint = {2306.16216},
 primaryClass = {astro-ph.HE},
       adsurl = {https://ui.adsabs.harvard.edu/abs/2023RAA....23g5024X}
}

@ARTICLE{2012ApJS..201...23E,
       author = {{Eracleous}, Michael and {Boroson}, Todd A. and {Halpern}, Jules P. and {Liu}, Jia},
        title = "{A Large Systematic Search for Close Supermassive Binary and Rapidly Recoiling Black Holes}",
      journal = {\apjs},
         year = 2012,
        month = aug,
       volume = {201},
       number = {2},
          eid = {23},
        pages = {23},
          doi = {10.1088/0067-0049/201/2/23},
archivePrefix = {arXiv},
       eprint = {1106.2952},
 primaryClass = {astro-ph.CO},
       adsurl = {https://ui.adsabs.harvard.edu/abs/2012ApJS..201...23E}
}

@ARTICLE{2024ApJ...974..261C,
       author = {{Chen}, Yunfeng and {Yu}, Qingjuan and {Lu}, Youjun},
        title = "{Constraining the Origin of the Nanohertz Gravitational-wave Background by Pulsar Timing Array Observations of Both the Background and Individual Supermassive Binary Black Holes}",
      journal = {\apj},
         year = 2024,
        month = oct,
       volume = {974},
       number = {2},
          eid = {261},
        pages = {261},
          doi = {10.3847/1538-4357/ad7582},
archivePrefix = {arXiv},
       eprint = {2409.18029},
 primaryClass = {astro-ph.HE},
       adsurl = {https://ui.adsabs.harvard.edu/abs/2024ApJ...974..261C}
}

@ARTICLE{2018ApJ...856...42S,
       author = {{Sesana}, Alberto and {Haiman}, Zolt{\'a}n and {Kocsis}, Bence and {Kelley}, Luke Zoltan},
        title = "{Testing the Binary Hypothesis: Pulsar Timing Constraints on Supermassive Black Hole Binary Candidates}",
      journal = {\apj},
         year = 2018,
        month = mar,
       volume = {856},
       number = {1},
          eid = {42},
        pages = {42},
          doi = {10.3847/1538-4357/aaad0f},
archivePrefix = {arXiv},
       eprint = {1703.10611},
 primaryClass = {astro-ph.HE},
       adsurl = {https://ui.adsabs.harvard.edu/abs/2018ApJ...856...42S}
}

@ARTICLE{2016A&A...594A..13P,
       author = {{Planck Collaboration} and {Ade}, P.~A.~R. and {Aghanim}, N. and {Arnaud}, M. and {Ashdown}, M. and {Aumont}, J. and {Baccigalupi}, C. and {Banday}, A.~J. and {Barreiro}, R.~B. and {Bartlett}, J.~G. and {Bartolo}, N. and {Battaner}, E. and {Battye}, R. and {Benabed}, K. and {Beno{\^\i}t}, A. and {Benoit-L{\'e}vy}, A. and {Bernard}, J. -P. and {Bersanelli}, M. and {Bielewicz}, P. and {Bock}, J.~J. and {Bonaldi}, A. and {Bonavera}, L. and {Bond}, J.~R. and {Borrill}, J. and {Bouchet}, F.~R. and {Boulanger}, F. and {Bucher}, M. and {Burigana}, C. and {Butler}, R.~C. and {Calabrese}, E. and {Cardoso}, J. -F. and {Catalano}, A. and {Challinor}, A. and {Chamballu}, A. and {Chary}, R. -R. and {Chiang}, H.~C. and {Chluba}, J. and {Christensen}, P.~R. and {Church}, S. and {Clements}, D.~L. and {Colombi}, S. and {Colombo}, L.~P.~L. and {Combet}, C. and {Coulais}, A. and {Crill}, B.~P. and {Curto}, A. and {Cuttaia}, F. and {Danese}, L. and {Davies}, R.~D. and {Davis}, R.~J. and {de Bernardis}, P. and {de Rosa}, A. and {de Zotti}, G. and {Delabrouille}, J. and {D{\'e}sert}, F. -X. and {Di Valentino}, E. and {Dickinson}, C. and {Diego}, J.~M. and {Dolag}, K. and {Dole}, H. and {Donzelli}, S. and {Dor{\'e}}, O. and {Douspis}, M. and {Ducout}, A. and {Dunkley}, J. and {Dupac}, X. and {Efstathiou}, G. and {Elsner}, F. and {En{\ss}lin}, T.~A. and {Eriksen}, H.~K. and {Farhang}, M. and {Fergusson}, J. and {Finelli}, F. and {Forni}, O. and {Frailis}, M. and {Fraisse}, A.~A. and {Franceschi}, E. and {Frejsel}, A. and {Galeotta}, S. and {Galli}, S. and {Ganga}, K. and {Gauthier}, C. and {Gerbino}, M. and {Ghosh}, T. and {Giard}, M. and {Giraud-H{\'e}raud}, Y. and {Giusarma}, E. and {Gjerl{\o}w}, E. and {Gonz{\'a}lez-Nuevo}, J. and {G{\'o}rski}, K.~M. and {Gratton}, S. and {Gregorio}, A. and {Gruppuso}, A. and {Gudmundsson}, J.~E. and {Hamann}, J. and {Hansen}, F.~K. and {Hanson}, D. and {Harrison}, D.~L. and {Helou}, G. and {Henrot-Versill{\'e}}, S. and {Hern{\'a}ndez-Monteagudo}, C. and {Herranz}, D. and {Hildebrandt}, S.~R. and {Hivon}, E. and {Hobson}, M. and {Holmes}, W.~A. and {Hornstrup}, A. and {Hovest}, W. and {Huang}, Z. and {Huffenberger}, K.~M. and {Hurier}, G. and {Jaffe}, A.~H. and {Jaffe}, T.~R. and {Jones}, W.~C. and {Juvela}, M. and {Keih{\"a}nen}, E. and {Keskitalo}, R. and {Kisner}, T.~S. and {Kneissl}, R. and {Knoche}, J. and {Knox}, L. and {Kunz}, M. and {Kurki-Suonio}, H. and {Lagache}, G. and {L{\"a}hteenm{\"a}ki}, A. and {Lamarre}, J. -M. and {Lasenby}, A. and {Lattanzi}, M. and {Lawrence}, C.~R. and {Leahy}, J.~P. and {Leonardi}, R. and {Lesgourgues}, J. and {Levrier}, F. and {Lewis}, A. and {Liguori}, M. and {Lilje}, P.~B. and {Linden-V{\o}rnle}, M. and {L{\'o}pez-Caniego}, M. and {Lubin}, P.~M. and {Mac{\'\i}as-P{\'e}rez}, J.~F. and {Maggio}, G. and {Maino}, D. and {Mandolesi}, N. and {Mangilli}, A. and {Marchini}, A. and {Maris}, M. and {Martin}, P.~G. and {Martinelli}, M. and {Mart{\'\i}nez-Gonz{\'a}lez}, E. and {Masi}, S. and {Matarrese}, S. and {McGehee}, P. and {Meinhold}, P.~R. and {Melchiorri}, A. and {Melin}, J. -B. and {Mendes}, L. and {Mennella}, A. and {Migliaccio}, M. and {Millea}, M. and {Mitra}, S. and {Miville-Desch{\^e}nes}, M. -A. and {Moneti}, A. and {Montier}, L. and {Morgante}, G. and {Mortlock}, D. and {Moss}, A. and {Munshi}, D. and {Murphy}, J.~A. and {Naselsky}, P. and {Nati}, F. and {Natoli}, P. and {Netterfield}, C.~B. and {N{\o}rgaard-Nielsen}, H.~U. and {Noviello}, F. and {Novikov}, D. and {Novikov}, I. and {Oxborrow}, C.~A. and {Paci}, F. and {Pagano}, L. and {Pajot}, F. and {Paladini}, R. and {Paoletti}, D. and {Partridge}, B. and {Pasian}, F. and {Patanchon}, G. and {Pearson}, T.~J. and {Perdereau}, O. and {Perotto}, L. and {Perrotta}, F. and {Pettorino}, V. and {Piacentini}, F. and {Piat}, M. and {Pierpaoli}, E. and {Pietrobon}, D. and {Plaszczynski}, S. and {Pointecouteau}, E. and {Polenta}, G. and {Popa}, L. and {Pratt}, G.~W. and {Pr{\'e}zeau}, G.},
        title = "{Planck 2015 results. XIII. Cosmological parameters}",
      journal = {\aap},
         year = 2016,
        month = sep,
       volume = {594},
          eid = {A13},
        pages = {A13},
          doi = {10.1051/0004-6361/201525830},
archivePrefix = {arXiv},
       eprint = {1502.01589},
 primaryClass = {astro-ph.CO},
       adsurl = {https://ui.adsabs.harvard.edu/abs/2016A&A...594A..13P}
}

@ARTICLE{2025arXiv250506656F,
       author = {{Foustoul}, Vincent and {Webb}, Natalie A. and {Mignon-Risse}, Rapha{\"e}l and {Kammoun}, Elias and {Volonteri}, Marta and {Dong-P{\'a}ez}, Chi An},
        title = "{A catalogue of candidate milli-parsec separation massive black hole binaries from long term optical photometric monitoring}",
      journal = {arXiv e-prints},
         year = 2025,
        month = may,
          eid = {arXiv:2505.06656},
        pages = {arXiv:2505.06656},
          doi = {10.48550/arXiv.2505.06656},
archivePrefix = {arXiv},
       eprint = {2505.06656},
 primaryClass = {astro-ph.GA},
       adsurl = {https://ui.adsabs.harvard.edu/abs/2025arXiv250506656F}
}

@ARTICLE{2025arXiv250516884H,
       author = {{Huijse}, Pablo and {Davelaar}, Jordy and {De Ridder}, Joris and {Jannsen}, Nicholas and {Aerts}, Conny},
        title = "{Periodic Variability in Space Photometry of 181 New Supermassive Black Hole Binary Candidates}",
      journal = {arXiv e-prints},
         year = 2025,
        month = may,
          eid = {arXiv:2505.16884},
        pages = {arXiv:2505.16884},
          doi = {10.48550/arXiv.2505.16884},
archivePrefix = {arXiv},
       eprint = {2505.16884},
 primaryClass = {astro-ph.HE},
       adsurl = {https://ui.adsabs.harvard.edu/abs/2025arXiv250516884H}
}

@ARTICLE{2023A&A...677A.184W,
       author = {{Weaver}, J.~R. and {Davidzon}, I. and {Toft}, S. and {Ilbert}, O. and {McCracken}, H.~J. and {Gould}, K.~M.~L. and {Jespersen}, C.~K. and {Steinhardt}, C. and {Lagos}, C.~D.~P. and {Capak}, P.~L. and {Casey}, C.~M. and {Chartab}, N. and {Faisst}, A.~L. and {Hayward}, C.~C. and {Kartaltepe}, J.~S. and {Kauffmann}, O.~B. and {Koekemoer}, A.~M. and {Kokorev}, V. and {Laigle}, C. and {Liu}, D. and {Long}, A. and {Magdis}, G.~E. and {McPartland}, C.~J.~R. and {Milvang-Jensen}, B. and {Mobasher}, B. and {Moneti}, A. and {Peng}, Y. and {Sanders}, D.~B. and {Shuntov}, M. and {Sneppen}, A. and {Valentino}, F. and {Zalesky}, L. and {Zamorani}, G.},
        title = "{COSMOS2020: The galaxy stellar mass function. The assembly and star formation cessation of galaxies at 0.2< z {\ensuremath{\leq}} 7.5}",
      journal = {\aap},
         year = 2023,
        month = sep,
       volume = {677},
          eid = {A184},
        pages = {A184},
          doi = {10.1051/0004-6361/202245581},
archivePrefix = {arXiv},
       eprint = {2212.02512},
 primaryClass = {astro-ph.GA},
       adsurl = {https://ui.adsabs.harvard.edu/abs/2023A&A...677A.184W}
}

@ARTICLE{2007ApJ...658..884C,
       author = {{Choi}, Yun-Young and {Park}, Changbom and {Vogeley}, Michael S.},
        title = "{Internal and Collective Properties of Galaxies in the Sloan Digital Sky Survey}",
      journal = {\apj},
         year = 2007,
        month = apr,
       volume = {658},
       number = {2},
        pages = {884-897},
          doi = {10.1086/511060},
archivePrefix = {arXiv},
       eprint = {astro-ph/0611607},
 primaryClass = {astro-ph},
       adsurl = {https://ui.adsabs.harvard.edu/abs/2007ApJ...658..884C}
}

@ARTICLE{2013JCAP...10..022P,
       author = {{Pi{\'o}rkowska}, Aleksandra and {Biesiada}, Marek and {Zhu}, Zong-Hong},
        title = "{Strong gravitational lensing of gravitational waves in Einstein Telescope}",
      journal = {\jcap},
         year = 2013,
        month = oct,
       volume = {2013},
       number = {10},
          eid = {022},
        pages = {022},
          doi = {10.1088/1475-7516/2013/10/022},
archivePrefix = {arXiv},
       eprint = {1309.5731},
 primaryClass = {astro-ph.CO},
       adsurl = {https://ui.adsabs.harvard.edu/abs/2013JCAP...10..022P}
}

@ARTICLE{2024ApJ...977...64C,
       author = {{Chen}, Zhiwei and {Xie}, Yushan and {Lu}, Youjun and {Shan}, Huanyuan and {Li}, Nan and {Luo}, Yuchao and {Guo}, Xiao},
        title = "{Detection Rate of Galaxy Cluster-lensed Stellar Binary Black Hole Mergers by the Third-generation Gravitational-wave Detectors}",
      journal = {\apj},
         year = 2024,
        month = dec,
       volume = {977},
       number = {1},
          eid = {64},
        pages = {64},
          doi = {10.3847/1538-4357/ad8b44},
archivePrefix = {arXiv},
       eprint = {2410.20701},
 primaryClass = {astro-ph.HE},
       adsurl = {https://ui.adsabs.harvard.edu/abs/2024ApJ...977...64C}
}

@ARTICLE{2022ApJ...940...17C,
       author = {{Chen}, Zhiwei and {Lu}, Youjun and {Zhao}, Yuetong},
        title = "{Constraining the Origin of Stellar Binary Black Hole Mergers by Detections of Their Lensed Host Galaxies and Gravitational Wave Signals}",
      journal = {\apj},
         year = 2022,
        month = nov,
       volume = {940},
       number = {1},
          eid = {17},
        pages = {17},
          doi = {10.3847/1538-4357/ac98b7},
archivePrefix = {arXiv},
       eprint = {2210.09892},
 primaryClass = {astro-ph.HE},
       adsurl = {https://ui.adsabs.harvard.edu/abs/2022ApJ...940...17C}
}

@ARTICLE{2015ApJ...811...20C,
       author = {{Collett}, Thomas E.},
        title = "{The Population of Galaxy-Galaxy Strong Lenses in Forthcoming Optical Imaging Surveys}",
      journal = {\apj},
         year = 2015,
        month = sep,
       volume = {811},
       number = {1},
          eid = {20},
        pages = {20},
          doi = {10.1088/0004-637X/811/1/20},
archivePrefix = {arXiv},
       eprint = {1507.02657},
 primaryClass = {astro-ph.CO},
       adsurl = {https://ui.adsabs.harvard.edu/abs/2015ApJ...811...20C}
}

@ARTICLE{2018PDU....22..189B,
       author = {{Birrer}, Simon and {Amara}, Adam},
        title = "{lenstronomy: Multi-purpose gravitational lens modelling software package}",
      journal = {Physics of the Dark Universe},
         year = 2018,
        month = dec,
       volume = {22},
        pages = {189-201},
          doi = {10.1016/j.dark.2018.11.002},
archivePrefix = {arXiv},
       eprint = {1803.09746},
 primaryClass = {astro-ph.CO},
       adsurl = {https://ui.adsabs.harvard.edu/abs/2018PDU....22..189B}
}

@ARTICLE{2024MNRAS.533.1960C,
       author = {{Cao}, Xiaoyue and {Li}, Ran and {Li}, Nan and {Li}, Rui and {Chen}, Yun and {Ding}, Keyi and {Shan}, Huanyuan and {Zhan}, Hu and {Zhang}, Xin and {Du}, Wei and {Cao}, Shuo},
        title = "{CSST strong lensing preparation: forecasting the galaxy-galaxy strong lensing population for the China space station telescope}",
      journal = {\mnras},
         year = 2024,
        month = sep,
       volume = {533},
       number = {2},
        pages = {1960-1975},
          doi = {10.1093/mnras/stae1865},
archivePrefix = {arXiv},
       eprint = {2312.06239},
 primaryClass = {astro-ph.GA},
       adsurl = {https://ui.adsabs.harvard.edu/abs/2024MNRAS.533.1960C}
}

@ARTICLE{2025arXiv250821510C,
       author = {{Chiesa}, Alfredo and {Izquierdo-Villalba}, David and {Sesana}, Alberto and {Cocchiararo}, Fabiola and {Franchini}, Alessia and {Lupi}, Alessandro and {Spinoso}, Daniele and {Bonoli}, Silvia},
        title = "{Identifying massive black hole binaries via light curve variability in optical time-domain surveys}",
      journal = {arXiv e-prints},
         year = 2025,
        month = aug,
          eid = {arXiv:2508.21510},
        pages = {arXiv:2508.21510},
          doi = {10.48550/arXiv.2508.21510},
archivePrefix = {arXiv},
       eprint = {2508.21510},
 primaryClass = {astro-ph.HE},
       adsurl = {https://ui.adsabs.harvard.edu/abs/2025arXiv250821510C}
}

@ARTICLE{2025arXiv250910601E,
       author = {{El-Badry}, Kareem and {Hogg}, David W. and {Rix}, Hans-Walter},
        title = "{Active galactic nuclei do not exhibit stably periodic brightness variations}",
      journal = {arXiv e-prints},
         year = 2025,
        month = sep,
          eid = {arXiv:2509.10601},
        pages = {arXiv:2509.10601},
          doi = {10.48550/arXiv.2509.10601},
archivePrefix = {arXiv},
       eprint = {2509.10601},
 primaryClass = {astro-ph.GA},
       adsurl = {https://ui.adsabs.harvard.edu/abs/2025arXiv250910601E}
}

@ARTICLE{2018MNRAS.474.2975D,
       author = {{D'Orazio}, Daniel J. and {Di Stefano}, Rosanne},
        title = "{Periodic self-lensing from accreting massive black hole binaries}",
      journal = {\mnras},
         year = 2018,
        month = mar,
       volume = {474},
       number = {3},
        pages = {2975-2986},
          doi = {10.1093/mnras/stx2936},
archivePrefix = {arXiv},
       eprint = {1707.02335},
 primaryClass = {astro-ph.HE},
       adsurl = {https://ui.adsabs.harvard.edu/abs/2018MNRAS.474.2975D}
}

@ARTICLE{2010MNRAS.405.2579O,
       author = {{Oguri}, Masamune and {Marshall}, Philip J.},
        title = "{Gravitationally lensed quasars and supernovae in future wide-field optical imaging surveys}",
      journal = {\mnras},
         year = 2010,
        month = jul,
       volume = {405},
       number = {4},
        pages = {2579-2593},
          doi = {10.1111/j.1365-2966.2010.16639.x},
archivePrefix = {arXiv},
       eprint = {1001.2037},
 primaryClass = {astro-ph.CO},
       adsurl = {https://ui.adsabs.harvard.edu/abs/2010MNRAS.405.2579O}
}

@ARTICLE{2006ApJ...648..523W,
       author = {{Watarai}, Ken-ya},
        title = "{New Analytical Formulae for Supercritical Accretion Flows}",
      journal = {\apj},
         year = 2006,
        month = sep,
       volume = {648},
       number = {1},
        pages = {523-533},
          doi = {10.1086/505854},
archivePrefix = {arXiv},
       eprint = {astro-ph/0605248},
 primaryClass = {astro-ph},
       adsurl = {https://ui.adsabs.harvard.edu/abs/2006ApJ...648..523W}
}

@ARTICLE{2017MNRAS.467..898Q,
       author = {{Qiao}, Erlin and {Liu}, B.~F.},
        title = "{The condensation of the corona for the correlation between the hard X-ray photon index {\ensuremath{\Gamma}} and the reflection scaling factor ℜ in active galactic nuclei}",
      journal = {\mnras},
         year = 2017,
        month = may,
       volume = {467},
       number = {1},
        pages = {898-905},
          doi = {10.1093/mnras/stx121},
archivePrefix = {arXiv},
       eprint = {1701.04211},
 primaryClass = {astro-ph.HE},
       adsurl = {https://ui.adsabs.harvard.edu/abs/2017MNRAS.467..898Q}
}

@ARTICLE{2020A&A...640A.105M,
       author = {{Millon}, M. and {Courbin}, F. and {Bonvin}, V. and {Paic}, E. and {Meylan}, G. and {Tewes}, M. and {Sluse}, D. and {Magain}, P. and {Chan}, J.~H.~H. and {Galan}, A. and {Joseph}, R. and {Lemon}, C. and {Tihhonova}, O. and {Anderson}, R.~I. and {Marmier}, M. and {Chazelas}, B. and {Lendl}, M. and {Triaud}, A.~H.~M.~J. and {Wyttenbach}, A.},
        title = "{COSMOGRAIL. XIX. Time delays in 18 strongly lensed quasars from 15 years of optical monitoring}",
      journal = {\aap},
         year = 2020,
        month = aug,
       volume = {640},
          eid = {A105},
        pages = {A105},
          doi = {10.1051/0004-6361/202037740},
archivePrefix = {arXiv},
       eprint = {2002.05736},
 primaryClass = {astro-ph.CO},
       adsurl = {https://ui.adsabs.harvard.edu/abs/2020A&A...640A.105M}
}

@ARTICLE{2023MNRAS.526.1588Z,
       author = {{Zhang}, XueGuang},
        title = "{A sub-pc BBH system in SDSS J1609+1756 through optical QPOs in ZTF light curves}",
      journal = {\mnras},
         year = 2023,
        month = nov,
       volume = {526},
       number = {1},
        pages = {1588-1600},
          doi = {10.1093/mnras/stad2848},
archivePrefix = {arXiv},
       eprint = {2309.08078},
 primaryClass = {astro-ph.GA},
       adsurl = {https://ui.adsabs.harvard.edu/abs/2023MNRAS.526.1588Z}
}

@ARTICLE{2011ApJ...738...96M,
       author = {{Mosquera}, A.~M. and {Kochanek}, C.~S.},
        title = "{The Microlensing Properties of a Sample of 87 Lensed Quasars}",
      journal = {\apj},
         year = 2011,
        month = sep,
       volume = {738},
       number = {1},
          eid = {96},
        pages = {96},
          doi = {10.1088/0004-637X/738/1/96},
archivePrefix = {arXiv},
       eprint = {1104.2356},
 primaryClass = {astro-ph.CO},
       adsurl = {https://ui.adsabs.harvard.edu/abs/2011ApJ...738...96M}
}

@ARTICLE{1986ApJ...301..503P,
       author = {{Paczynski}, B.},
        title = "{Gravitational Microlensing at Large Optical Depth}",
      journal = {\apj},
         year = 1986,
        month = feb,
       volume = {301},
        pages = {503},
          doi = {10.1086/163919},
       adsurl = {https://ui.adsabs.harvard.edu/abs/1986ApJ...301..503P}
}

@ARTICLE{2024SSRv..220...14V,
       author = {{Vernardos}, G. and {Sluse}, D. and {Pooley}, D. and {Schmidt}, R.~W. and {Millon}, M. and {Weisenbach}, L. and {Motta}, V. and {Anguita}, T. and {Saha}, P. and {O'Dowd}, M. and {Peel}, A. and {Schechter}, P.~L.},
        title = "{Microlensing of Strongly Lensed Quasars}",
      journal = {\ssr},
         year = 2024,
        month = feb,
       volume = {220},
       number = {1},
          eid = {14},
        pages = {14},
          doi = {10.1007/s11214-024-01043-8},
archivePrefix = {arXiv},
       eprint = {2312.00931},
 primaryClass = {astro-ph.GA},
       adsurl = {https://ui.adsabs.harvard.edu/abs/2024SSRv..220...14V}
}

@ARTICLE{2012A&A...544A..62S,
       author = {{Sluse}, D. and {Hutsem{\'e}kers}, D. and {Courbin}, F. and {Meylan}, G. and {Wambsganss}, J.},
        title = "{Microlensing of the broad line region in 17 lensed quasars}",
      journal = {\aap},
         year = 2012,
        month = aug,
       volume = {544},
          eid = {A62},
        pages = {A62},
          doi = {10.1051/0004-6361/201219125},
archivePrefix = {arXiv},
       eprint = {1206.0731},
 primaryClass = {astro-ph.CO},
       adsurl = {https://ui.adsabs.harvard.edu/abs/2012A&A...544A..62S}
}

@ARTICLE{1990ApJ...358L..33W,
       author = {{Wambsganss}, J. and {Paczynski}, B. and {Schneider}, P.},
        title = "{Interpretation of the Microlensing Event in QSO 2237+0305}",
      journal = {\apjl},
         year = 1990,
        month = aug,
       volume = {358},
        pages = {L33},
          doi = {10.1086/185773},
       adsurl = {https://ui.adsabs.harvard.edu/abs/1990ApJ...358L..33W}
}

@ARTICLE{2004ApJ...605...58K,
       author = {{Kochanek}, C.~S.},
        title = "{Quantitative Interpretation of Quasar Microlensing Light Curves}",
      journal = {\apj},
         year = 2004,
        month = apr,
       volume = {605},
       number = {1},
        pages = {58-77},
          doi = {10.1086/382180},
archivePrefix = {arXiv},
       eprint = {astro-ph/0307422},
 primaryClass = {astro-ph},
       adsurl = {https://ui.adsabs.harvard.edu/abs/2004ApJ...605...58K}
}

@ARTICLE{2006MNRAS.367.1241O,
       author = {{Oguri}, Masamune},
        title = "{The image separation distribution of strong lenses: halo versus subhalo populations}",
      journal = {\mnras},
         year = 2006,
        month = apr,
       volume = {367},
       number = {3},
        pages = {1241-1250},
          doi = {10.1111/j.1365-2966.2006.10043.x},
archivePrefix = {arXiv},
       eprint = {astro-ph/0508528},
 primaryClass = {astro-ph},
       adsurl = {https://ui.adsabs.harvard.edu/abs/2006MNRAS.367.1241O}
}

@ARTICLE{2024SSRv..220...23L,
       author = {{Lemon}, Cameron and {Courbin}, Fr{\'e}d{\'e}ric and {More}, Anupreeta and {Schechter}, Paul and {Ca{\~n}ameras}, Raoul and {Delchambre}, Ludovic and {Leung}, Calvin and {Shu}, Yiping and {Spiniello}, Chiara and {Hezaveh}, Yashar and {Kl{\"u}ter}, Jonas and {McMahon}, Richard},
        title = "{Searching for Strong Gravitational Lenses}",
      journal = {\ssr},
         year = 2024,
        month = feb,
       volume = {220},
       number = {2},
          eid = {23},
        pages = {23},
          doi = {10.1007/s11214-024-01042-9},
archivePrefix = {arXiv},
       eprint = {2310.13466},
 primaryClass = {astro-ph.GA},
       adsurl = {https://ui.adsabs.harvard.edu/abs/2024SSRv..220...23L}
}

@ARTICLE{2020MNRAS.495.3727R,
       author = {{Robertson}, Andrew and {Smith}, Graham P. and {Massey}, Richard and {Eke}, Vincent and {Jauzac}, Mathilde and {Bianconi}, Matteo and {Ryczanowski}, Dan},
        title = "{What does strong gravitational lensing? The mass and redshift distribution of high-magnification lenses}",
      journal = {\mnras},
         year = 2020,
        month = jul,
       volume = {495},
       number = {4},
        pages = {3727-3739},
          doi = {10.1093/mnras/staa1429},
archivePrefix = {arXiv},
       eprint = {2002.01479},
 primaryClass = {astro-ph.CO},
       adsurl = {https://ui.adsabs.harvard.edu/abs/2020MNRAS.495.3727R}
}

@ARTICLE{2012ApJ...749...38M,
       author = {{More}, A. and {Cabanac}, R. and {More}, S. and {Alard}, C. and {Limousin}, M. and {Kneib}, J.-P. and {Gavazzi}, R. and {Motta}, V.},
        title = "{The CFHTLS-Strong Lensing Legacy Survey (SL2S): Investigating the Group-scale Lenses with the SARCS Sample}",
      journal = {\apj},
         year = 2012,
        month = apr,
       volume = {749},
       number = {1},
          eid = {38},
        pages = {38},
          doi = {10.1088/0004-637X/749/1/38},
archivePrefix = {arXiv},
       eprint = {1109.1821},
 primaryClass = {astro-ph.CO},
       adsurl = {https://ui.adsabs.harvard.edu/abs/2012ApJ...749...38M}
}

@ARTICLE{2011ApJ...737..103S,
       author = {{Schlafly}, Edward F. and {Finkbeiner}, Douglas P.},
        title = "{Measuring Reddening with Sloan Digital Sky Survey Stellar Spectra and Recalibrating SFD}",
      journal = {\apj},
         year = 2011,
        month = aug,
       volume = {737},
       number = {2},
          eid = {103},
        pages = {103},
          doi = {10.1088/0004-637X/737/2/103},
archivePrefix = {arXiv},
       eprint = {1012.4804},
 primaryClass = {astro-ph.GA},
       adsurl = {https://ui.adsabs.harvard.edu/abs/2011ApJ...737..103S}
}

@ARTICLE{1999ApJ...523..617F,
       author = {{Falco}, E.~E. and {Impey}, C.~D. and {Kochanek}, C.~S. and {Leh{\'a}r}, J. and {McLeod}, B.~A. and {Rix}, H.-W. and {Keeton}, C.~R. and {Mu{\~n}oz}, J.~A. and {Peng}, C.~Y.},
        title = "{Dust and Extinction Curves in Galaxies with z>0: The Interstellar Medium of Gravitational Lens Galaxies}",
      journal = {\apj},
         year = 1999,
        month = oct,
       volume = {523},
       number = {2},
        pages = {617-632},
          doi = {10.1086/307758},
archivePrefix = {arXiv},
       eprint = {astro-ph/9901037},
 primaryClass = {astro-ph},
       adsurl = {https://ui.adsabs.harvard.edu/abs/1999ApJ...523..617F}
}

@ARTICLE{2006ApJS..166..443E,
       author = {{El{\'\i}asd{\'o}ttir}, {\'A}. and {Hjorth}, J. and {Toft}, S. and {Burud}, I. and {Paraficz}, D.},
        title = "{Extinction Curves of Lensing Galaxies out to z = 1}",
      journal = {\apjs},
         year = 2006,
        month = oct,
       volume = {166},
       number = {2},
        pages = {443-469},
          doi = {10.1086/507131},
archivePrefix = {arXiv},
       eprint = {astro-ph/0606649},
 primaryClass = {astro-ph},
       adsurl = {https://ui.adsabs.harvard.edu/abs/2006ApJS..166..443E}
}

@ARTICLE{2021A&A...646A..83R,
       author = {{Richard}, Johan and {Claeyssens}, Ad{\'e}la{\"\i}de and {Lagattuta}, David and {Guaita}, Lucia and {Bauer}, Franz Erik and {Pello}, Roser and {Carton}, David and {Bacon}, Roland and {Soucail}, Genevi{\`e}ve and {Lyon}, Gonzalo Prieto and {Kneib}, Jean-Paul and {Mahler}, Guillaume and {Cl{\'e}ment}, Benjamin and {Mercier}, Wilfried and {Variu}, Andrei and {Tamone}, Am{\'e}lie and {Ebeling}, Harald and {Schmidt}, Kasper B. and {Nanayakkara}, Themiya and {Maseda}, Michael and {Weilbacher}, Peter M. and {Bouch{\'e}}, Nicolas and {Bouwens}, Rychard J. and {Wisotzki}, Lutz and {de la Vieuville}, Geoffroy and {Martinez}, Johany and {Patr{\'\i}cio}, Vera},
        title = "{An atlas of MUSE observations towards twelve massive lensing clusters}",
      journal = {\aap},
         year = 2021,
        month = feb,
       volume = {646},
          eid = {A83},
        pages = {A83},
          doi = {10.1051/0004-6361/202039462},
archivePrefix = {arXiv},
       eprint = {2009.09784},
 primaryClass = {astro-ph.GA},
       adsurl = {https://ui.adsabs.harvard.edu/abs/2021A&A...646A..83R}
}

@ARTICLE{2022A&A...662A.112E,
       author = {{Euclid Collaboration} and {Scaramella}, R. and {Amiaux}, J. and {Mellier}, Y. and {Burigana}, C. and {Carvalho}, C.~S. and {Cuillandre}, J.-C. and {Da Silva}, A. and {Derosa}, A. and {Dinis}, J. and {Maiorano}, E. and {Maris}, M. and {Tereno}, I. and {Laureijs}, R. and {Boenke}, T. and {Buenadicha}, G. and {Dupac}, X. and {Gaspar Venancio}, L.~M. and {G{\'o}mez-{\'A}lvarez}, P. and {Hoar}, J. and {Lorenzo Alvarez}, J. and {Racca}, G.~D. and {Saavedra-Criado}, G. and {Schwartz}, J. and {Vavrek}, R. and {Schirmer}, M. and {Aussel}, H. and {Azzollini}, R. and {Cardone}, V.~F. and {Cropper}, M. and {Ealet}, A. and {Garilli}, B. and {Gillard}, W. and {Granett}, B.~R. and {Guzzo}, L. and {Hoekstra}, H. and {Jahnke}, K. and {Kitching}, T. and {Maciaszek}, T. and {Meneghetti}, M. and {Miller}, L. and {Nakajima}, R. and {Niemi}, S.~M. and {Pasian}, F. and {Percival}, W.~J. and {Pottinger}, S. and {Sauvage}, M. and {Scodeggio}, M. and {Wachter}, S. and {Zacchei}, A. and {Aghanim}, N. and {Amara}, A. and {Auphan}, T. and {Auricchio}, N. and {Awan}, S. and {Balestra}, A. and {Bender}, R. and {Bodendorf}, C. and {Bonino}, D. and {Branchini}, E. and {Brau-Nogue}, S. and {Brescia}, M. and {Candini}, G.~P. and {Capobianco}, V. and {Carbone}, C. and {Carlberg}, R.~G. and {Carretero}, J. and {Casas}, R. and {Castander}, F.~J. and {Castellano}, M. and {Cavuoti}, S. and {Cimatti}, A. and {Cledassou}, R. and {Congedo}, G. and {Conselice}, C.~J. and {Conversi}, L. and {Copin}, Y. and {Corcione}, L. and {Costille}, A. and {Courbin}, F. and {Degaudenzi}, H. and {Douspis}, M. and {Dubath}, F. and {Duncan}, C.~A.~J. and {Dusini}, S. and {Farrens}, S. and {Ferriol}, S. and {Fosalba}, P. and {Fourmanoit}, N. and {Frailis}, M. and {Franceschi}, E. and {Franzetti}, P. and {Fumana}, M. and {Gillis}, B. and {Giocoli}, C. and {Grazian}, A. and {Grupp}, F. and {Haugan}, S.~V.~H. and {Holmes}, W. and {Hormuth}, F. and {Hudelot}, P. and {Kermiche}, S. and {Kiessling}, A. and {Kilbinger}, M. and {Kohley}, R. and {Kubik}, B. and {K{\"u}mmel}, M. and {Kunz}, M. and {Kurki-Suonio}, H. and {Lahav}, O. and {Ligori}, S. and {Lilje}, P.~B. and {Lloro}, I. and {Mansutti}, O. and {Marggraf}, O. and {Markovic}, K. and {Marulli}, F. and {Massey}, R. and {Maurogordato}, S. and {Melchior}, M. and {Merlin}, E. and {Meylan}, G. and {Mohr}, J.~J. and {Moresco}, M. and {Morin}, B. and {Moscardini}, L. and {Munari}, E. and {Nichol}, R.~C. and {Padilla}, C. and {Paltani}, S. and {Peacock}, J. and {Pedersen}, K. and {Pettorino}, V. and {Pires}, S. and {Poncet}, M. and {Popa}, L. and {Pozzetti}, L. and {Raison}, F. and {Rebolo}, R. and {Rhodes}, J. and {Rix}, H.-W. and {Roncarelli}, M. and {Rossetti}, E. and {Saglia}, R. and {Schneider}, P. and {Schrabback}, T. and {Secroun}, A. and {Seidel}, G. and {Serrano}, S. and {Sirignano}, C. and {Sirri}, G. and {Skottfelt}, J. and {Stanco}, L. and {Starck}, J.~L. and {Tallada-Cresp{\'\i}}, P. and {Tavagnacco}, D. and {Taylor}, A.~N. and {Teplitz}, H.~I. and {Toledo-Moreo}, R. and {Torradeflot}, F. and {Trifoglio}, M. and {Valentijn}, E.~A. and {Valenziano}, L. and {Verdoes Kleijn}, G.~A. and {Wang}, Y. and {Welikala}, N. and {Weller}, J. and {Wetzstein}, M. and {Zamorani}, G. and {Zoubian}, J. and {Andreon}, S. and {Baldi}, M. and {Bardelli}, S. and {Boucaud}, A. and {Camera}, S. and {Di Ferdinando}, D. and {Fabbian}, G. and {Farinelli}, R. and {Galeotta}, S. and {Graci{\'a}-Carpio}, J. and {Maino}, D. and {Medinaceli}, E. and {Mei}, S. and {Neissner}, C. and {Polenta}, G. and {Renzi}, A. and {Romelli}, E. and {Rosset}, C. and {Sureau}, F. and {Tenti}, M. and {Vassallo}, T. and {Zucca}, E. and {Baccigalupi}, C. and {Balaguera-Antol{\'\i}nez}, A. and {Battaglia}, P. and {Biviano}, A. and {Borgani}, S. and {Bozzo}, E. and {Cabanac}, R. and {Cappi}, A.},
        title = "{Euclid preparation. I. The Euclid Wide Survey}",
      journal = {\aap},
         year = 2022,
        month = jun,
       volume = {662},
          eid = {A112},
        pages = {A112},
          doi = {10.1051/0004-6361/202141938},
archivePrefix = {arXiv},
       eprint = {2108.01201},
 primaryClass = {astro-ph.CO},
       adsurl = {https://ui.adsabs.harvard.edu/abs/2022A&A...662A.112E}
}

@ARTICLE{2025CQGra..42k3001L,
       author = {{Lu}, Youjun},
        title = "{Supermassive binary black holes and current status of their multimessenger observations}",
      journal = {Classical and Quantum Gravity},
         year = 2025,
        month = jun,
       volume = {42},
       number = {11},
          eid = {113001},
        pages = {113001},
          doi = {10.1088/1361-6382/adc4b0},
       adsurl = {https://ui.adsabs.harvard.edu/abs/2025CQGra..42k3001L}
}

@ARTICLE{2021MNRAS.503.1319G,
       author = {{Geng}, Shuaibo and {Cao}, Shuo and {Liu}, Yuting and {Liu}, Tonghua and {Biesiada}, Marek and {Lian}, Yujie},
        title = "{The velocity dispersion function of early-type galaxies and its redshift evolution: the newest results from lens redshift test}",
      journal = {\mnras},
         year = 2021,
        month = may,
       volume = {503},
       number = {1},
        pages = {1319-1326},
          doi = {10.1093/mnras/stab519},
archivePrefix = {arXiv},
       eprint = {2102.12140},
 primaryClass = {astro-ph.GA},
       adsurl = {https://ui.adsabs.harvard.edu/abs/2021MNRAS.503.1319G}
}

@ARTICLE{2023MNRAS.520.4463L,
       author = {{Liao}, Shihong and {Johansson}, Peter H. and {Mannerkoski}, Matias and {Irodotou}, Dimitrios and {Rizzuto}, Francesco Paolo and {McAlpine}, Stuart and {Rantala}, Antti and {Rawlings}, Alexander and {Sawala}, Till},
        title = "{Modelling the accretion and feedback of supermassive black hole binaries in gas-rich galaxy mergers}",
      journal = {\mnras},
         year = 2023,
        month = apr,
       volume = {520},
       number = {3},
        pages = {4463-4489},
          doi = {10.1093/mnras/stad412},
archivePrefix = {arXiv},
       eprint = {2211.11788},
 primaryClass = {astro-ph.GA},
       adsurl = {https://ui.adsabs.harvard.edu/abs/2023MNRAS.520.4463L}
}

@ARTICLE{2023MNRAS.518.5059S,
       author = {{Siwek}, Magdalena and {Weinberger}, Rainer and {Mu{\~n}oz}, Diego J. and {Hernquist}, Lars},
        title = "{Preferential accretion and circumbinary disc precession in eccentric binary systems}",
      journal = {\mnras},
         year = 2023,
        month = feb,
       volume = {518},
       number = {4},
        pages = {5059-5071},
          doi = {10.1093/mnras/stac3263},
archivePrefix = {arXiv},
       eprint = {2203.02514},
 primaryClass = {astro-ph.HE},
       adsurl = {https://ui.adsabs.harvard.edu/abs/2023MNRAS.518.5059S}
}

@ARTICLE{2010ApJ...715L..30L,
       author = {{Liu}, Xin and {Greene}, Jenny E. and {Shen}, Yue and {Strauss}, Michael A.},
        title = "{Discovery of Four kpc-scale Binary Active Galactic Nuclei}",
      journal = {\apjl},
         year = 2010,
        month = may,
       volume = {715},
       number = {1},
        pages = {L30-L34},
          doi = {10.1088/2041-8205/715/1/L30},
archivePrefix = {arXiv},
       eprint = {1003.3467},
 primaryClass = {astro-ph.CO},
       adsurl = {https://ui.adsabs.harvard.edu/abs/2010ApJ...715L..30L}
}

@ARTICLE{2014ApJ...789..140L,
       author = {{Liu}, Xin and {Shen}, Yue and {Bian}, Fuyan and {Loeb}, Abraham and {Tremaine}, Scott},
        title = "{Constraining Sub-parsec Binary Supermassive Black Holes in Quasars with Multi-epoch Spectroscopy. II. The Population with Kinematically Offset Broad Balmer Emission Lines}",
      journal = {\apj},
         year = 2014,
        month = jul,
       volume = {789},
       number = {2},
          eid = {140},
        pages = {140},
          doi = {10.1088/0004-637X/789/2/140},
archivePrefix = {arXiv},
       eprint = {1312.6694},
 primaryClass = {astro-ph.CO},
       adsurl = {https://ui.adsabs.harvard.edu/abs/2014ApJ...789..140L}
}

@ARTICLE{2003ApJ...582L..15K,
       author = {{Komossa}, S. and {Burwitz}, V. and {Hasinger}, G. and {Predehl}, P. and {Kaastra}, J.~S. and {Ikebe}, Y.},
        title = "{Discovery of a Binary Active Galactic Nucleus in the Ultraluminous Infrared Galaxy NGC 6240 Using Chandra}",
      journal = {\apjl},
         year = 2003,
        month = jan,
       volume = {582},
       number = {1},
        pages = {L15-L19},
          doi = {10.1086/346145},
archivePrefix = {arXiv},
       eprint = {astro-ph/0212099},
 primaryClass = {astro-ph},
       adsurl = {https://ui.adsabs.harvard.edu/abs/2003ApJ...582L..15K}
}

@ARTICLE{2011ApJ...735...48S,
       author = {{Shen}, Yue and {Liu}, Xin and {Greene}, Jenny E. and {Strauss}, Michael A.},
        title = "{Type 2 Active Galactic Nuclei with Double-peaked [O III] Lines. II. Single AGNs with Complex Narrow-line Region Kinematics are More Common than Binary AGNs}",
      journal = {\apj},
         year = 2011,
        month = jul,
       volume = {735},
       number = {1},
          eid = {48},
        pages = {48},
          doi = {10.1088/0004-637X/735/1/48},
archivePrefix = {arXiv},
       eprint = {1011.5246},
 primaryClass = {astro-ph.CO},
       adsurl = {https://ui.adsabs.harvard.edu/abs/2011ApJ...735...48S}
}

@ARTICLE{2019BAAS...51c.341D,
       author = {{Dore}, Olivier and {Hirata}, C. and {Wang}, Y. and {Weinberg}, D. and {Eifler}, Tim and {Foley}, R.~J. and {Heinrich}, C. He and {Krause}, E. and {Perlmutter}, S. and {Pisani}, A. and {Scolnic}, D. and {Spergel}, D.~N. and {Suntzeff}, N. and {Aldering}, G. and {Baltay}, C. and {Capak}, P. and {Choi}, A. and {Dvorkin}, C. and {Fall}, S.~M. and {Fang}, X. and {Fruchter}, A. and {Galbany}, L. and {Ho}, S. and {Hounsell}, R. and {Izard}, A. and {Jain}, B. and {Koekemoer}, A.~M. and {Kruk}, J. and {Leauthaud}, A. and {Malhotra}, S. and {Mandelbaum}, R. and {Massara}, E. and {Masters}, D. and {Miyatake}, H. and {Plazas}, A. and {Rhoads}, J. and {Rhodes}, J. and {Rose}, B. and {Rubin}, D. and {Sako}, M. and {Samushia}, L. and {Shirasaki}, M. and {Simet}, M. and {Takada}, M. and {Troxel}, M.~A. and {Wu}, H. and {Yoshida}, N. and {Zhai}, Z.},
        title = "{WFIRST: The Essential Cosmology Space Observatory for the Coming Decade}",
      journal = {\baas},
         year = 2019,
        month = may,
       volume = {51},
       number = {3},
          eid = {341},
        pages = {341},
          doi = {10.48550/arXiv.1904.01174},
archivePrefix = {arXiv},
       eprint = {1904.01174},
 primaryClass = {astro-ph.CO},
       adsurl = {https://ui.adsabs.harvard.edu/abs/2019BAAS...51c.341D}
}

@ARTICLE{2026MNRAS.tmp..388Y,
       author = {{Yan}, Changshuo and {Lu}, Youjun and {Ge}, Junqian and {Qiao}, Erlin},
        title = "{An equal mass ratio supermassive binary black holes in Q J0158-4325 with periodic microlensing signature?}",
      journal = {\mnras},
         year = 2026,
        month = mar,
          doi = {10.1093/mnras/stag402},
       adsurl = {https://ui.adsabs.harvard.edu/abs/2026MNRAS.tmp..388Y}
}

@ARTICLE{1988ApJ...325..628S,
       author = {{Sillanpaa}, A. and {Haarala}, S. and {Valtonen}, M.~J. and {Sundelius}, B. and {Byrd}, G.~G.},
        title = "{OJ 287: Binary Pair of Supermassive Black Holes}",
      journal = {\apj},
         year = 1988,
        month = feb,
       volume = {325},
        pages = {628},
          doi = {10.1086/166033},
       adsurl = {https://ui.adsabs.harvard.edu/abs/1988ApJ...325..628S}
}

@ARTICLE{2008Natur.452..851V,
       author = {{Valtonen}, M.~J. and {Lehto}, H.~J. and {Nilsson}, K. and {Heidt}, J. and {Takalo}, L.~O. and {Sillanp{\"a}{\"a}}, A. and {Villforth}, C. and {Kidger}, M. and {Poyner}, G. and {Pursimo}, T. and {Zola}, S. and {Wu}, J.-H. and {Zhou}, X. and {Sadakane}, K. and {Drozdz}, M. and {Koziel}, D. and {Marchev}, D. and {Ogloza}, W. and {Porowski}, C. and {Siwak}, M. and {Stachowski}, G. and {Winiarski}, M. and {Hentunen}, V.-P. and {Nissinen}, M. and {Liakos}, A. and {Dogru}, S.},
        title = "{A massive binary black-hole system in OJ287 and a test of general relativity}",
      journal = {\nat},
         year = 2008,
        month = apr,
       volume = {452},
       number = {7189},
        pages = {851-853},
          doi = {10.1038/nature06896},
archivePrefix = {arXiv},
       eprint = {0809.1280},
 primaryClass = {astro-ph},
       adsurl = {https://ui.adsabs.harvard.edu/abs/2008Natur.452..851V}
}

@ARTICLE{2023Galax..11...71H,
       author = {{Hamolli}, Lindita and {Hafizi}, Mimoza and {De Paolis}, Francesco and {Guliqani}, Esmeralda},
        title = "{Investigating Gravitationally Lensed Quasars Observable by Nancy Grace Roman Space Telescope}",
      journal = {Galaxies},
         year = 2023,
        month = jun,
       volume = {11},
       number = {3},
          eid = {71},
        pages = {71},
          doi = {10.3390/galaxies11030071},
       adsurl = {https://ui.adsabs.harvard.edu/abs/2023Galax..11...71H}
}

\clearpage
\begin{appendix}

\section{Other lensing properties}
\label{app:lensing_properties}

\begin{figure}
\includegraphics[width=0.48\textwidth]{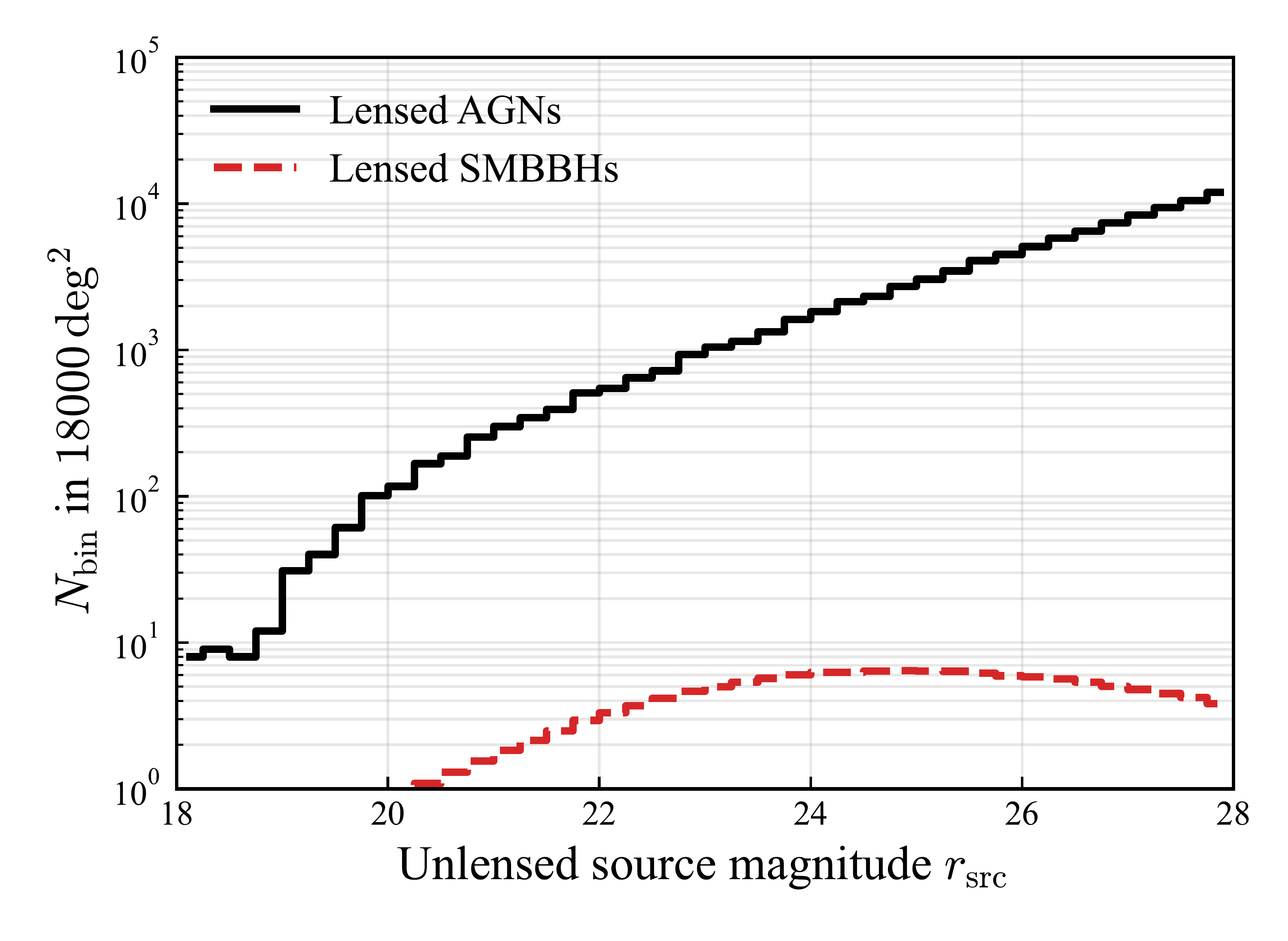}  
\caption{
Magnitude distribution of the AGN and close-SMBBH populations statistically expected to undergo strong lensing as a function of the unlensed source $r$-band magnitude. The vertical axis shows the number of systems per magnitude bin, with $\Delta r=0.25$ mag, in a survey area of $18000~{\rm deg}^2$. The lensed AGN population is computed from the adopted $r$-band QLF in Section~\ref{sec:lenBBHs} weighted by the strong-lensing optical depth, while the lensed close-SMBBH counts are averaged from 1000 Monte Carlo realizations and rescaled to the same survey area. No magnification bias or image-level selection is included in these distributions.
} 
\label{fig:app1}
\end{figure}

\begin{figure}
\includegraphics[width=0.48\textwidth]{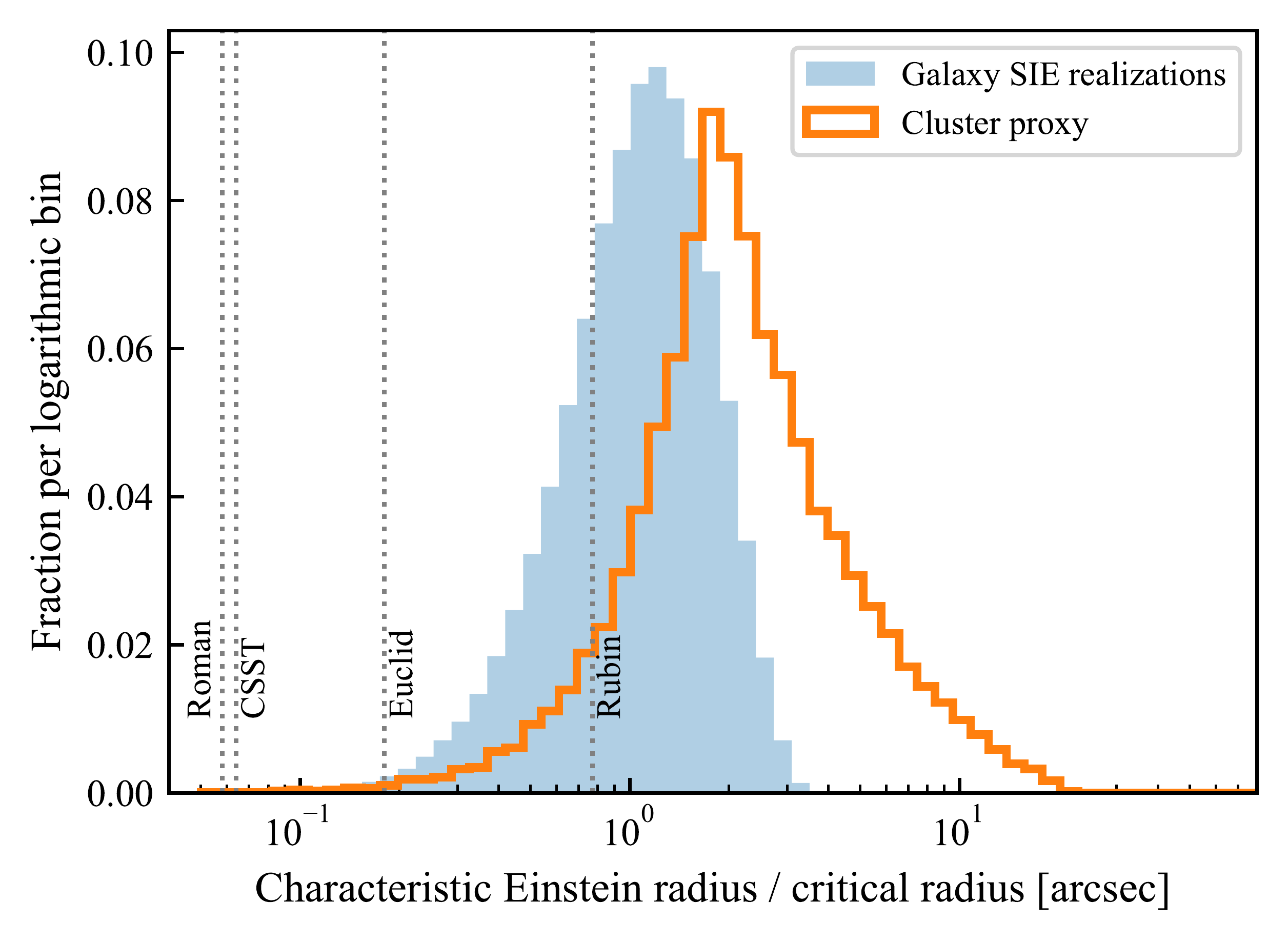}  
\caption{
Comparison of the characteristic angular scales of galaxy and cluster lenses. The galaxy distribution uses the Einstein radii of the SIE lenses in our mock strongly lensed SMBBH sample (Section~\ref{sec:lenBBHs}), while the cluster distribution is a proxy based on spherical NFW+SIS tangential critical radii sampled from the halo mass function (utilizing the Python package \textit{hmf} \cite{2013A&C.....3...23M}) at the lens and source redshifts of our statistical cluster-lensing sample, following Ref.~\cite{2024ApJ...977...64C}. Vertical dotted lines show the ${\rm FWHM}$ angular scales of the surveys considered in this work. Since the characteristic image separation of a double lens is approximately $2\theta_E$, the criterion $\Delta\theta_{\rm img}>2{\rm FWHM}$ corresponds approximately to $\theta_E>{\rm FWHM}$ here. The cluster curve is intended only to illustrate the characteristic angular scale of the statistical cluster-lensing contribution, not to represent a full image-level cluster-lens selection function.
}
\label{fig:app2}
\end{figure}

This appendix presents two supplementary lensing-related distributions used to support the statistical estimates in the main text. Figure~\ref{fig:app1} shows the magnitude distribution of the AGN/SMBBH populations statistically expected to undergo strong lensing, while Figure~\ref{fig:app2} provides an angular-scale validation for the galaxy/cluster lensing contribution.

\end{appendix}




\end{document}